\documentclass[preprint,12pt]{elsarticle}

\usepackage{amssymb}
\usepackage{amsmath}

\usepackage{subfigure}
\usepackage{multirow}
\usepackage{multicol}
\usepackage{subcaption}
\usepackage{url}
\usepackage[dvipsnames]{xcolor}
\usepackage{algorithm}
\usepackage{algpseudocode}

\algnewcommand\algorithmicforeach{\textbf{foreach}}
\algdef{S}[FOR]{ForEach}[1]{\algorithmicforeach\ #1\ \algorithmicdo}

\journal{Journal of Parallel and Distributed Computing}

\begin{document}

\begin{frontmatter}



\title{Near Real-time Adaptive Isotropic and Anisotropic Image-to-mesh Conversion for Numerical Simulations Involving Cerebral Aneurysms}


\author[ODU]{Kevin Garner
} 
\author[ODU]{Fotis Drakopoulos} 
\author[SBU]{Chander Sadasivan} 
\author[ODU]{Nikos Chrisochoides
\corref{corresponding-author}
} 

\affiliation[ODU]{organization={Center for Real-time Computing, Old Dominion University},
            addressline={4700 Elkhorn Ave}, 
            city={Norfolk},
            postcode={23529}, 
            state={VA},
            country={USA}}
\affiliation[SBU]{organization={Cerebrovascular Center for Research, Stony Brook University},
            addressline={101 Nicolls Rd}, 
            city={Stony Brook},
            postcode={11794}, 
            state={NY},
            country={USA}}

\cortext[corresponding-author]{Corresponding author. \textit{E-mail address:} nikos@cs.odu.edu (N. Chrisochoides). Phone number: 1-757-753-3242 (N. Chrisochoides)}

\begin{abstract}
Presented are two techniques that are designed to help streamline the discretization of complex vascular geometries within the numerical modeling process. The first method integrates multiple software tools into a single pipeline which can generate adaptive anisotropic meshes from segmented medical images. The pipeline is shown to satisfy quality, fidelity, smoothness, and robustness requirements while providing near real-time performance for medical image-to-mesh conversion. The second method approximates a user-defined sizing function to generate adaptive isotropic meshes of good quality and fidelity in real-time. Tested with two brain aneurysm cases and utilizing up to 96 CPU cores within a single, multicore node on Purdue University's Anvil supercomputer, the parallel adaptive anisotropic meshing method utilizes a hierarchical load balancing model (designed for large, cc-NUMA shared memory architectures) and contains an optimized local reconnection operation that performs three times faster than its original implementation from previous studies. The adaptive isotropic method is shown to generate a mesh of up to approximately 50 million elements in less than a minute while the adaptive anisotropic method is shown to generate approximately the same number of elements in about 5 minutes.
\end{abstract}



\begin{keyword}
image-to-mesh conversion \sep high-performance computing \sep Anvil supercomputer \sep aneurysms \sep numerical simulations



\end{keyword}

\end{frontmatter}



\section{Introduction}
\label{introduction}
Interventional radiology involves minimally invasive, catheter-based, deployment of medical devices to treat several vascular diseases throughout the body. This field has seen an explosion of device technology over the past decade ranging from stent-grafts for aortic aneurysms to valve replacement for diseased cardiac valves. Neuro-interventional radiology, which deals with diseases of the cerebral vasculature, has expanded from the treatment of brain aneurysms to removal of blood clots in large vessel occlusion strokes to embolization of the middle meningeal artery for treatment of chronic subdural hematomas to venous sinus stenting for treatment of idiopathic intracranial hypertension and pulsatile tinnitus. As hemodynamics are an intrinsic factor of vascular disease, there has been a concomitant increase in the number of studies using Computational Fluid Dynamics (CFD) and/or fluid-structure interactions to better understand disease prognosis or treatment outcomes after deployment of the pertinent medical device. For example, the transport of cardiogenic emboli causing ischemic stroke can be evaluated using imaging scans of the entire vascular branching network in each patient \cite{RoleOfCircle2018}. The recent introduction of miniaturized optical coherence tomography allows for high-resolution imaging of medical device features as well as the vascular wall in any given patient \cite{VolumetricMicroscopy2024}, which in turn facilitates high-resolution computational studies of the pathology. CFD may be able to provide clarity in the device treatment of cerebral venous disease, which has irregular morphologies and complex flow patterns \cite{SoVein2023}. 

Brain aneurysms are abnormal focal dilations of intracranial arteries which when left untreated may rupture, resulting in fatal outcomes for patients. Aneurysms have been treated by neuro-interventional devices for nearly three decades, and hence, numerous numerical modeling studies have been conducted on this disease. High resolution CFD simulations provide the opportunity to study hemodynamics in arteries, which can assist clinicians in determining an aneurysm's potential to rupture and possible methods of treatment (such as the design of neurovascular stents) \cite{CFDAneurysmRupture2019, HemodynamicReview2019, EfficientSimulation2005, ComputationalVascular2010, OptimizationOfCardiovascular2012, CBC3DYixun}. Brain aneurysms that form at vascular bifurcations may be subjected to flow instabilities that can only be captured by high-fidelity CFD \cite{OnThePrevalence2021}. Stents used to treat brain aneurysms have wire sizes that can be as small as 25-30 microns with aneurysm sizes ranging from millimeters to centimeters. CFD studies on the effect of such devices (including overlapping devices) on intraneurysmal hemodynamics involve generation of volumetric meshes with elements at multiple sizes \cite{CompactingSingle2017}. Processing patient-specific medical imaging data of cerebral aneurysms involves the discretization of these complex bodies in the form of smoothed free-form anisotropic body-conforming meshes. Mesh generation methods utilized in commercially available products work well when processing patient-specific cerebral aneurysm geometries; however, these methods do not always satisfy the time constraints of high resolution CFD simulations.

These types of medical simulations can be classified into two categories: interactive and predictive. Interactive simulations require real-time mesh generation which can be used to allow surgical residents to practice performing a procedure within a virtual environment. Real-time mesh generation is also required for image registration applications, as they provide guidance to surgeons during operations \cite{ComparisonPhysicsBased2023}. Predictive simulations predict and optimize the outcome of an intervention using patient-specific pre-operative image data. Although predictive simulations don't necessarily require real-time mesh generation, both types of simulations are time-sensitive given their impact on a patient's treatment and associated costs.

This paper presents two techniques that are designed to help streamline the discretization of complex vascular geometries 
within the numerical modeling process. Several requirements should be addressed with regards to medical image-to-mesh (I2M) conversion: real-time mesh generation, robustness, fidelity, quality, and smoothness. Robustness concerns the software's ability to process different input types such as CAD models (for example, of stents) and patient-specific medical images. Utilizing both is essential when running a medical simulation meant to aid in determining potential methods of treatment \cite{MelinaNumericalSimulation2016, InVitro2015, HighFidelity2015}. Fidelity measures the degree to which the mesh surface aligns with an image boundary. Quality (determined by the shape and size of mesh elements) and smoothness affect the accuracy of solutions for CFD simulations \cite{meshEvaluationNRR}. Each of these requirements are independently challenging problems. The first proposed method integrates multiple software tools into a single pipeline to address the robustness, fidelity, quality, and (to a degree) smoothness requirements (providing a smoothness of $C^0$, as opposed to other available real-time I2M methods that do not address smoothness \cite{foteinos2014high}). Given that the image-to-mesh conversion and boundary layer generation steps are sequential, the pipeline only addresses the real-time requirement with its parallel adaptive anisotropic mesh generation. After this feasibility study, the next steps will be to fully parallelize all components of the I2M pipeline in order to maximize potential performance. The I2M pipeline involves several steps: (1) utilize a method known as CBC3D \cite{EwCCBC3D} to discretize a segmented image using a Body-Centered Cubic lattice of high-quality tetrahedra and deform the generated mesh surfaces to their corresponding tissue boundaries to improve fidelity while maintaining quality, (2) construct an anisotropic metric tensor field, (3) generate a boundary layer grid from the high-fidelity CBC3D mesh surface using a method known as Advancing Front Local Reconnection (AFLR) \cite{AFLRBoundaryLayer2013}, and (4) utilize a multicore cc-NUMA-based mesh adaptation method known as CDT3D \cite{EwCCDT3D} to generate an anisotropic mesh volume using the metric tensor field and CBC3D isotropic mesh as a background mesh. To help satisfy real-time requirements, CDT3D utilizes an optimized local reconnection operation and leverages a hierarchical model to balance workloads between threads. 

The second method extends the functionality of a Delaunay-based image-to-mesh conversion software called PODM \cite{foteinos2014high}. The method can now utilize a user-defined sizing function to adapt volume elements when generating an isotropic mesh from a segmented image, while satisfying fidelity, quality, and real-time requirements. We evaluate these methods using two aneurysm cases (a carotid cavernous aneurysm and middle cerebral artery bifurcation aneurysm) obtained from the Cerebrovascular Center for Research at Stony Brook University under Institutional Review Board approval. We show that the proposed approaches meet the aforementioned I2M conversion requirements. Additionally, when utilizing 96 CPU cores on a single, multicore node on Purdue University's Anvil supercomputer \cite{Anvil2022}, the adaptive isotropic method generates about 50 million elements in less than a minute while the adaptive anisotropic CDT3D method generates approximately the same amount in about 5 minutes. Given that CDT3D also exhibited excellent performance and accurately captured features of underlying aerospace simulations when coupled with a solver (also processing CAD data to perform adaptation \cite{EwCCDT3D}), the end goal is to utilize this method within a vascular flow simulation involving stents. This is however outside the scope of this paper and we first test the feasibility of utilizing the methods to help satisfy the aforementioned image-to-mesh conversion requirements.

Ultimately, this paper contributes the following:
\begin{enumerate}
    \item an image-to-mesh conversion pipeline for adaptive anisotropic mesh generation that satisfies robustness, quality, fidelity, smoothness, and near real-time medical I2M conversion requirements,
    \item a hierarchical load balancing model (designed to target large cc-NUMA shared memory architectures) for speculative execution applied to an adaptive anisotropic mesh generation method that only utilized 40 CPU cores for aerospace cases in earlier studies \cite{EwCCDT3D, TsolakisEvaluation} and is now tested up to 96 for medical cases,
    \item an optimized local reconnection algorithm for anisotropic adaptation, that when compared to its implementation in earlier studies \cite{EwCCDT3D, TsolakisEvaluation}, is now three times faster, 
    \item and a new sizing function approximation technique to generate adaptive isotropic meshes from segmented medical images.
\end{enumerate}

Section \ref{sec:lit_review} presents an overview of related work. Section \ref{sec:background} describes each of the tools utilized in the I2M pipeline. Section \ref{sec:method} describes each of the above contributions in detail. Section \ref{pipeline_results} shows how the aforementioned I2M conversion requirements are satisfied by both the I2M pipeline and PODM. Section \ref{HLB_results} shows the impact of the hierarchical load balancing model as it is tested on different supercomputers. Section \ref{local_reconnection_results} shows the performance improvement from the optimized local reconnection algorithm and the performance of both CDT3D and PODM when generating large-size meshes (up to about 100 million elements). Section \ref{sec:discussion} provides an analysis of our results while section \ref{i2m_future_work} discusses needed future work and section \ref{sec:conclusion} concludes our paper.

\section{Related Work}
\label{sec:lit_review}
There are numerous methods that process medical data as input to create meshes. The type of meshes generated must be taken into account when considering the target application. Meshes can be classified into two categories - structured and unstructured. While some structured mesh generation methods have been shown to provide accurate results within CFD simulations \cite{AcceleratedEstimation2021, DynamicOfBloodFlows2021, Enhanced4D2022}, these types of methods can sometimes fail to capture features within complex geometries \cite{SAMRBook1999}. Unstructured meshes can be further classified into two categories - isotropic and anisotropic. Isotropic mesh generation methods are well suited for high curvature geometries. Anisotropic meshes contain high aspect ratio elements that include directional information (e.g., hemodynamics). Additionally, they may contain fewer elements in some cases compared to isotropic meshes generated from the same image. A mesh should ideally be as small as possible to reduce the memory and CPU requirements for the solver. These benefits make anisotropic meshes suitable for blood flow problems.  

There are many methods that convert 3D images into 3D meshes. For a comprehensive overview of these methods, see \cite{ZhangBook2016}. Many 3D isotropic methods utilize Delaunay refinement techniques \cite{Boltcheva, CGAL2007, Pons2007, AnisotropicFluidSolidI2M2010, foteinos2014high, PDRPODMExperience}. A challenge with Delaunay refinement, which still remains an open problem, is that almost flat tetrahedra (slivers) can survive known heuristics designed to remove them \cite{Chew:1997, cheng2000sliver, Boltcheva}. These low quality elements introduce error when processed by a solver. Additional I2M methods utilize lattice space-tree (octree) decomposition techniques \cite{LatticeCleaving, ZhangAutomaticMultimaterial2010, Zhang3DFinite2005, LD_Chernikov, ZhangChallenges2013} to provide meshes with adaptively-sized, high quality elements. Some also utilize the Dual Contouring technique to improve element quality \cite{DualContour3DSegmentationAnisotropic2018, ParallelDC2013}. Others warp the surfaces of the lattice to the boundaries of the object using either a mass-spring system, an FE constitutive model, or an optimization scheme \cite{Molino03acrystalline, CBC3DYixun, Fuchs20011400}. Some methods focus on processing surface geometries of medical data to create structured Cartesian cut-cell meshes \cite{SolutionAdaptive2021} and isotropic tetrahedral meshes \cite{AssessmentMeshGen2008}. The study in \cite{MeshGenNRR2008} evaluates several of these sequential isotropic-based methods' suitability for surgical simulations and non-rigid registration applications. There are methods which utilize parallel mesh generation techniques (with either CPUs \cite{ParallelDC2013, AcceleratedEstimation2021, DynamicOfBloodFlows2021, Enhanced4D2022} or GPUs \cite{ParallelDC2013}). The methods in \cite{AcceleratedEstimation2021, DynamicOfBloodFlows2021, Enhanced4D2022} all focus on generating Cartesian cut-cell meshes while \cite{ParallelDC2013} focuses on generating isotropic tetrahedral meshes from biomolecular complexes (as opposed to 3D images). \cite{DualContour3DSegmentationAnisotropic2018} utilizes an anisotropic Giaquinta–Hildebrandt operator-based geometric flow method to smooth mesh surfaces while preserving surface and volume features. However, this smoothing is applied to isotropic meshes.

Some methods focus on converting medical images into anisotropic meshes. Although potentially useful for blood flow simulations, \cite{ClericiAnisotropic2020} focuses on using an anisotropic mesh adaptation procedure to construct a high-fidelity virtual model of a fractured pelvis, and then to build a physical replica (by means of additive manufacturing techniques) to represent the anatomical object from a segmented image. The method in \cite{AnisotropicFluidSolidI2M2010} generates high-quality adaptive hybrid meshes for fluid-solid geometries within cardiac environments. Following segmentation, surfaces between heart tissue and blood (and other non-cardiovascular tissue/fluids) are extracted and adapted. Anisotropic tetrahedra are generated for the interior of heart tissue, a prismatic boundary layer is generated for the interface between tissue and blood, and TetGen (an isotropic boundary-constrained Delaunay-based method \cite{TetGen2015}) is used to tetrahedralize the remainder of the blood region. Boundary layers in meshes are useful when calculating metrics such as pressure or wall shear stress (WSS) caused by blood flow \cite{CFDAneurysmRupture2019,AnisotropicFluidSolidI2M2010}. Although the method generated high-quality hybrid meshes, the Delaunay-based isotropic elements in fluid flow regions (such as in the coronary arteries) were identified as a drawback, with anisotropy suggested as an improvement. Additionally, Delaunay tetrahedralization limits the quality of the tetrahedral core for complex biological geometries \cite{AnisotropicFluidSolidI2M2010} and can fail to terminate for certain cases \cite{EwCCBC3D}. Another method was developed to model cerebral aneurysms for fluid-structure interaction simulations \cite{AnisotropicCerebralAneurysms2013}. The surface mesh of a healthy blood vessel is deformed onto an aneurysm model, simulating the formation of the aneurysm. While the simulation results demonstrated the benefit of utilizing anisotropic meshes rather than isotropic meshes, the mesh generation method is sequential, as with all the aforementioned adaptive anisotropic methods.

Two approaches are proposed to help streamline the discretization of complex vascular geometries within the numerical modeling process. The first combines multiple software tools into a single pipeline to provide the following: (1) image-to-mesh conversion which satisfies the quality, fidelity, and smoothness ($C^0$) requirements, (2) the generation of a boundary layer grid over the high fidelity surface, (3) a parallel adaptive anisotropic meshing procedure which provides near real-time performance, and (4) robustness, which is satisfied by the pipeline's ability to process segmented images and CAD models \cite{EwCCDT3D}. To the best of our knowledge, there is no single method that provides all of the above. Although it is a Delaunay-based method, the second approach was previously tested to generate a mesh of reasonably good fidelity and quality when converting a segmented image of an aneurysm into an isotropic mesh \cite{EwCCBC3D}. Consequently, we extend its functionality to not only adapt a mesh based on the curvature of the image boundary but to also approximate a user-defined sizing function to allow adaptivity within the volume when generating isotropic meshes (potentially making this method suitable for vascular flow simulations, where the next step is to test the method with error-based sizing criteria within a simulation). While some of the aforementioned methods are capable of generating adaptive (within the volume) isotropic meshes of good quality, these methods are sequential \cite{CGAL2007, Pons2007, ZhangAutomaticMultimaterial2010, AnisotropicFluidSolidI2M2010, LatticeCleaving} (as opposed to our second method's parallel speculative execution model that offers real-time performance even when generating large meshes). 
We test both techniques' (converting segmented medical images to either isotropic or anisotropic meshes) feasibility as options to satisfy medical image-to-mesh conversion requirements when generating small-size meshes (e.g., 1 million elements) and much larger meshes (up to 100 million elements).

\section{Background for Tools in Adaptive Anisotropic Image-to-mesh Conversion Pipeline} \label{sec:background}
Details of each tool utilized in the I2M pipeline to generate adaptive anisotropic meshes are given. Section \ref{Pipeline_I2M} describes the lattice-based I2M software CBC3D, which generates isotropic meshes of high quality, good fidelity, and smoothness. Section \ref{aflr_I2M} describes the Advancing Front Local Reconnection (AFLR) software utilized to generate a boundary layer grid over the CBC3D-generated mesh surface. Section \ref{CDT3D} gives details about the parallel adaptive anisotropic mesh generation software CDT3D.
\subsection{Image-to-mesh Conversion} \label{Pipeline_I2M}
Image-to-mesh conversion in the adaptive anisotropic pipeline is handled by a software called CBC3D \cite{EwCCBC3D}. 
CBC3D converts a segmented, multi-labeled image into an isotropic tetrahedral (or mixed element) mesh. The segmented image is discretized using a regular Body-Centered Cubic lattice. The user has the option of adjusting the distance between the vertices of the lattice. Any resulting tetrahedra located outside the object are discarded. Adaptive refinement commences over these high quality elements using red (regular) - green (irregular) subdivisions. A Euclidean Distance Transform is used to determine if an element should be subdivided. CBC3D allows the user to specify an input fidelity parameter which influences how fine the mesh will be near the boundaries. A label-specific fidelity can also be specified to allow for localized refinement in materials of interest (reducing overall element count). Additionally, a mixed element mesh can be created where tetrahedra are merged into hexahedra. The topology of generated surfaces or interfaces between labels is preserved by only merging within homogenous regions where the lattice is uniform. While useful for further reducing element count, the proposed approach only utilizes tetrahedral meshes generated by CBC3D given that CDT3D (the adaptive anisotropic method) only processes tetrahedral background meshes. As seen in our evaluation (see section \ref{results_section}), CDT3D meets the real-time requirement when processing tetrahedral background meshes to generate small-size anisotropic meshes (and near real-time when generating large meshes, e.g., 50 million elements). The adaptive refinement approach utilized by CBC3D guarantees an output mesh with high quality elements \cite{EwCCBC3D}. CBC3D finally deforms the mesh surfaces to their corresponding image boundaries. 
The final mesh has a smoothness of $C^0$ while maintaining the same high quality. Consequently, CBC3D was selected to be the image-to-mesh conversion tool within the pipeline of the proposed approach because it: (1) generates meshes of high quality and fidelity, (2) provides smoothness, and (3) was previously used to discretize patient-specific medical imaging data and STLs of CAD-based stent geometries for use within CFD simulations \cite{MelinaNumericalSimulation2016, InVitro2015, HighFidelity2015}.

\subsection{Boundary Layer Generation} \label{aflr_I2M}
The generation of boundary layers over the high fidelity CBC3D mesh surface is carried out by a software known as AFLR. AFLR is an unstructured mesh generation method used extensively in industry. While it is capable of both isotropic \cite{MarcumAFLR} and anisotropic mesh generation \cite{AnisotropicAFLR}, it is sequential. Consequently, its use within the proposed approach is only focused on its boundary layer generation capability \cite{AFLRBoundaryLayer2013}. AFLR generates boundary layers using a spacing that is normal to the boundary and that grows geometrically. As new points are generated, quality is maintained by checking their distribution (i.e., distance from one another according to a local element length scale). New points are only accepted and inserted into the grid if they produce elements that satisfy dihedral angle and element aspect ratio thresholds. The user has numerous options for controlling the spacing distribution and may also specify the number of layers generated.

\subsection{Adaptive Anisotropic Mesh Generation} \label{CDT3D}
The final adaptive anisotropic mesh is generated using a software tool called CDT3D. CDT3D is a multicore cc-NUMA-based mesh generation method that exploits fine-grain parallelism at the cavity level using data decomposition. It is capable of both isotropic \cite{DrakopoulosCDT3D} and adaptive anisotropic mesh generation \cite{EwCCDT3D}. A speculative execution model is implemented, which serves to exploit parallelism ``everywhere possible" throughout the mesh generation procedure. Atomic lock instructions are utilized by several of CDT3D's meshing operations so that different data can be modified concurrently by threads while guaranteeing correctness (i.e., conformity). A lock attempts to acquire the necessary dependencies for a corresponding operation. If unable to do so, any acquired resources are released and the operation is applied on a different set of data. This is repeated until the mesh satisfies spacing and qualitative criteria. There are a number of parallel mesh operations in CDT3D, including: edge collapse, point creation/insertion (which by default uses an advancing front technique for isotropic mesh generation and a centroid-based technique for anisotropic adaptation), CAD projection, local reconnection (edge/face swapping), and point smoothing. See \cite{DrakopoulosCDT3D} and \cite{EwCCDT3D} for more information on their design and implementation. CDT3D's adaptation pipeline is organized into two phases - mesh adaptation and quality improvement. A subset of the operations are utilized in each phase \cite{EwCCDT3D}. The mesh adaptation phase is designed to focus on modifying elements so that they conform to point spacing as required by the metric tensor field. The quality improvement phase focuses on improving element shape (i.e., improving element mean ratio). See section \ref{pipeline_results} for details regarding how qualitative criteria are measured in the metric space for our evaluation.

CDT3D accepts an analytic or discrete metric field as input when performing metric-based anisotropic mesh adaptation. CAD geometries can also be processed and used to perform adaptation (thus allowing the proposed approach to handle both medical I2M conversion and potentially the future processing of CAD models for CFD simulations involving stents). CDT3D has been quantitatively and qualitatively evaluated with several benchmarks based on aerospace cases \cite{TsolakisEvaluation,EwCCDT3D}. Not only was it shown to be stable in terms of metric conformity (element shape size and edge length), it also showcased excellent performance when utilizing up to 40 cores on a single multicore node. When coupled with a solver in a CFD simulation pipeline, it captured features of the underlying simulation and occupied only a small fraction of the pipeline's runtime \cite{EwCCDT3D}.

\section{Method} \label{sec:method}
We present several contributions to address some of the challenges of medical image-to-mesh conversion.
\subsection{I2M Pipeline} \label{method_i2m_pipeline}
\begin{figure}[!htb]
\begin{center}
\includegraphics[width=0.6\textwidth]{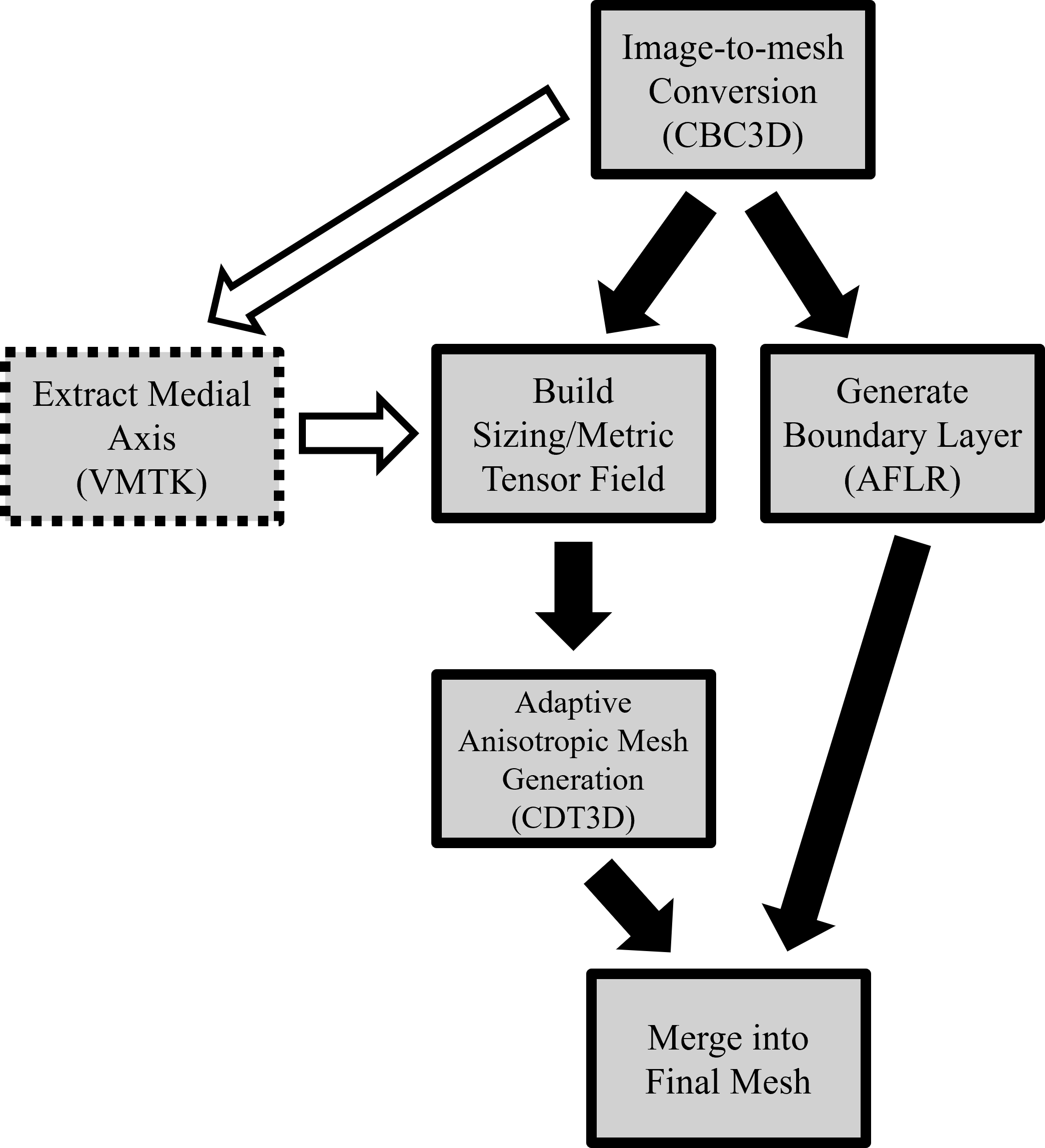}
\end{center}
\caption{Presented is the pipeline of software tools used to satisfy the medical image-to-mesh conversion requirements for use within a medical numerical simulation.}
\label{fig:pipeline}
\end{figure}

The proposed approach involves several steps: (1) utilize CBC3D to convert a segmented image into a high quality and high fidelity background mesh (for anisotropic mesh adaptation), (2) construct an anisotropic metric tensor field, (3) generate a boundary layer grid outside the CBC3D mesh surface using AFLR, and (4) utilize CDT3D to generate an anisotropic mesh volume using the metric tensor field. The boundary layer grid and anisotropic mesh are merged into one final mesh. A visual overview of this pipeline of tools can be seen in Figure \ref{fig:pipeline}. A script is utilized, supplying the output of one program as input to the next. Extracting the medial axis with VMTK \cite{VMTK2008, Centerline2003} is optional, as this technique utilized to build an anisotropic metric tensor field only serves as an example for the purpose of testing the pipeline's robustness.

The proposed approach is not bound to any particular method for constructing the metric tensor field. As stated previously, the pipeline is evaluated based on the aforementioned requirements to gauge its suitability for medical numerical simulations. Consequently, our method is tested using two different techniques for constructing a metric tensor field, where each is used with a different aneurysm case. For the first case, the metric is constructed based on a method also utilized for non-rigid registration applications intended for image-guided neurosurgery \cite{ComparisonPhysicsBased2023}. This method focuses on capturing registration points within an adaptive mesh. A metric tensor field is built by evaluating the k-closest registration points to each mesh vertex using a k-nearest neighbor search from the VTK library \cite{VTKWebsite}. k is the number of points to find that are closest to a given point. The Khachiyan algorithm \cite{Khachiyan2007} is then used to construct a minimum volume bounding ellipsoid which encompasses the given point set. This ellipsoid has a natural mapping to a 3 x 3 positive definite matrix that can be used as a metric to guide adaptivity. See \cite{ComparisonPhysicsBased2023} for more details. To exploit parallelism, an OpenMP construct is utilized when processing the mesh points. For our first test case, we utilize this method using the points of an approximate medial axis of the generated CBC3D mesh (in place of registration points). The approximate medial axis (also called a centerline) is obtained for our first test case using VMTK \cite{VMTK2008, Centerline2003} and can be seen in Figure \ref{fig:artery1} of our results. 

The anisotropic metric tensor field and CBC3D background volume mesh (converted from the segmented image) are provided as input to CDT3D, which generates an adaptive anisotropic volume mesh. In this study, surface adaptation is turned off to ensure that the high fidelity surface generated by CBC3D remains intact. The surface is extracted from the CBC3D mesh and is given as input to AFLR to generate a boundary layer grid. AFLR is set to only generate a boundary layer grid, designating the input as an open domain so that only boundary layer elements are generated on the outside of the surface (as opposed to within the artery lumen). Such a technique may be useful for fluid-structure interaction (FSI) simulations that couple arterial wall-motion to pulsatile blood flow through the lumen. Mechanical properties of arterial wall tissue (even aneurysmal wall tissue) could be measured and used for the outer layers generated by AFLR. This would allow for calculating metrics such as peak stress within the arterial wall \cite{WallTensionCerebralAneurysms2008, WallsAbdominalAneurysms2025}. While AFLR is used to generate a grid external to the artery lumen for this feasibility study, it can easily be used to generate a grid internal to the lumen for blood flow simulations in future work. The internal AFLR boundary elements can provide increased accuracy of wall shear stress or wall shear stress gradients in the fluid domain. Surface refinement is turned off for AFLR (again, for this particular study) to ensure that the high fidelity CBC3D surface remains intact and so that both the boundary layer grid and CDT3D mesh are conforming. The boundary layer grid (generated outside the artery surface) and the CDT3D mesh (generated within the manifold artery surface) are merged to create one final output mesh. This is done by reading both generated meshes and merging their elements/points to then be output as a single mesh (again, using the VTK library \cite{VTKWebsite}).

For our second aneurysm test case, we utilize streamline data of the velocity field from a CFD simulation executed outside of this study \cite{CFDDrivenTopology2024} (seen in Figure \ref{fig:artery2_streamlines} of section \ref{results_section}). The velocity at each point along these streamlines is used to construct a metric tensor field to guide adaptivity for anisotropic mesh generation. Because this aneurysm case was only provided to us as a surface mesh, an isotropic volume mesh is first generated using the surface mesh. A k-nn search is utilized to find the 3 closest velocity points to each mesh volume vertex. Using the 3 closest points, a velocity gradient tensor is calculated at each mesh vertex which can be mapped to a 3x3 positive definite matrix. The magnitude (eigenvalue) of each velocity gradient tensor is used to apply a stretching factor to the positive definite matrix which ensures that elements are stretched in the dominant direction of the velocity. OpenMP is again utilized to parallelize the processing of the mesh points. The isotropic volume mesh is processed as a background mesh with the velocity-based metric tensor field to generate anisotropic meshes. A qualitative and performance evaluation of the pipeline is given in section \ref{results_section}.

\subsection{Hierarchical Load Balancing}
A hierarchical load balancing technique was previously seen to enhance the performance of the I2M Delaunay-based multi-threaded software PODM, which also utilizes a speculative execution model \cite{foteinos2014high} (discussed in more detail in section \ref{function_approximation}). Consequently, we introduce a cc-NUMA-based hierarchical load balancing technique for the adaptive anisotropic parallel mesh generation software CDT3D. Before adaptation, CDT3D organizes tetrahedra into "buckets" that are all assigned to pinned threads (i.e., each thread is pinned to a CPU core) \cite{EwCCDT3D, DrakopoulosCDT3D}. When a thread finishes processing its list of buckets during adaptation, it is inserted into an idle list. Originally, after processing a bucket, a busy thread (those that still have buckets to process) would give a fraction of its buckets to the first thread in the idle list (if the list is not empty) to balance workloads. CDT3D has been updated so that a busy thread will now search the list of idle threads to find the pinned thread that is located within the shortest numa node distance among all the idle threads. cc-NUMA-based shared memory architectures organize CPU cores into numa nodes, where one numa node will consist of multiple CPU cores. When a thread creates new elements, their respective data structures are allocated in memory that is local to the numa node within which that pinned thread's CPU core is located. Any CPU core (and its pinned thread) within that same numa node will relatively have the same memory access time to those data structures. Remote memory accesses (by CPU cores located in other numa nodes on the shared memory system) require greater access time. Consequently, it is ideal for busy threads to give workloads to threads that are pinned to CPU cores within the shortest numa node distance to reduce the number of remote memory accesses and limit the latency of such accesses during adaptation. Both the syscall\footnote{\url{https://man7.org/linux/man-pages/man2/syscall.2.html}} and numa\footnote{\url{https://man7.org/linux/man-pages/man3/numa.3.html}} libraries are utilized to get the numa node id of a CPU core and the distance between two numa nodes (given their ids), respectively. Algorithm \ref{alg:nws} gives an overview of CDT3D's hierarchical load balancing. We evaluate the effectiveness of such a technique compared to CDT3D's original load balancing method in section \ref{HLB_results}.

\begin{algorithm} 
\caption{CDT3D's Hierarchical Load Balancing}
\label{alg:nws}
\scriptsize{
\begin{flushleft}
NUMA\_LOAD\_BALANCE($threads_1...threads_N$, $buckets_1...buckets_N$) \\
\textbf{Input}: $threads_1...threads_N$ are the threads \\
\hspace*{\algorithmicindent} \ \ \ \ \ $buckets_1...buckets_N$ is the list of buckets per thread in $threads_1...threads_N$ \\
\textbf{Output}: Parallel adapted mesh
\end{flushleft}
\begin{algorithmic}[1]
\State idle\_thread\_list = empty list
\ForEach{$thread_i \in threads_1...threads_N$}
    \State PIN\_THREAD($thread_i$)
    \State $numa\_node\_id_{thread_i}$ = GET\_NUMA\_NODE\_ID($thread_i$)
\EndFor
\ForEach{$thread_i \in threads_1...threads_N$} \Comment{Executed in parallel}
    \While{$buckets_i$ is NOT empty}
        \If{idle\_thread\_list is NOT empty}
            \State closest\_thread = idle thread w/ minimum numa node distance from $numa\_node\_id_{thread_i}$
            \State Assign fraction of $thread_i$'s buckets to closest\_thread and remove them from $buckets_i$
            \State WAKE\_UP(closest\_thread)
        \EndIf
        \State bucket = first bucket in $buckets_i$
        \State APPLY\_MESH\_OPERATIONS(bucket)
        \State Remove bucket from $buckets_i$
    \EndWhile
    \State idle\_thread\_list->insert($thread_i$)
    \State SLEEP($thread_i$)
\EndFor
\State TERMINATE
\end{algorithmic}
}
\end{algorithm}

\subsection{Optimized Local Reconnection} \label{local_reconnection}
Upon further study, we have found that a percentage of the elements within meshes generated by CDT3D may not fully satisfy qualitative criteria required by the target metric field. Despite this phenomena, CDT3D's meshes have been compared to those generated by several state-of-the-art methods, exhibiting comparable overall quality \cite{TsolakisEvaluation} and excellent performance when integrated into an adaptive CFD simulation pipeline \cite{EwCCDT3D}. An example of this phenomena is that approximately 6-8\% of elements generated for our second aneurysm case (where its metric field is scaled to 50 million complexity to generate about 100 million elements, seen in section \ref{results_section}) do not satisfy qualitative criteria by the end of execution. Furthermore, we found that the local reconnection operation wastes a significant amount of time attempting to improve these elements. The local reconnection operation is utilized within both phases of CDT3D (mesh adaptation and quality improvement), consisting of topological transformations (also called flips) \cite{DrakopoulosCDT3D, EwCCDT3D}. Operations within each phase are executed repeatedly over numerous iterations until little to no new points are inserted (for mesh adaptation) or a user-defined threshold is reached. If CDT3D detects that elements do not satisfy qualitative criteria, then it will utilize the local reconnection operation to attempt to improve those elements. A problem is that it will attempt to improve these elements during every iteration, regardless of whether or not that element or its neighbors were modified in a previous iteration. This means that CDT3D will waste time attempting to improve the same elements repeatedly over numerous iterations. We found that on average, the local reconnection operation simply repeats itself (for more than one iteration) attempting to improve approximately 80\% of the elements (the 6-8\%) that do not satisfy qualitative criteria by the end of execution.

The local reconnection operation has now been optimized to avoid unnecessary repetition when attempting to improve element quality. It keeps track of whether or not flips have already been attempted for an element, and recognizes that it should not re-attempt improvement if the element or its neighbors have not been modified since the last attempt (from a previous iteration). The method also recognizes if flips should be re-attempted due to unsuccessful dependency locking (when attempting to improve the element during a previous iteration) under CDT3D's speculative execution model. Algorithm \ref{alg:local_reconnection} shows this new functionality. The amount of time saved and the impact of this optimization can be seen in section \ref{local_reconnection_results}.

\begin{algorithm} 
\caption{Optimized Local Reconnection}
\label{alg:local_reconnection}
\scriptsize{
\begin{flushleft}
LOCAL\_RECONNECTION(tetrahedra, current\_iteration) \\
\textbf{Input}: tetrahedra is the set of tetrahedra upon which to attempt flips \\
\hspace*{\algorithmicindent} \ \ \ \ \ current\_iteration is the iteration number in which this operation is being executed (among the iterations for either mesh adaptation or quality improvement) \\
\textbf{Output}: tetrahedra with potentially improved quality
\end{flushleft}
\begin{algorithmic}[1]
\ForEach{tet $\in$ tetrahedra}
    \State neighborElements = set of tetrahedra that share a face with tet
    \If{(current\_iteration $\neq$ 1 AND tet has NOT encountered unsuccessful dependency locking)}
        \State mainModified = false
        \State neighborsModified = false
        \If{(tet was created during current\_iteration OR \\ 
        \hspace*{\algorithmicindent} \ \ \ \ tet was modified by another operation during current\_iteration)}
            \State mainModified = true
        \EndIf
        \If{NOT mainModified}
            \ForEach{neighbor $\in$ neighborElements}
                \If{(neighbor was created during current\_iteration OR \\ \hspace*{\algorithmicindent} \ \ \ \ \ \ \ \ \ \ \ \ \ neighbor was modified by another operation during current\_iteration)}
                    \State neighborsModified = true
                \EndIf
            \EndFor
        \EndIf
        \If{(NOT mainModified AND NOT neighborsModified)}
            \State Skip tet and continue to next iteration of for loop (line 1)
        \EndIf
    \EndIf
    \If{neighborElements are successfully locked}
        \State ATTEMPT\_FLIPS(tet, neighborElements)
        \State Unmark tet for unsuccessful dependency locking
    \Else
        \State Mark tet with unsuccessful dependency locking
    \EndIf
\EndFor
\end{algorithmic}

}
\end{algorithm}

\subsection{Function Approximation for Medical Imaging} \label{function_approximation}
Although it may be inadequate in resolving features of some stent geometries and its output mesh quality is sometimes inferior to that of CBC3D's \cite{EwCCBC3D}, we gauge the feasibility of utilizing the image-to-mesh conversion software PODM \cite{foteinos2014high} to satisfy the I2M requirements for high resolution numerical simulations involving aneurysms. Despite the drawbacks of its Delaunay-based meshing techniques, PODM has been shown to generate good quality meshes for a segmented aneurysm image with reasonably good fidelity \cite{EwCCBC3D}. The method also utilizes a parallel speculative execution model, exhibiting excellent performance when generating large meshes (e.g., 1 billion elements) on a large shared memory architecture (achieving a speedup of approximately 123 when utilizing 144 cores \cite{foteinos2014high}). Consequently, we extend the method's function approximation capability (introduced in \cite{CNFPaper} to specifically address nuclear femtography problems) to now generate an adaptive isotropic mesh for an aneurysm case. To guarantee quality and fidelity, PODM adheres to specific criteria when determining if elements should be adapted (i.e., split). The user can specify their own criteria through a sizing function that can be loaded as a shared library, which PODM will utilize while generating a mesh in parallel. For our first test case (the segmented image), we construct a sizing function in a similar manner as described in section \ref{method_i2m_pipeline}, based on \cite{ComparisonPhysicsBased2023}. Algorithm \ref{alg:function_approximation} shows the sizing function that is utilized to determine if an element should be adapted. 3x3 positive definite metric tensor matrices are calculated for each of the element's points based on their distance to the closest landmark \cite{ComparisonPhysicsBased2023} (i.e., the closest point along the medial axis). Next, the sizing function checks if any of the element's edges have a "metric length" that is larger than a specified threshold. The metric length is a measure of how long an edge is in the space defined by the metric tensor \cite{PLGeorgeMeshGeneration2008}. The average of the tensors at the two endpoints of an edge is calculated to obtain an interpolated tensor for that edge. If the calculated metric length exceeds the threshold, the element is marked for refinement. PODM's output mesh quality and performance when it uses this sizing function for volume adaptivity (and when it does not, i.e., only boundary adaptivity based on the curvature) is evaluated in section \ref{results_section}.

\begin{algorithm} 
\caption{PODM's Function Approximation for an Aneurysm}
\label{alg:function_approximation}
\scriptsize{
\begin{flushleft}
SHOULD\_BE\_ADAPTED($P_1...P_N$, $L_1...L_K$, adapt\_const) \\
\textbf{Input}: $P_1...P_N$ is the list of an element's $N$ points \\
\hspace*{\algorithmicindent} \ \ \ \ \ $L_1...L_K$ is the list of landmark points that affect adaptivity of elements throughout the mesh \\
\hspace*{\algorithmicindent} \ \ \ \ \ adapt\_const is a heuristic value that controls how aggressive adaptation should be \\
\textbf{Output}: Returns either true or false, designating if the element should be adapted
\end{flushleft}
\begin{algorithmic}[1]
\State $MT_1,_9...MT_N,_9$ = 2D list of metric tensor 3x3 matrices, where each 1D list is of size 9 and initialized to 0, corresponding to each point in $P_1...P_N$
\ForEach{p $\in$ $P_1...P_N$}
    \State dist = distance between p and the closest landmark point among $L_1...L_K$
    \State diag = 1 / (dist * dist)
    \State $MT_p,_1$ = diag
    \State $MT_p,_5$ = diag
    \State $MT_p,_9$ = diag
\EndFor
\State edges = list of edge combinations among the points in $P_1...P_N$
\ForEach{edge $\in$ edges}
    \State $edge\_point_1$ = first point in edge
    \State $edge\_point_2$ = second point in edge
    \State edge\_vector.x = $edge\_point_2$.x - $edge\_point_1$.x
    \State edge\_vector.y = $edge\_point_2$.y - $edge\_point_1$.y
    \State edge\_vector.z = $edge\_point_2$.z - $edge\_point_1$.z
    \ForEach{i $\in$ 1...9}
        \State $avgTensor_i$ = ($MT_{edge\_point_1,_i}$ + $MT_{edge\_point_2,_i}$) / 2
    \EndFor
    \State metric\_length = $\sqrt{GET\_METRIC\_LENGTH\_SQUARED(edge\_vector, avgTensor)}$
    \If{metric\_length > adapt\_const}
        \State return true
    \EndIf
\EndFor
\State return false
\end{algorithmic}

\begin{flushleft}
$GET\_METRIC\_LENGTH\_SQUARED(edge\_vector, metric\_tensor)$ \\
\end{flushleft}
\begin{algorithmic}[1]
\State return $edge\_vector.x$ * ($metric\_tensor_1$ * $edge\_vector.x$ + $metric\_tensor_2$ * $edge\_vector.y$ + $metric\_tensor_3$ * $edge\_vector.z$) + $edge\_vector.y$ * ($metric\_tensor_4$ * $edge\_vector.x$ + $metric\_tensor_5$ * $edge\_vector.y$ + $metric\_tensor_6$ * $edge\_vector.z$) + $edge\_vector.z$ * ($metric\_tensor_7$ * $edge\_vector.x$ + $metric\_tensor_8$ * $edge\_vector.y$ + $metric\_tensor_9$ * $edge\_vector.z$)
\end{algorithmic}

}
\end{algorithm}

\section{Results} \label{results_section}
We first describe how our experiment is set up in terms of the test cases and the supercomputers utilized to test the proposed methods in section \ref{I2M_experimental_setup}. We then evaluate the adaptive anisotropic I2M pipeline and adaptive isotropic I2M method in section \ref{pipeline_results}, followed by a study of the hierarchical load balancing model in section \ref{HLB_results} and the optimized local reconnection algorithm when generating large meshes in section \ref{local_reconnection_results}.

\subsection{Experimental Setup} \label{I2M_experimental_setup}
\begin{figure}[htb]
\centering
\subfigure[]{\label{fig:artery2_original_image_1}\includegraphics[width=0.49\textwidth]{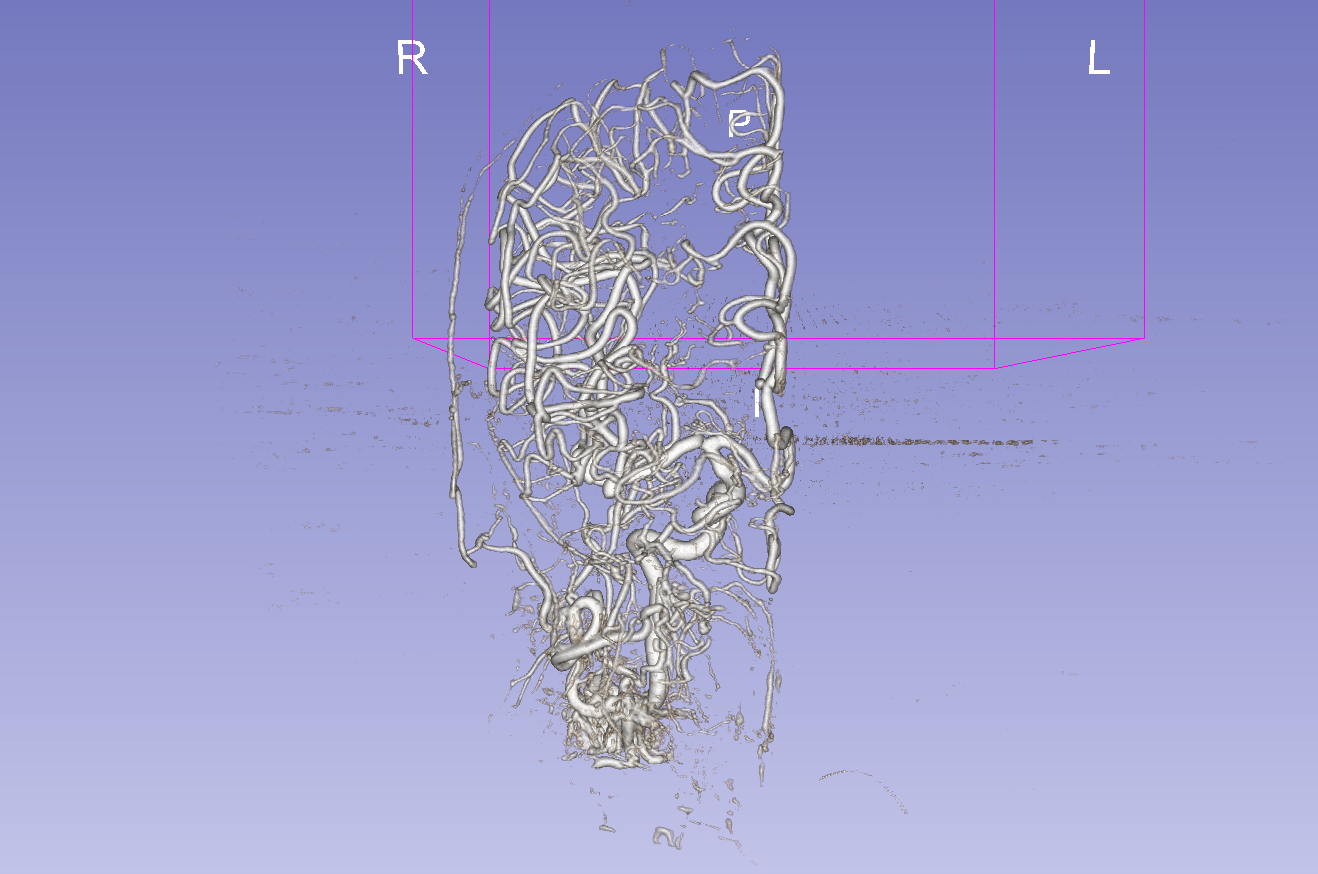}}
\subfigure[]{\label{fig:artery2_original_image_2}\includegraphics[width=0.49\textwidth]{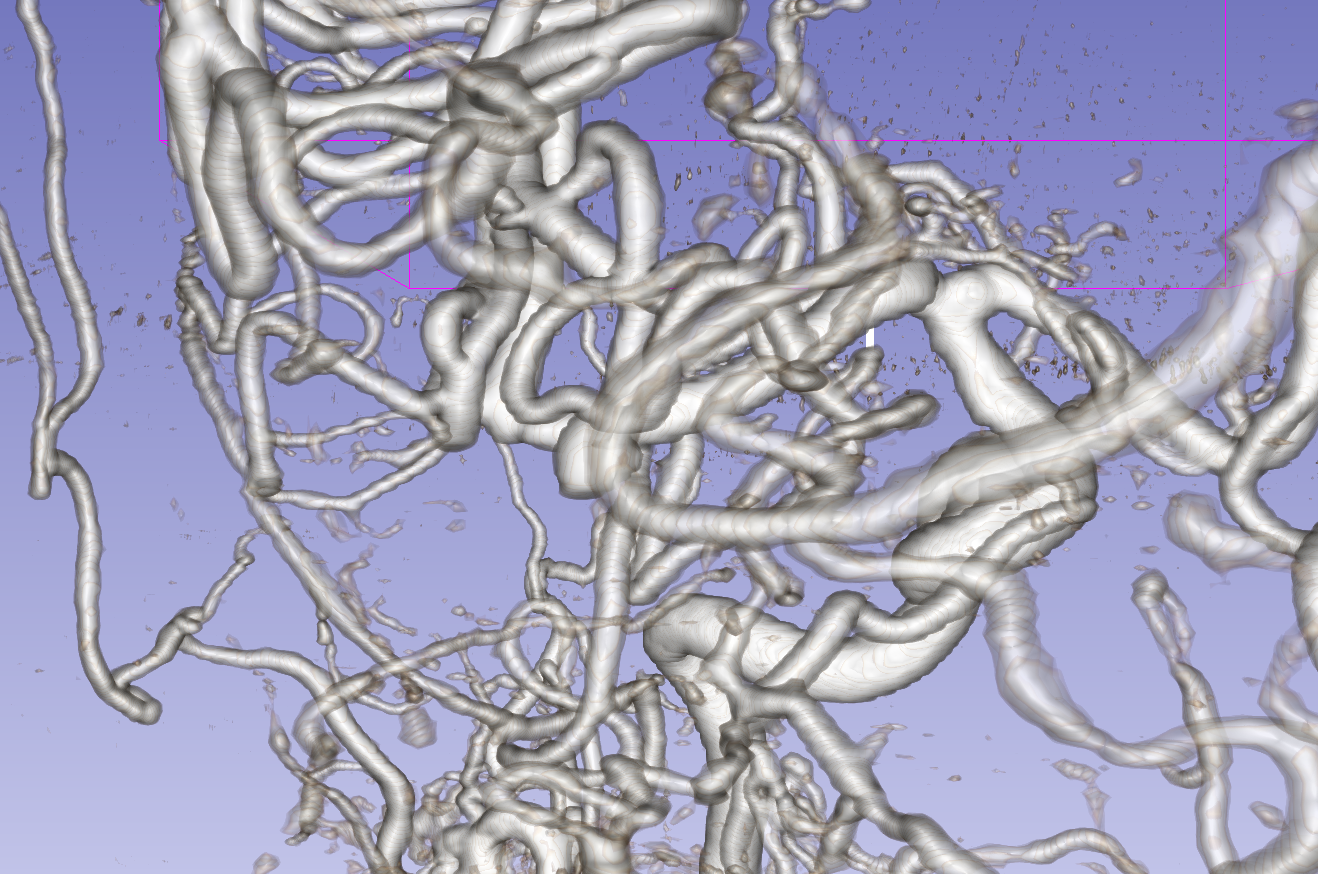}}
\subfigure[]
{\label{fig:artery2_original_image_4}\includegraphics[width=0.49\textwidth]{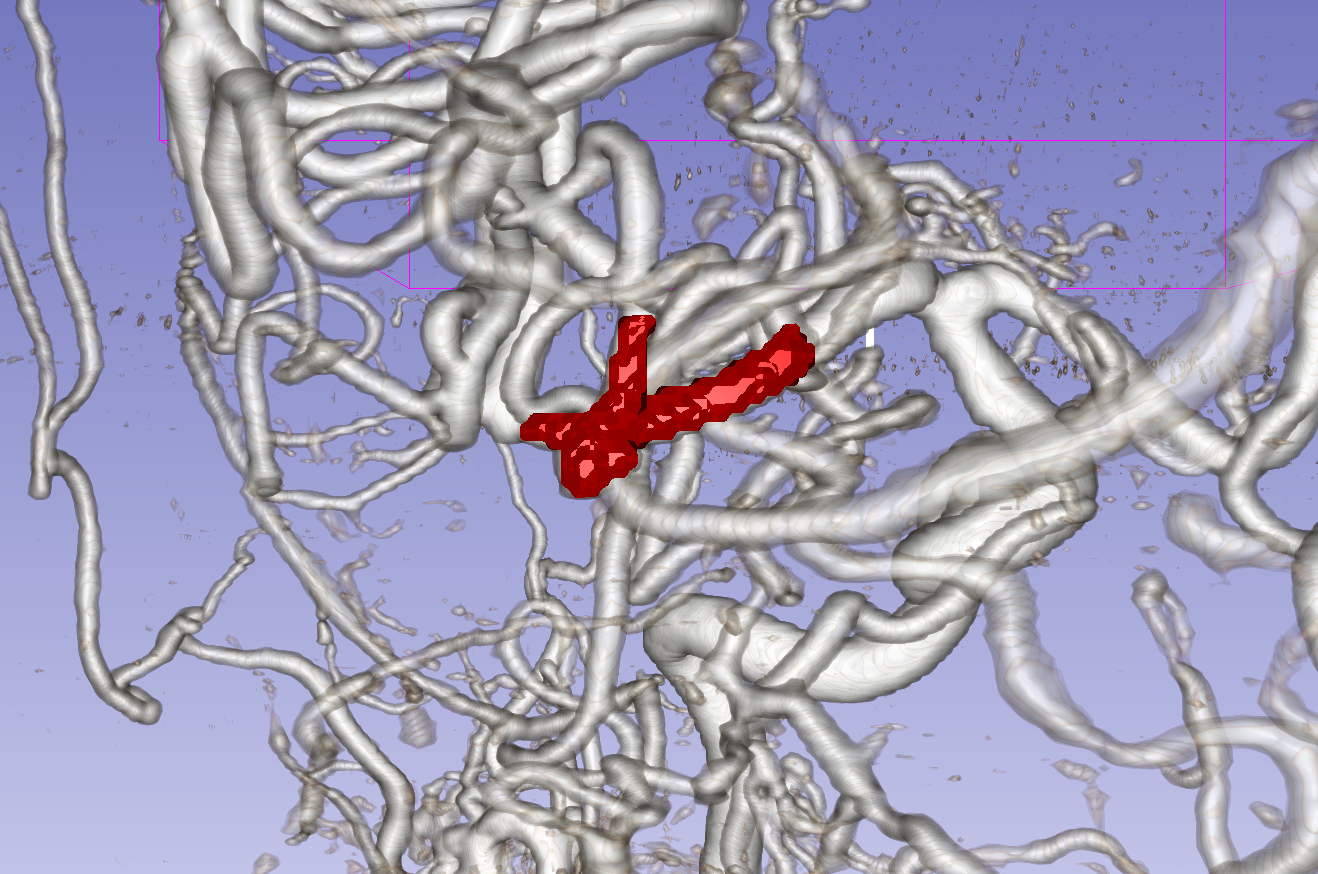}}
\subfigure[]
{\label{fig:artery2_plc_2}\includegraphics[scale=0.165]{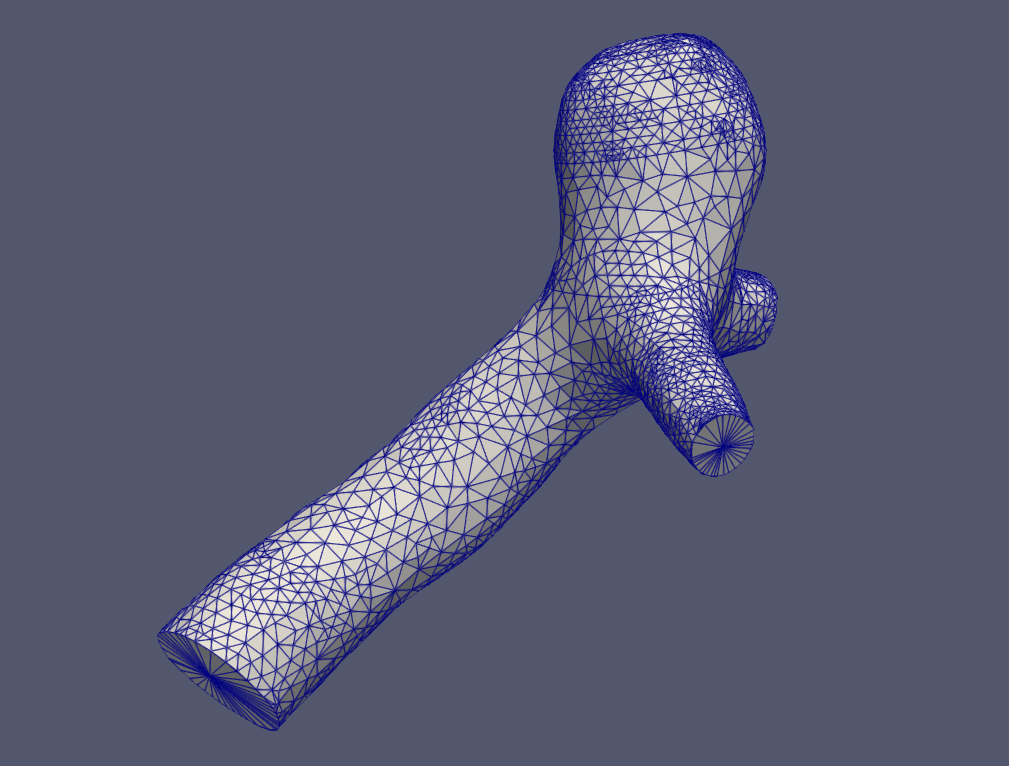}}
\caption{The original image of the second (middle cerebral artery bifurcation) aneurysm case is shown from an anteroposterior view. (a) shows the vascular structure imaged by rotational angiography of the right carotid artery. (b) zooms in to the region of interest while (c) highlights the PLC obtained from the segmentation of the aneurysm. (d) shows only the resulting PLC.}
\label{fig:artery2_images}
\end{figure}

\begin{figure}[htbp]
\centering
\subfigure[]{\label{fig:artery2_streamlines_1}\includegraphics[width=0.4\textwidth]{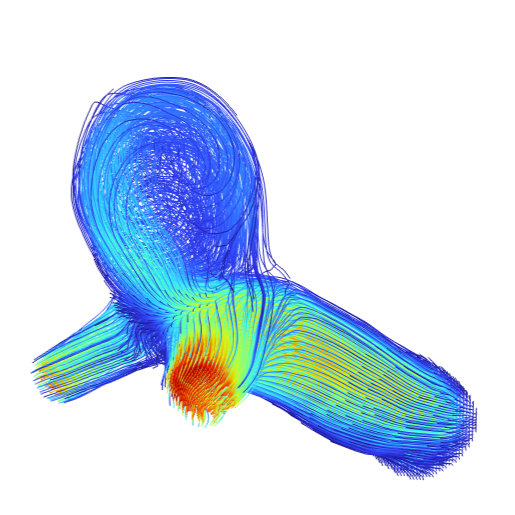}}
\subfigure[]{\label{fig:artery2_streamlines_2}\includegraphics[width=0.59\textwidth]{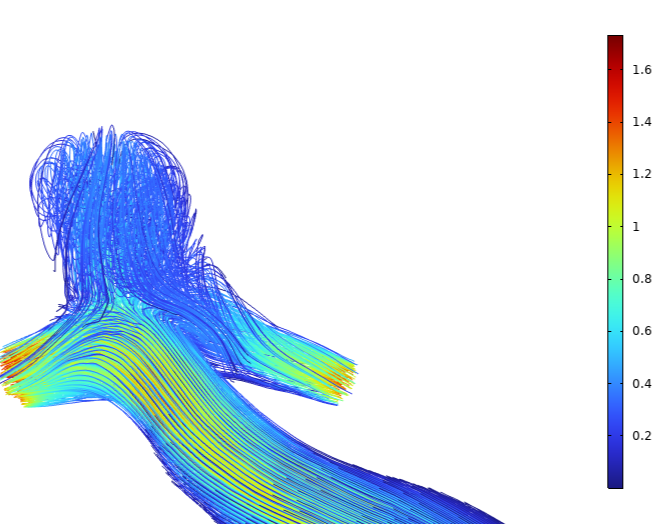}}
\caption{Shown is streamline data of the velocity field from a CFD simulation with the second (middle cerebral artery bifurcation) aneurysm case. (a) and (b) show different viewpoints. The velocity color legend is in m/s \cite{CFDDrivenTopology2024}.}
\label{fig:artery2_streamlines}
\end{figure}

The first aneurysm case is based on a rotational angiography scan of a carotid cavernous aneurysm with image spacing of 1.00 x 1.00 x 1.00 mm$^3$ and an image size of 512 x 512 x 508 voxels$^3$. The CBC3D mesh generated from it can be seen in Figure \ref{fig:artery1}, in addition to its approximate medial axis. The second case is a surface mesh of a middle cerebral artery bifurcation aneurysm given as a Piecewise Linear Complex (PLC) in a .stl format (seen in Figure \ref{fig:artery2_plc_2}). Figure \ref{fig:artery2_images} shows the vascular structure imaged by rotational angiography of the right carotid artery for the second aneurysm case, in addition to the given surface mesh obtained from a segmentation of the region of interest. Figure \ref{fig:artery2_streamlines} shows streamline data from a CFD simulation for this case (executed outside of this study \cite{CFDDrivenTopology2024}). Recall that we evaluate the adaptive anisotropic mesh generation method with metric tensor fields that are constructed using different techniques for each case. The full pipeline of the proposed approach is tested with the first case while some of the methods attempt to perform mesh generation/adaptation using only the surface mesh for the second case. All codes were compiled using GNU GCC 11.4.1. Data were collected on three supercomputers - Purdue University's Anvil \cite{AnvilCommunity2014, Anvil2022} and Old Dominion University's Wahab and Turing supercomputers. On Anvil, we utilized a dual socket node that features two AMD EPYC 7763 CPUs @ 2.45 GHz (64 slots each) and 256 GB of memory. On Wahab, we utilized a dual socket node that features two Intel Xeon Gold 6148 CPUs @ 2.40 GHz (20 slots each) and 384GB of memory. Finally, a dual socket node in the Turing supercomputer features two Intel Xeon E5-2698 v3 CPUs @ 2.30 GHz (16 slots each) and 128 GB of memory. For a performance evaluation, each run in the following experiments was executed three times and the results were averaged. 


For the generated isotropic meshes, qualitative results are examined with respect to element dihedral angles while quality is examined with respect to metric conformity for the adapted anisotropic meshes. The goal of metric conformity is to create a unit grid, where edges are unit-length and elements are unit-volume with respect to the target metric. For calculating edge length and element mean ratio, we adopted the same definitions that appear in \cite{ALAUZETSizeGradation}. Given that optimal edges should be unit-length, edges with length above or below one are considered to be sub-optimal. The measure for mean ratio is bounded between zero and one since it is normalized by the volume of an equilateral element. One is the optimal quality for an element's mean ratio shape. 





We also evaluate the performance of our parallel adaptive anisotropic mesh generation method by scaling the complexity of the metric tensor fields to generate much larger meshes (up to 100 million elements). The complexity \emph{C} of a continuous metric field \begin{math}\begin{mathcal}M\end{mathcal}\end{math} is defined as:

\begin{equation}
C(\begin{mathcal}M\end{mathcal}) = \int_\Omega \sqrt{det(\begin{mathcal}M\end{mathcal}(x))}dx.
\end{equation}

Complexity on the discrete grid is computed by sampling \(\begin{mathcal}M\end{mathcal}\) at each vertex \emph{i} as the discrete metric field \emph{M},

\begin{equation}
C(M) = \sum_{i=1}^{N} \sqrt{det(M_i)}V_i,
\end{equation}

\noindent where \(V_i\) is the volume of the Voronoi dual surrounding each node. The complexity of a grid is known to have a linear dependency with respect to the number of points and tetrahedra, shown theoretically in \cite{LoseilleAlauzetCMFI} and experimentally verified in \cite{LoseilleAlauzetCMFII, Park2015ComparingAO}. The number of vertices are approximately 2\emph{C} while the number of tetrahedra are approximately 12\emph{C}. As shown in \cite{LoseilleAlauzetCMFI, TsolakisEvaluation}, scaling the complexity of a metric can generate the same relative distribution of element density and shape over a uniformly refined grid compared to the original complexity. The metric tensor \(M_{C_r}\) that corresponds to the target complexity \(C_r\) is evaluated by \cite{LoseilleAlauzetCMFI}:

\begin{equation}
    M_{C_r} = \left(\frac{C_r}{C(M)}\right)^\frac{2}{3} M,
\end{equation}

\noindent where \emph{M} is the metric tensor before scaling and C(M) is the complexity of the discrete metric before scaling.

\subsection{Quantitative and Qualitative Analysis of I2M Pipeline} \label{pipeline_results}
The results obtained from CDT3D are also compared to PODM \cite{foteinos2014high} and TetGen \cite{TetGen2015} 
(replacing the adaptive anisotropic method with these adaptive isotropic methods). TetGen is a sequential, Delaunay-based method that generates isotropic tetrahedral meshes. In addition to an anisotropic metric field, an isotropic sizing function is also constructed (based on the aforementioned approach in \cite{ComparisonPhysicsBased2023}) specifically for PODM (also shown in algorithm \ref{alg:function_approximation}) and TetGen. We compare results between meshes generated by PODM and TetGen when they utilize this sizing function and when they do not (PODM's default image-to-mesh conversion for the first case and TetGen's tetrahedralization of the CBC3D surface). Additionally, we compare the PODM and TetGen isotropic meshes to isotropic meshes generated by CDT3D (again simply processing the same surface as a PLC - .stl).

Table \ref{table_input_parameters} shows the parameters used for each mesh generation method in the pipeline (except CBC3D). The first aneurysm case was also used to evaluate CBC3D in \cite{EwCCBC3D} (as it was compared to other isotropic I2M conversion methods from industry and academia). The same parameters utilized for CBC3D in that study are utilized in this evaluation. A wide range of values were tested for the value of k used in the k-nn search when building the sizing function and metric tensor fields for the first case. 
The optimal parameters were determined based on the best quality mesh that was generated. If certain parameters are not listed in Table \ref{table_input_parameters}, then their default value was utilized. The optimal value for k with the isotropic runs of TetGen and PODM is 30. The optimal value for k in the metric-adapted  
anisotropic runs of CDT3D is 40. 
The CDT3D parameters \textit{nqual} and \textit{nsmth} were both increased to 10 while \textit{cdfn} was changed to 0.8 to improve the final anisotropic mesh quality for the second aneurysm case (with no surface adaptation by CDT3D, explained later).

\begin{table}[htbp]\footnotesize
\caption{Input parameters for each mesh generation method utilized in the pipeline (after CBC3D). Val. is an abbreviation for value.}
\centering
\begin{tabular}{lcc}
\hline
Method & Parameter/Val. & Description \\
\hline
\multirow{6}{*}{TetGen (w/ adaptivity)} & r & reconstruct mesh \\
 & m & use sizing \\
 & q & optimize quality \\
 & R & reduce elements \\
 & O/9 & optimize mesh \\
 & Y & freeze surface \\
\hline
\multirow{5}{*}{TetGen (no adaptivity)} & p & mesh plc \\
 & q & improve quality \\
 & R & reduce elements \\
 & O/9 & optimize mesh \\
 & Y & freeze surface \\
\hline
\multirow{2}{*}{PODM} & $\delta$/1.2 & element size \\
& nthreads/* & parallel threads \\
& adapt\_const/2 & heuristic for level of adaptation \\
\hline
\multirow{4}{*}{AFLR (bl)} & mbl/1 & boundary layer \\
 & blds/1.2 & initial bl spacing \\
 & (for case 1) & \\
 & blds/0.12 & initial bl spacing \\
 & (for case 2) & \\
 & fints/0, & freeze surface \\
 & open & only bl \\
\hline
\multirow{2}{*}{CDT3D (iso)} & sref/0 & freeze surface \\
 & nthreads/* & parallel threads \\
\hline
CDT3D (aniso) & met/1 & use metric \\
 & meval/1 & metric-vertices \\
 & sref/0 & freeze surface \\
 & reconstruct/1 & reconstruct mesh \\
 & mpp/5 & point creation \\
 & & method - centroid \\
 & cdff/1.2 & edge length mult. \\
 & cdfn/0.6 & nearby node dist. \\
 & nqual/3 & optimize quality \\
 & nsmth/5 & smooth points \\
 & nthreads/* & parallel threads \\
\hline
\end{tabular}
\label{table_input_parameters}
\end{table}

\begin{figure}[htbp]
\centering
\subfigure[CBC3D mesh with an approximate medial axis]{\label{fig:artery1}\includegraphics[width=0.555\textwidth]{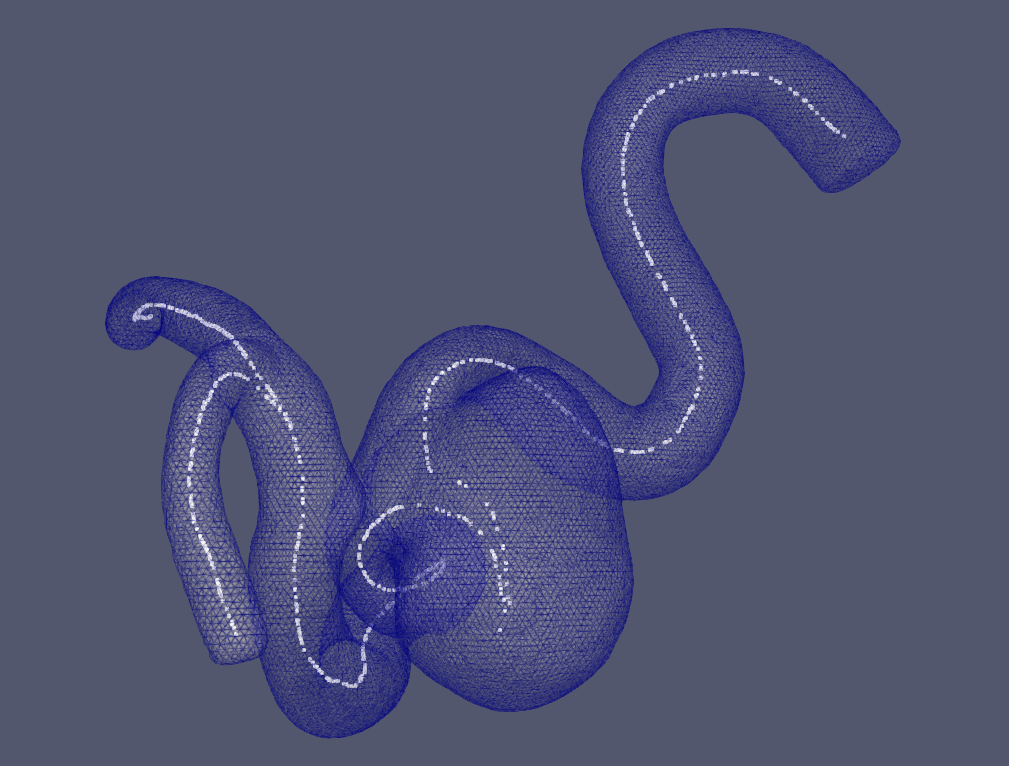}}
\subfigure[CBC3D Volume]{\label{fig:cbc3d_artery1_cut}\includegraphics[width=0.425\textwidth]{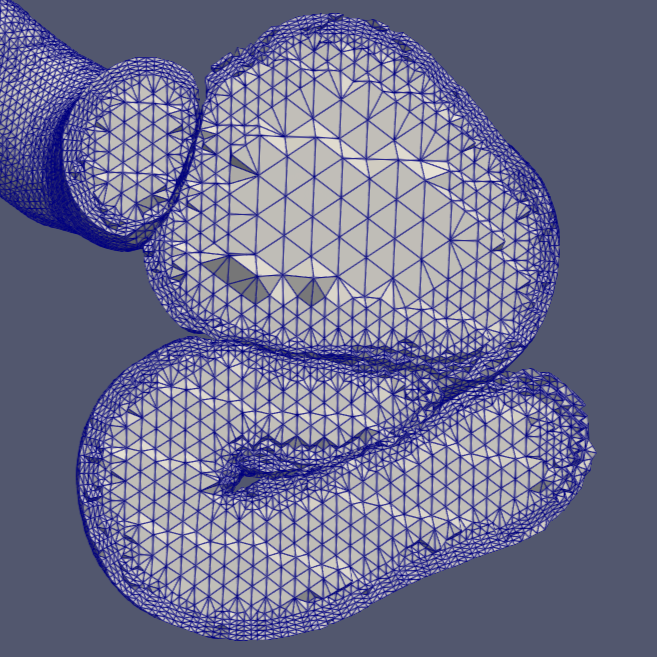}}
\subfigure[TetGen (isotropic from PLC)]{\label{fig:tetgen_iso_artery1_cut}\includegraphics[width=0.49\textwidth]{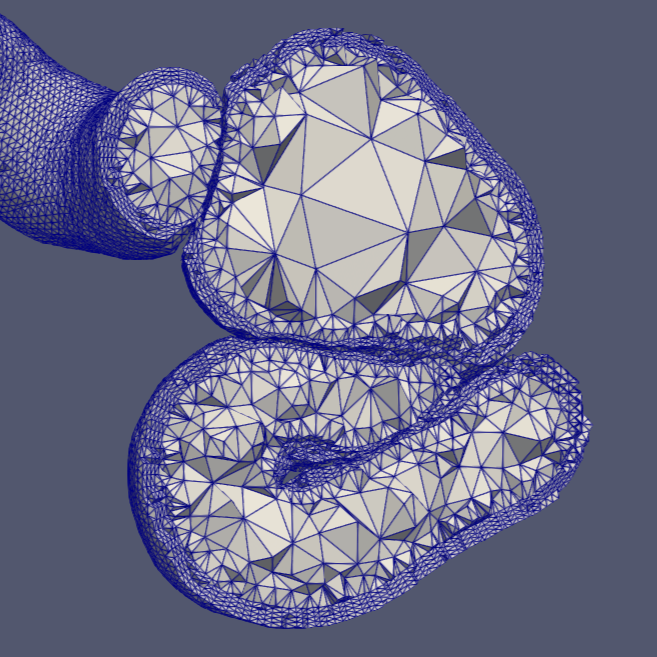}}
\subfigure[TetGen (isotropic with sizing function)]{\label{fig:tetgen_iso-sizing_artery1_cut}\includegraphics[width=0.49\textwidth]{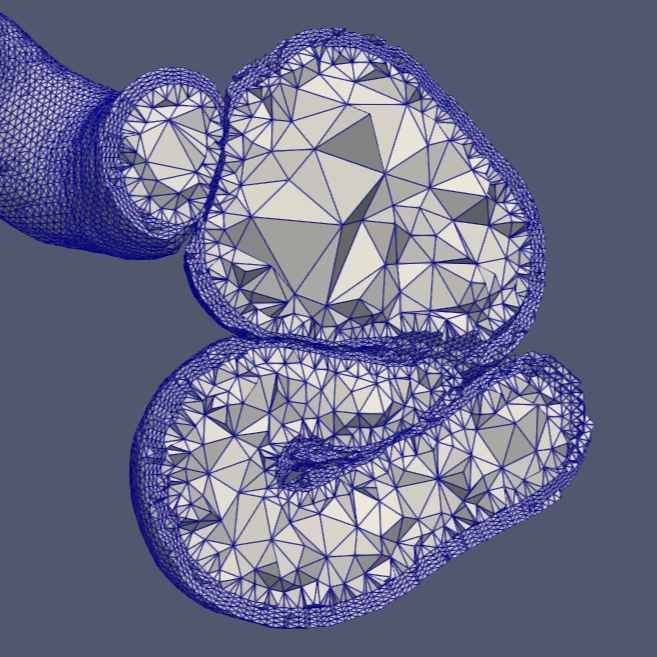}}
\caption{Cross sections are shown of the volume meshes generated by CBC3D and TetGen for the first aneurysm case, merged with the conforming boundary layer mesh generated by AFLR.}
\label{fig:Mesh_Cuts_1_artery1}
\end{figure}

\begin{figure}[htbp]
\centering
\subfigure[PODM (isotropic, no vol. adaptivity)]{\label{fig:podm_no_adapt_artery1_cut}\includegraphics[width=0.49\textwidth]{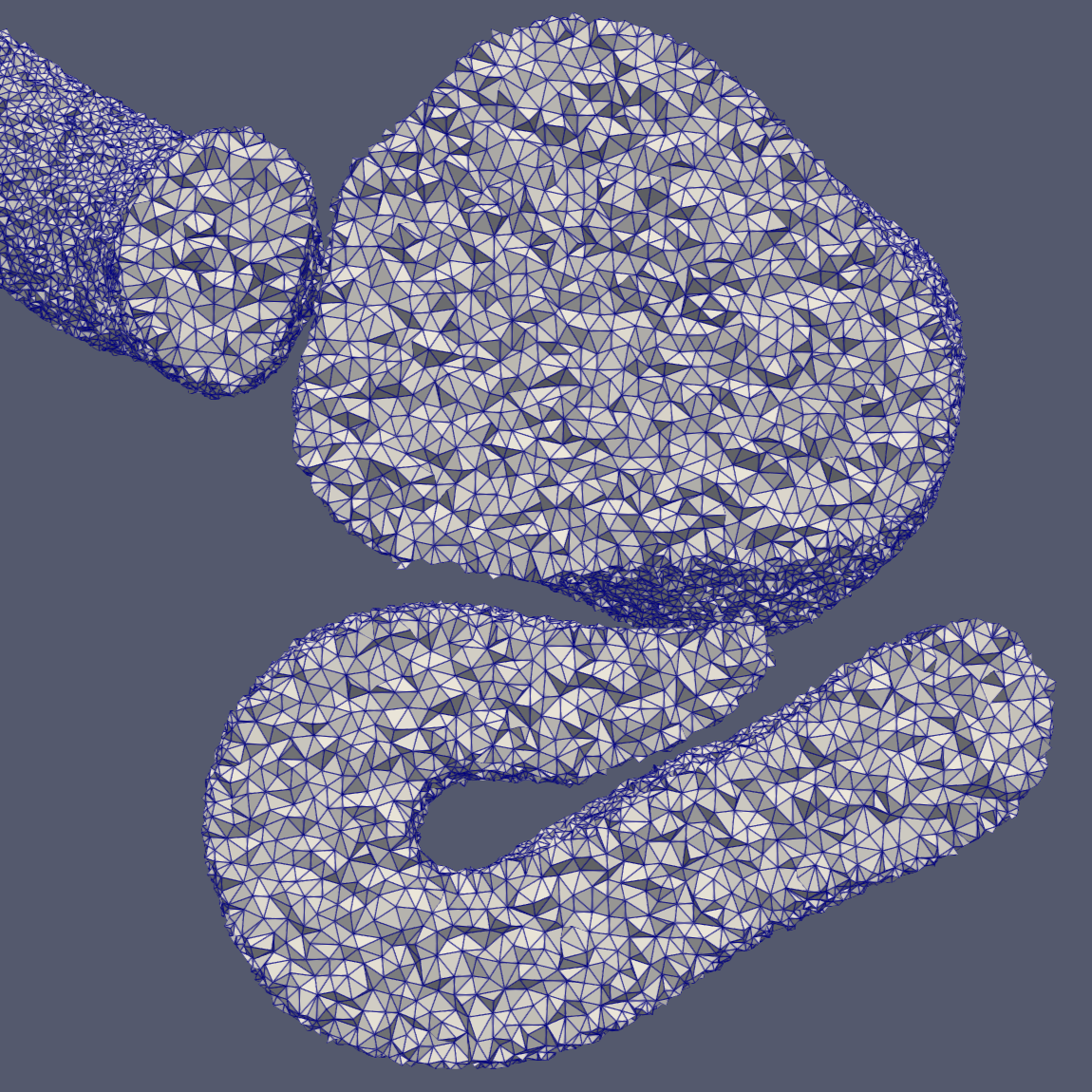}}
\subfigure[PODM (isotropic w/ vol. adaptivity)]{\label{fig:podm_adapt_artery1_cut}\includegraphics[width=0.49\textwidth]{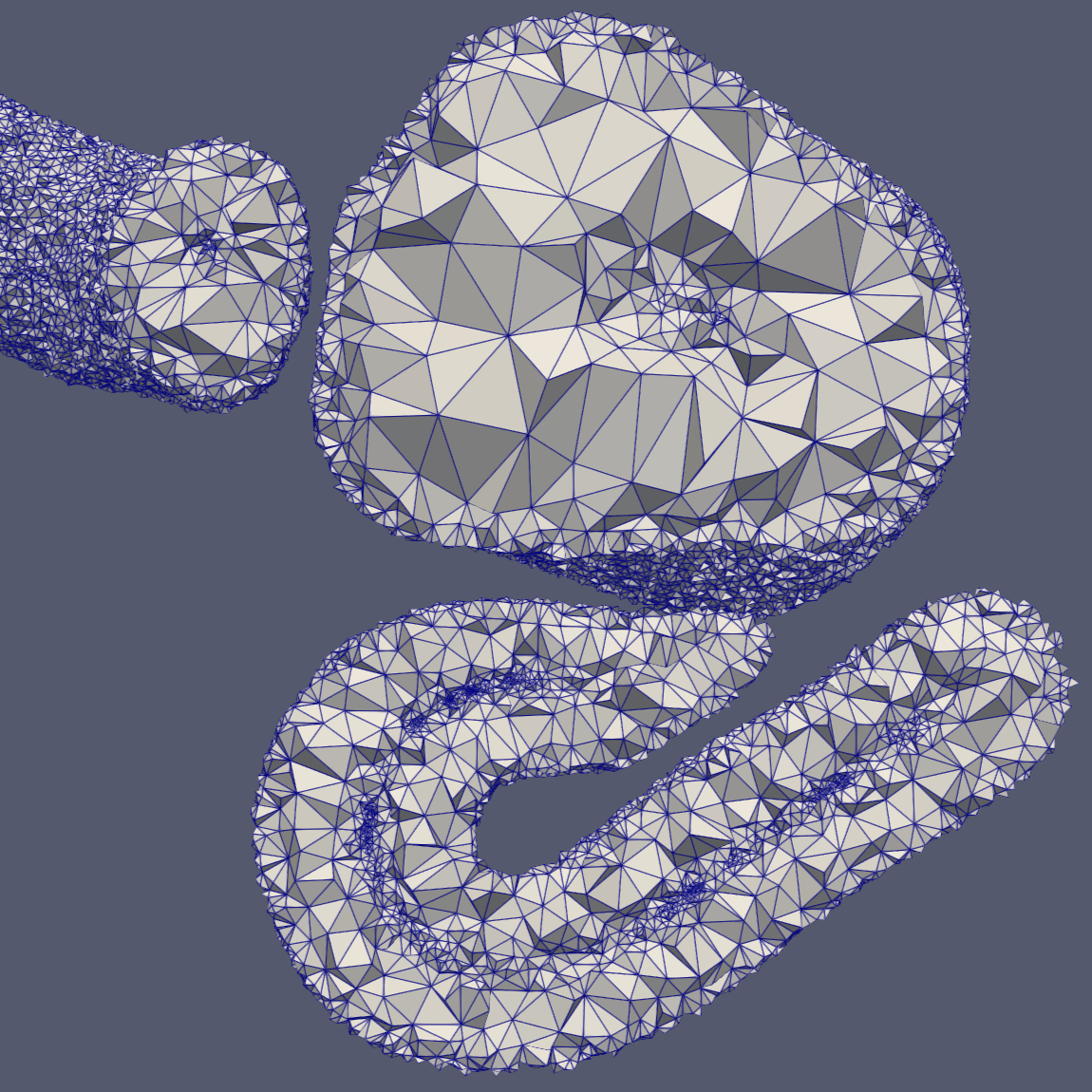}}\\
\subfigure[CDT3D (isotropic from PLC)]{\label{fig:cdt3d_iso_artery1_cut}\includegraphics[width=0.49\textwidth]{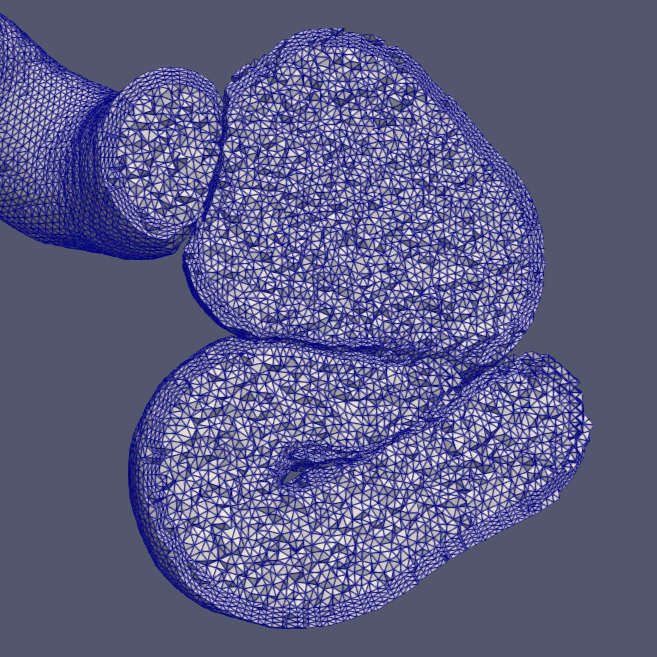}}
\subfigure[CDT3D (anisotropic with metric tensor field)]{\label{fig:cdt3d_aniso_artery1_cut}\includegraphics[width=0.49\textwidth]{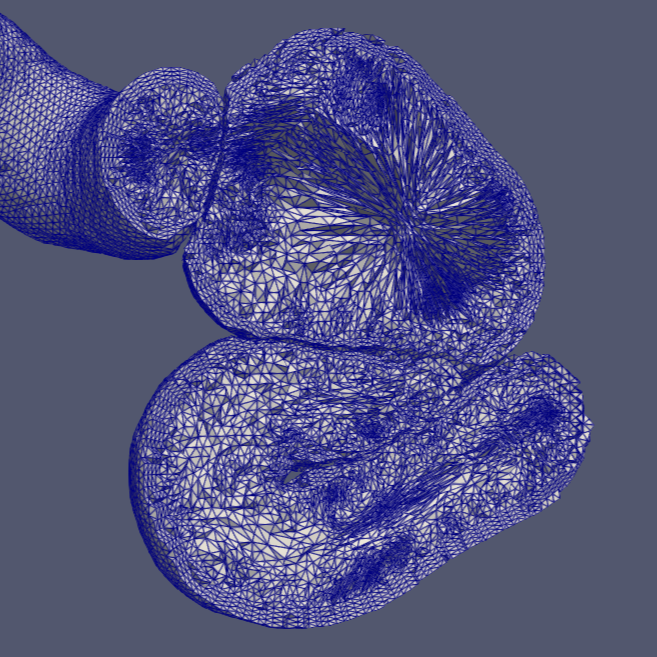}}
\caption{Cross sections are shown of the volume meshes generated by PODM and CDT3D for the first aneurysm case, where the CDT3D meshes are merged with the conforming boundary layer mesh generated by AFLR.}
\label{fig:Mesh_Cuts_2_artery1}
\end{figure}

\begin{figure}[htbp]
\centering
\subfigure[TetGen (isotropic from PLC)]{\label{fig:tetgen_artery2_cut}\includegraphics[width=0.49\textwidth]{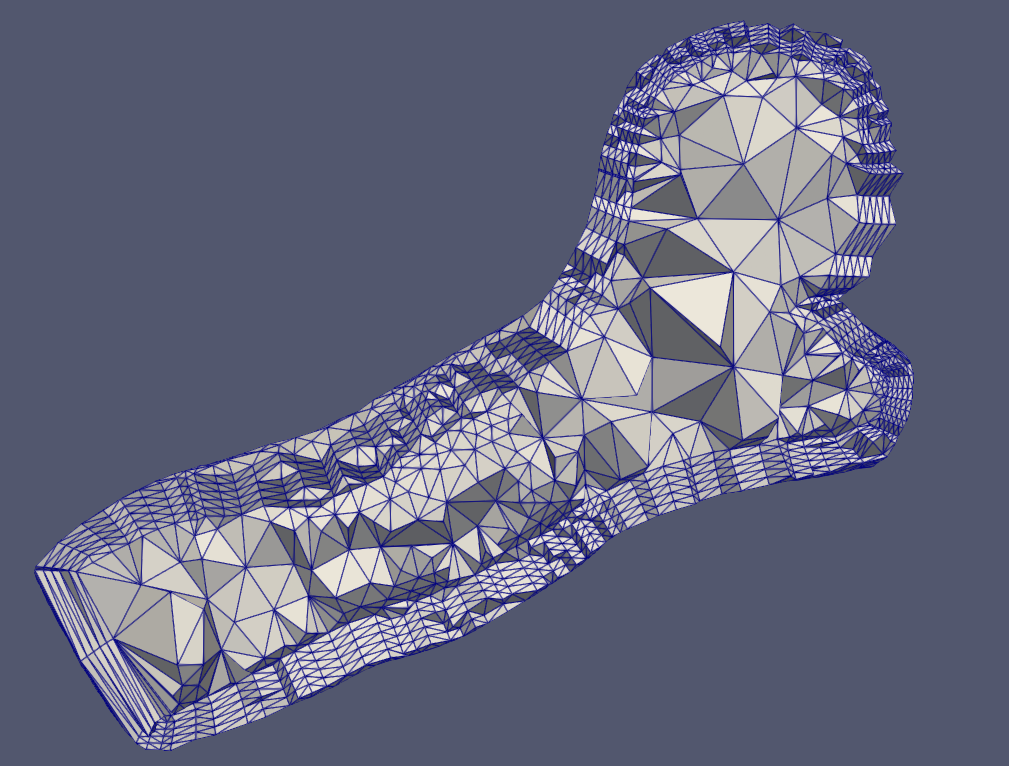}}
\subfigure[CDT3D (isotropic from PLC)]{\label{fig:cdt3d_artery2_cut_aflr_iso}\includegraphics[width=0.49\textwidth]{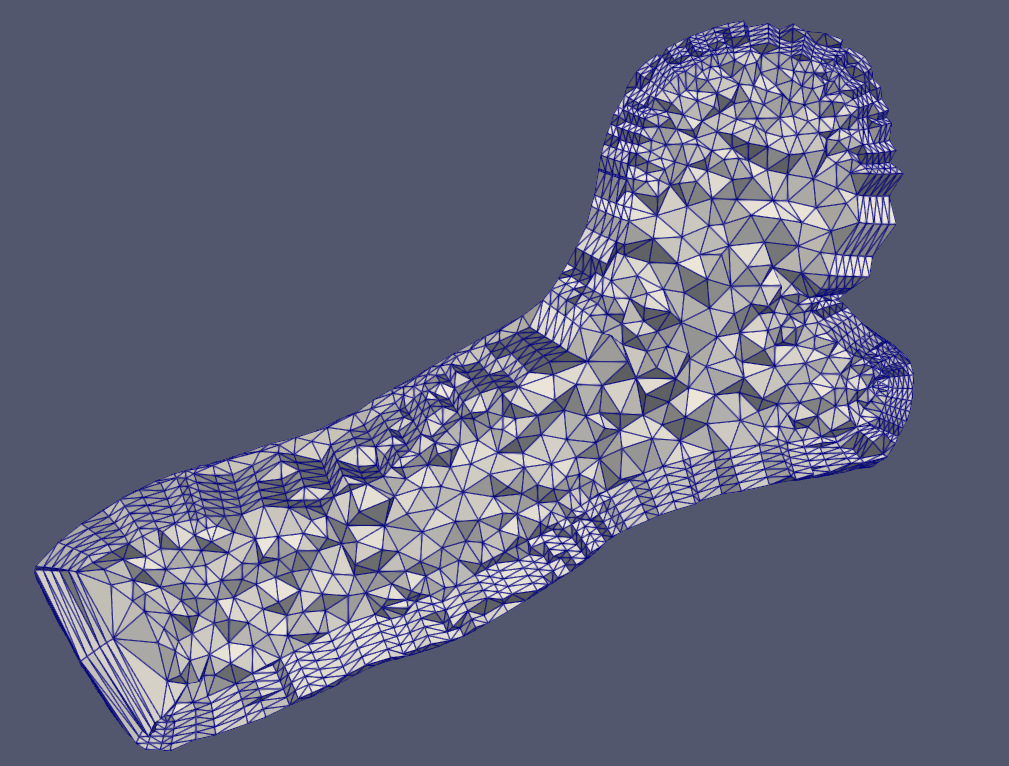}}\\
\subfigure[CDT3D (anisotropic with velocity-based metric tensor field) view 1]{\label{fig:cdt3d_artery2_cut_aniso_bottom}\includegraphics[width=0.43\textwidth]{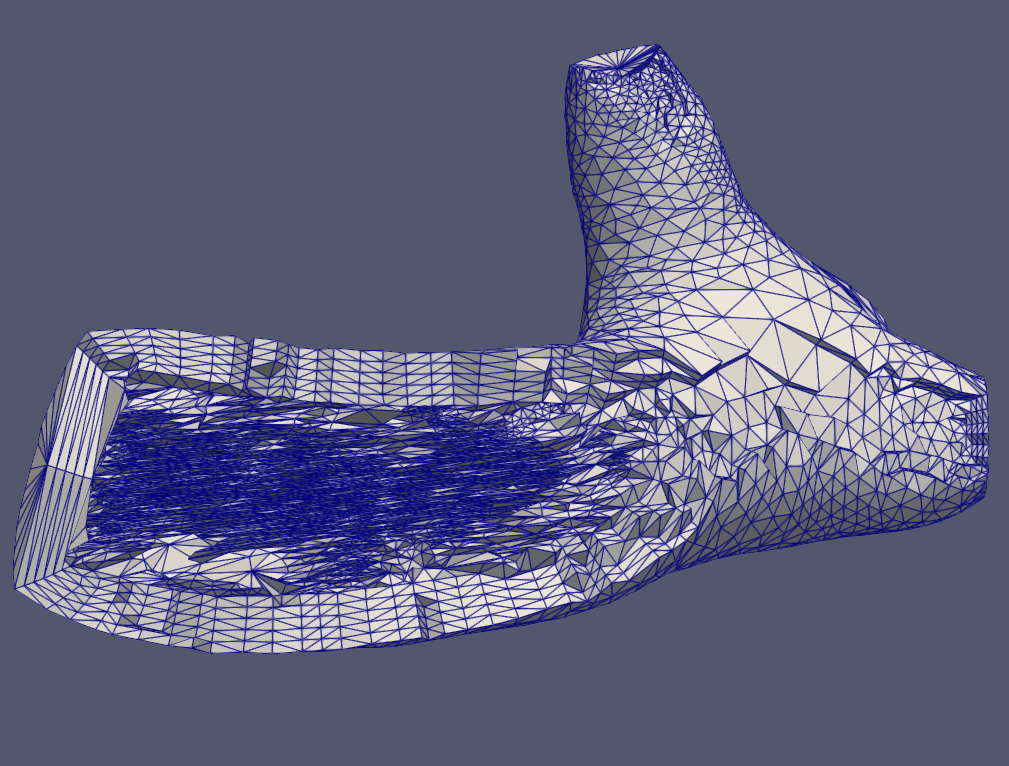}}
\subfigure[CDT3D (anisotropic with velocity-based metric tensor field) view 2]{\label{fig:cdt3d_artery2_cut_aniso_side}\includegraphics[width=0.492\textwidth]{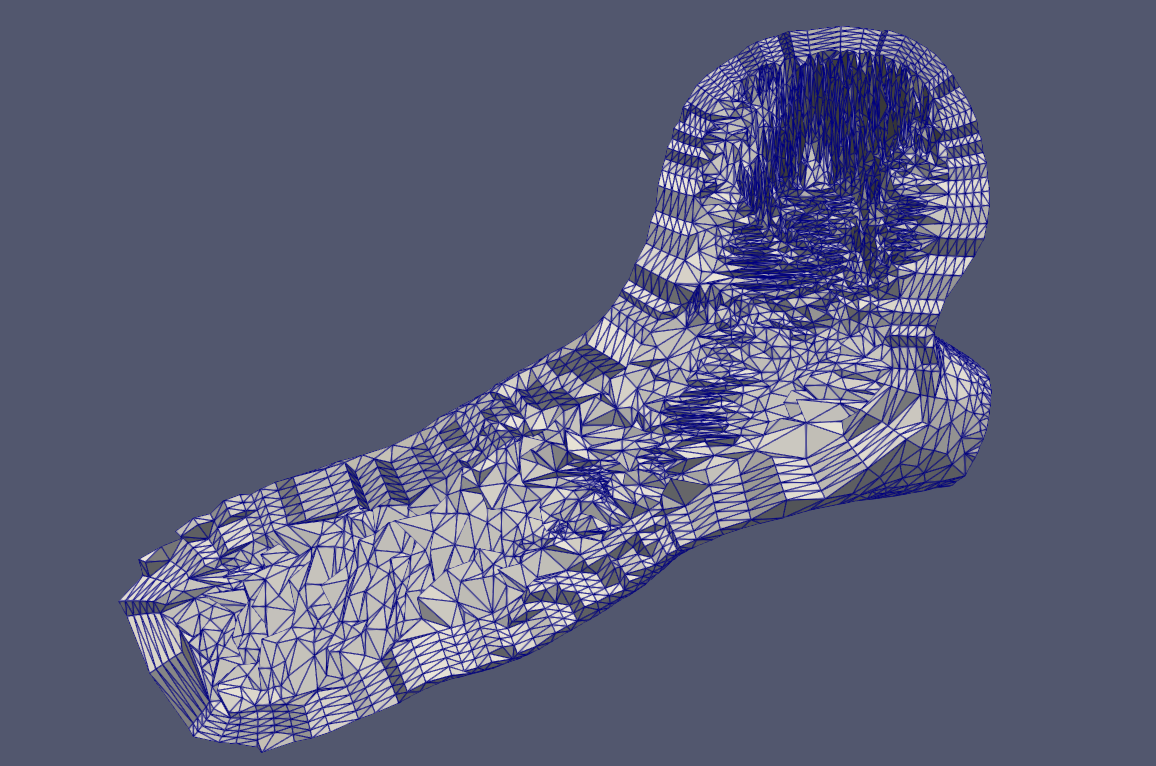}}
\caption{Cross sections are shown of the volume meshes generated by each method for the second aneurysm case, merged with the conforming boundary layer mesh generated by AFLR.}
\label{fig:Mesh_Cuts_artery2}
\end{figure}

CBC3D takes approximately 61.59 seconds (1 min) to convert the segmented image of the first aneurysm into a mesh of approximately 272K tetrahedra and 63K points. Figures \ref{fig:Mesh_Cuts_1_artery1} and \ref{fig:Mesh_Cuts_2_artery1} show cross sections of the volume meshes generated by each method, merged with the conforming AFLR boundary layer mesh. Figure \ref{fig:Mesh_Cuts_artery2} shows cross sections of the volume meshes generated for the second aneurysm case. Table \ref{table_number_of_elements} shows the number of elements generated by each method when processing the CBC3D mesh of aneurysm 1 and the surface mesh of aneurysm 2. The TetGen meshes contain fewer elements compared to the CDT3D meshes in both cases. Tables \ref{table_artery_1_performance} and \ref{table_artery_2_performance} show the runtime of each method for aneurysm cases 1 and 2, respectively. They also show the mesh generation/adaptation rates of each method (measured in elements generated per second). TetGen exhibits good performance when generating its small-size meshes. Although it is not listed in Table \ref{table_artery_1_performance}, it should be noted that CDT3D exhibits good scalability, as its runtime is reduced to approximately 1 minute and its anisotropic adaptation rate becomes 27K when utilizing 40 cores. When utilizing 10 cores, CDT3D's mesh generation/adaptation rate exceeds that of TetGen's. For the second case, CDT3D completes anisotropic adaptation in about 49 seconds when utilizing 40 cores. Using the same number of cores, PODM takes about 2 seconds (about 260K elements/sec) to generate an isotropic mesh for the first case without utilizing the sizing function and 4 seconds (about 135K elements/sec) when it does.

\begin{table}[htb]\footnotesize
\caption{Number of tetrahedra and points generated by each method when processing either the segmented image (PODM) or the CBC3D mesh converted from the segmented image of the first (carotid cavernous) aneurysm and the surface mesh of the second (middle cerebral artery bifurcation) aneurysm. `Tets' is short for tetrahedra. K means thousand and M means million.}
\centering
\begin{tabular}{lc|cc}
\hline
Methods & & Aneurysm 1 & Aneurysm 2 \\
\hline
\multirow{2}{*}{TetGen (w/ adaptivity)} & Tets & 120K & - \\
 & Points & 37K & - \\
\hline
\multirow{2}{*}{TetGen (no adaptivity)} & Tets & 118K & 12K \\
 & Points & 36K & 3K \\
\hline
\multirow{2}{*}{PODM (no vol. adaptivity)} & Tets & 599K & - \\
 & Points & 131K & - \\
 \hline
\multirow{2}{*}{PODM (w/ vol. adaptivity)} & Tets & 644K & - \\
 & Points & 141K & - \\
\hline
\multirow{2}{*}{AFLR (bl)} & Tets & 912K & 82K \\
 & Points & 183K & 16K \\
\hline
\multirow{2}{*}{CDT3D (iso)} & Tets & 631K & 31K \\
 & Points & 122K & 7K \\
\hline
\multirow{2}{*}{CDT3D (aniso)} & Tets & 1.59M & 80K \\
 & Points & 284K & 15K \\
\hline
\end{tabular}
\label{table_number_of_elements}
\end{table}

\begin{table}[htbp]\footnotesize
\caption{The time spent (approximately in \textbf{minutes}) and the number of elements generated per second by each method (in parentheses) for small-size meshes on the Wahab supercomputer is shown when processing either the segmented image (PODM) or the CBC3D mesh converted from the segmented image of the first (carotid cavernous) aneurysm. Instead of elements per second, the values in parentheses for Build\_Sizing/Metric represent the CBC3D volume mesh points processed per second. `K' means thousand.}
\centering
\begin{tabular}{l|ccc}
\hline
& \multicolumn{3}{c}{CPU Cores} \\
Methods & 1 & 10 & 20 \\\hline
Build\_Sizing/Metric & 191 (5K) & 19 (55K) & 9 (117K) \\
TetGen (w/ adaptivity) & 0.1 (20K) & - & - \\
TetGen (no adaptivity) & 0.06 (32K) & - & - \\
PODM (no vol. adaptivity) & 0.35 (28K) & 0.05 (167K) & 0.04 (217K) \\
PODM (w/ vol. adaptivity) & 1.1 (9K) & 0.15 (67K) & 0.1 (101K) \\
AFLR (bl) & 0.43 (35K) & - & - \\
CDT3D (iso) & 0.6 (17K) & 0.3 (42K) & 0.1 (78K) \\
CDT3D (aniso) & 20 (1K) & 3 (8K) & 2 (14K) \\
\hline
\end{tabular}
\label{table_artery_1_performance}
\end{table}

\begin{table}[htbp]\footnotesize
\caption{The time spent (approximately in \textbf{seconds}) and the number of elements generated per second by each method (in parentheses) for small-size meshes on the Wahab supercomputer is shown when processing the surface mesh of the second (middle cerebral artery bifurcation) aneurysm. `K' means thousand.}
\centering
\begin{tabular}{l|ccc}
\hline
& \multicolumn{3}{c}{CPU Cores} \\
Methods & 1 & 10 & 20 \\\hline
TetGen (no adaptivity) & 1.54 (8K) & - & - \\
AFLR (bl) & 3.2 (25K) & - & - \\
CDT3D (iso) & 2.2 (14K) & 1.73 (18K) & 1.24 (25K) \\
CDT3D (aniso) & 910 (87) & 105 (761) & 69 (1K) \\
\hline
\end{tabular}
\label{table_artery_2_performance}
\end{table}

Mesh fidelity is evaluated using a two-sided Hausdorff Distance (HD) metric (based on an open source implementation \cite{Commandeur}): $\text{HD} = \max\{ \text{HD}_{\text{I}\rightarrow \text{M}}, \text{HD}_{\text{M}\rightarrow \text{I}}\}$, where $\text{HD}_{\text{I}\rightarrow \text{M}}$ is the value of the metric from the image to the mesh, and $\text{HD}_{\text{M}\rightarrow \text{I}}$ is the value of the metric from the mesh to the image. A low HD error indicates a high fidelity. Hausdorff Distance is computed between two point sets. The first point set contains the vertices located on the mesh surface and the second point set contains the voxels located on the boundaries of the segmented material. The HD for the CBC3D-generated mesh of the first aneurysm case (in Figure \ref{fig:artery1}) is 4.51. The meshes generated by PODM (Figures \ref{fig:podm_no_adapt_artery1_cut} and \ref{fig:podm_adapt_artery1_cut}) also exhibit good fidelity, as their HD are both 2.42. The fidelity of the meshes generated by CBC3D and PODM were also evaluated for the first aneurysm case in \cite{EwCCBC3D}. They were both compared to several highly used I2M methods, where CBC3D exhibited good fidelity while maintaining better smoothness compared to the other methods' meshes (including PODM). Irrespective of PODM, surface refinement/adaptation is turned off for all mesh generation methods tested in the pipeline (except CBC3D). This ensures that the high fidelity and smooth surface generated by CBC3D remains intact. 

\begin{figure}[!htb]
\centering
\subfigure[]{\label{fig:artery1_dihedral_overall}\includegraphics[width=0.9\textwidth]{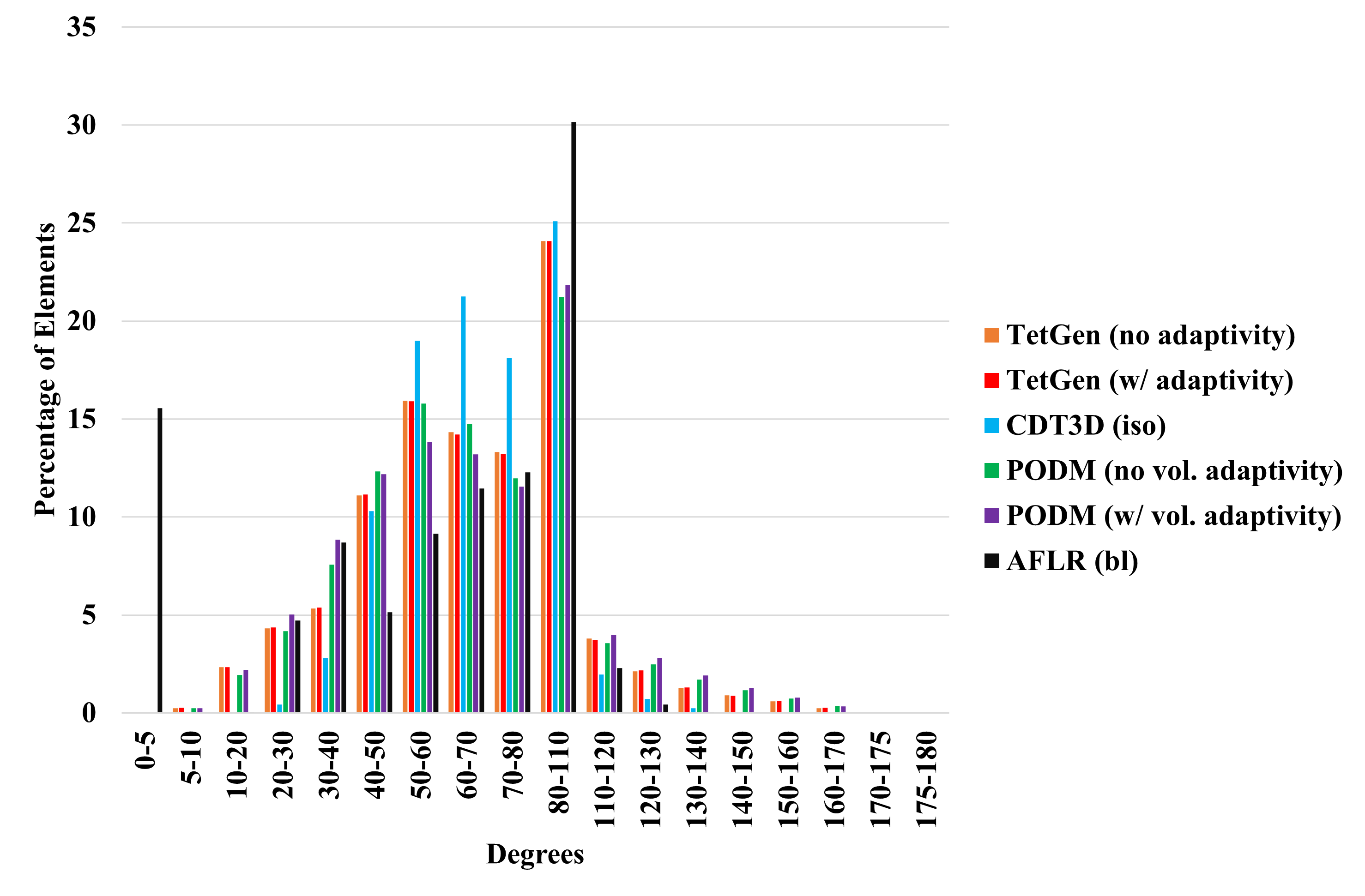}}\\
\subfigure[]{\label{fig:artery1_dihedral_small}\includegraphics[width=0.49\textwidth]{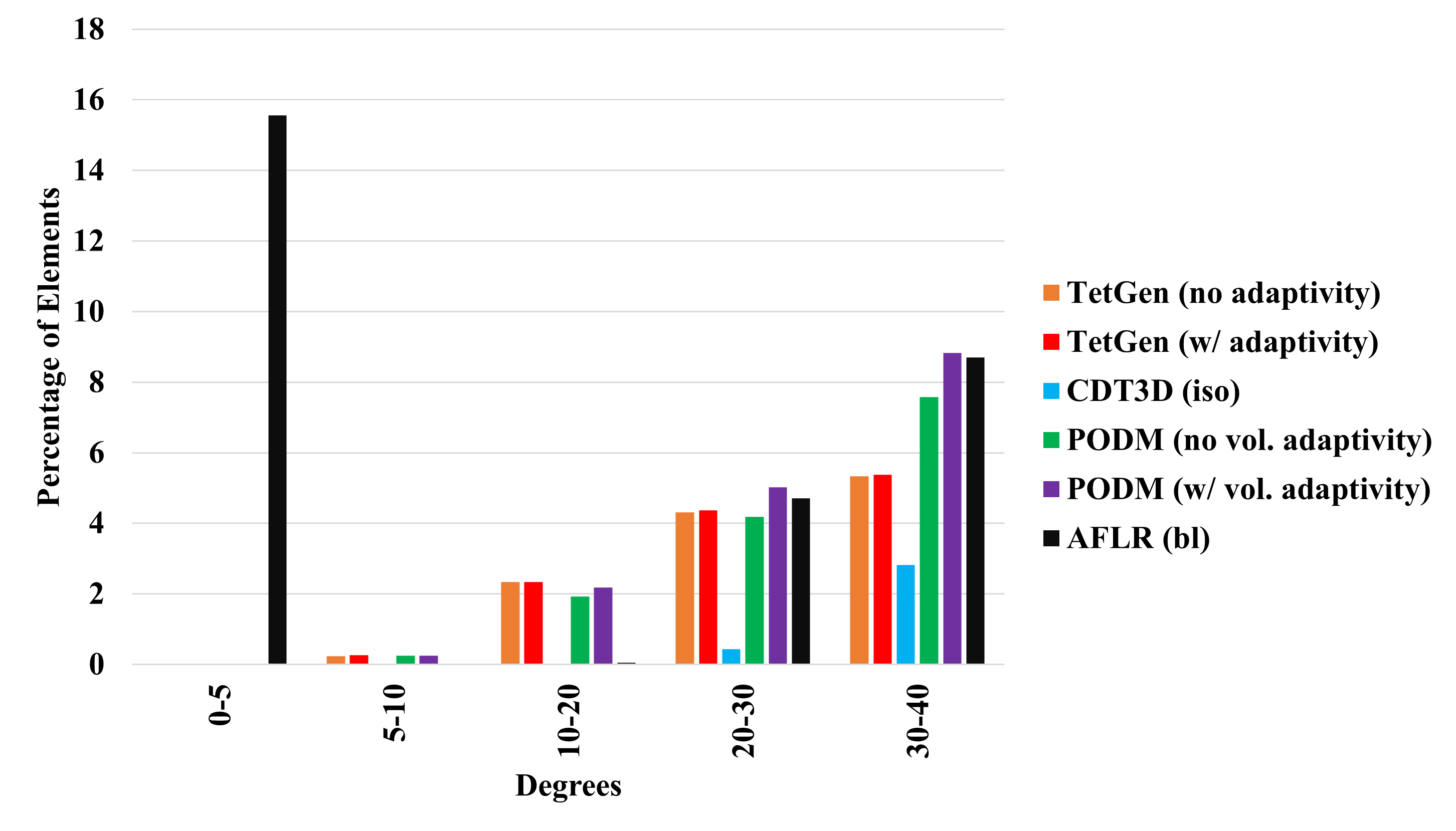}}
\subfigure[]{\label{fig:artery1_dihedral_large}\includegraphics[width=0.49\textwidth]{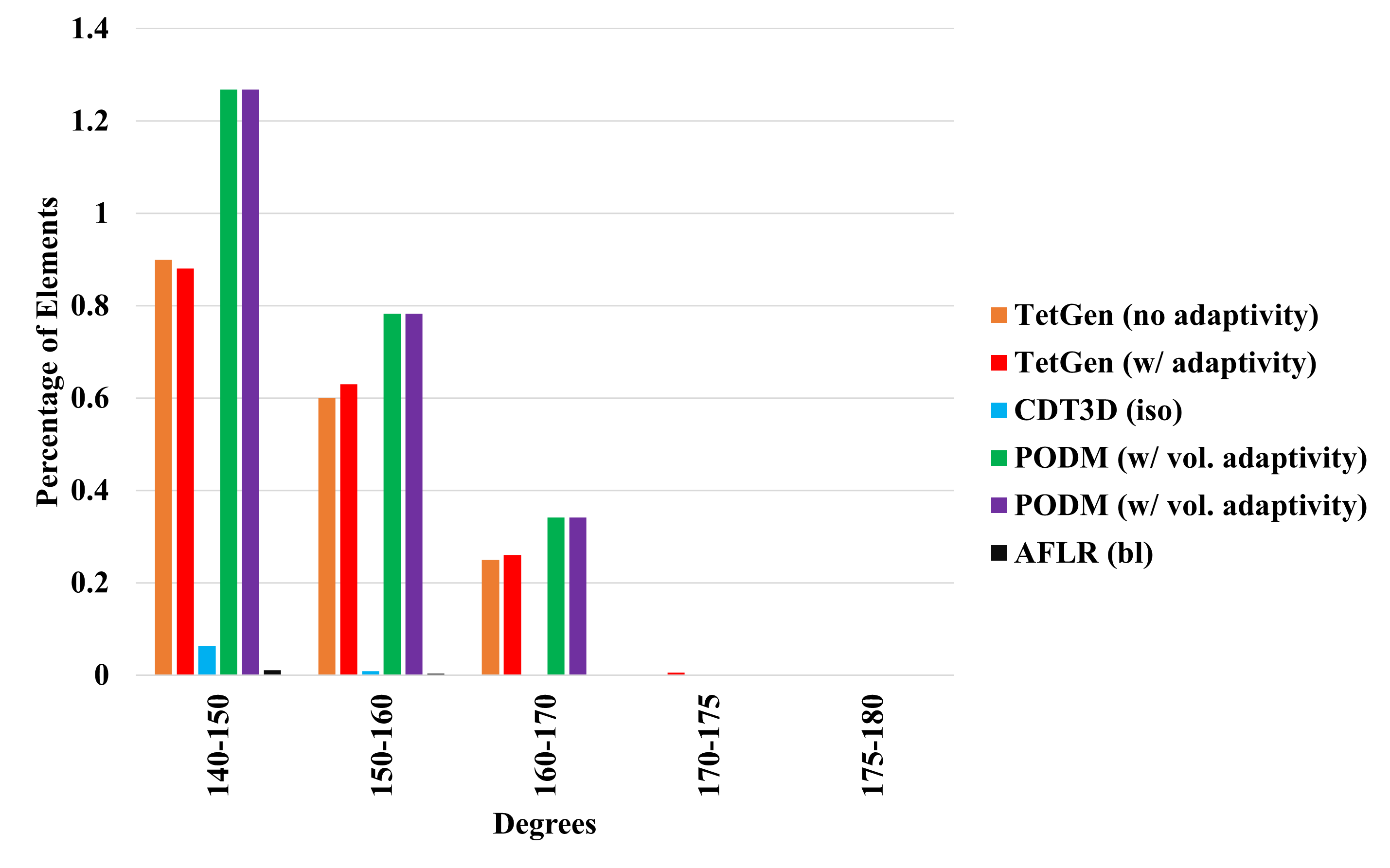}}
\caption{Quality statistics are shown comparing the dihedral angles of the isotropic meshes generated by TetGen, CDT3D, PODM, and AFLR (boundary layer) for the first (carotid cavernous) aneurysm case. (a) shows the full distribution of element dihedral angles. (b) shows the distribution of elements with dihedral angles between 0 and 40 degrees. (c) shows the distribution of elements with dihedral angles between 140 and 180 degrees.}
\label{fig:artery1_dihedrals}
\end{figure}

\begin{figure}[!htb]
\centering
\subfigure[]{\label{fig:artery2_dihedral_overall}\includegraphics[width=0.9\textwidth]{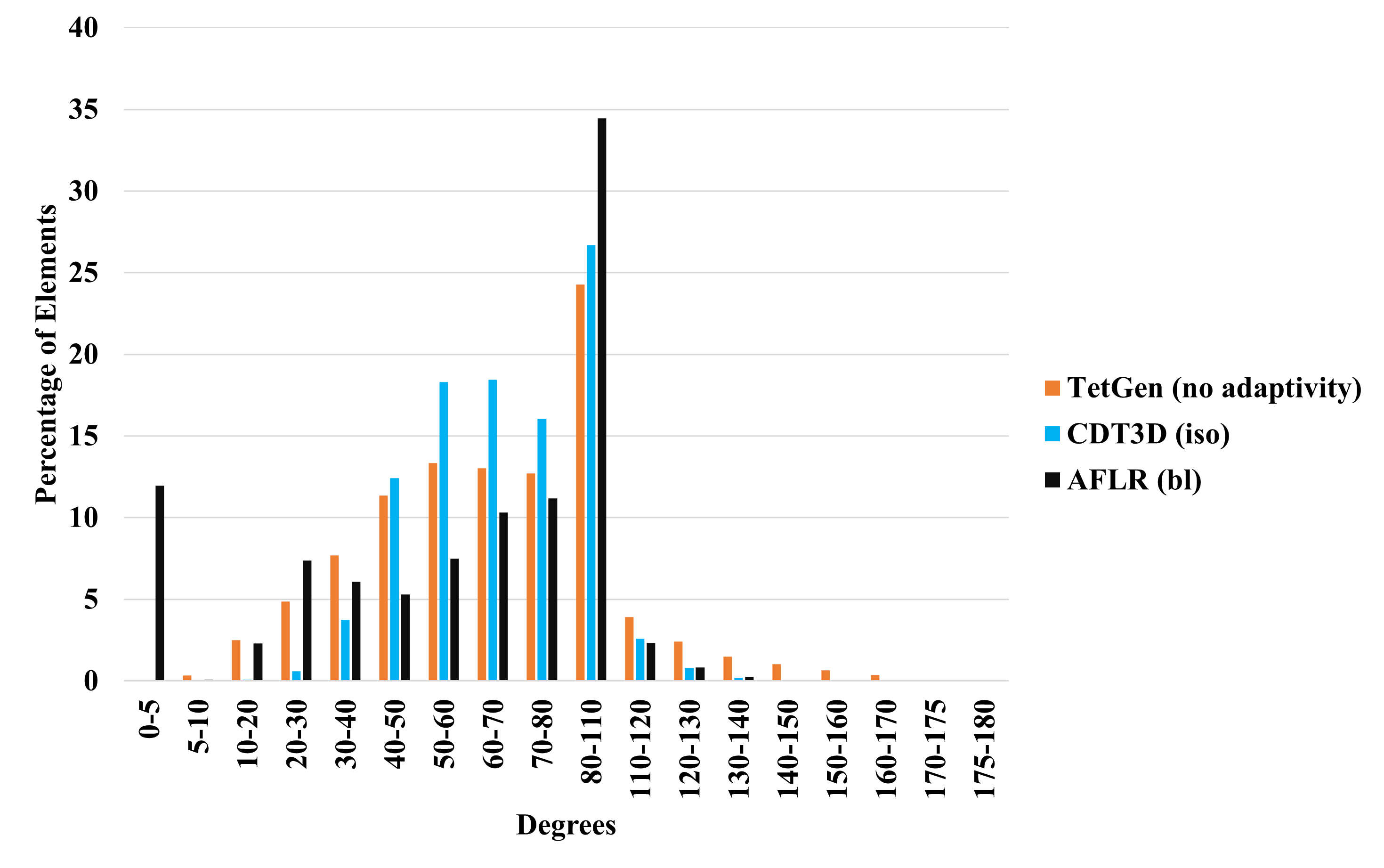}}
\subfigure[]{\label{fig:artery2_dihedral_small}\includegraphics[width=0.49\textwidth]{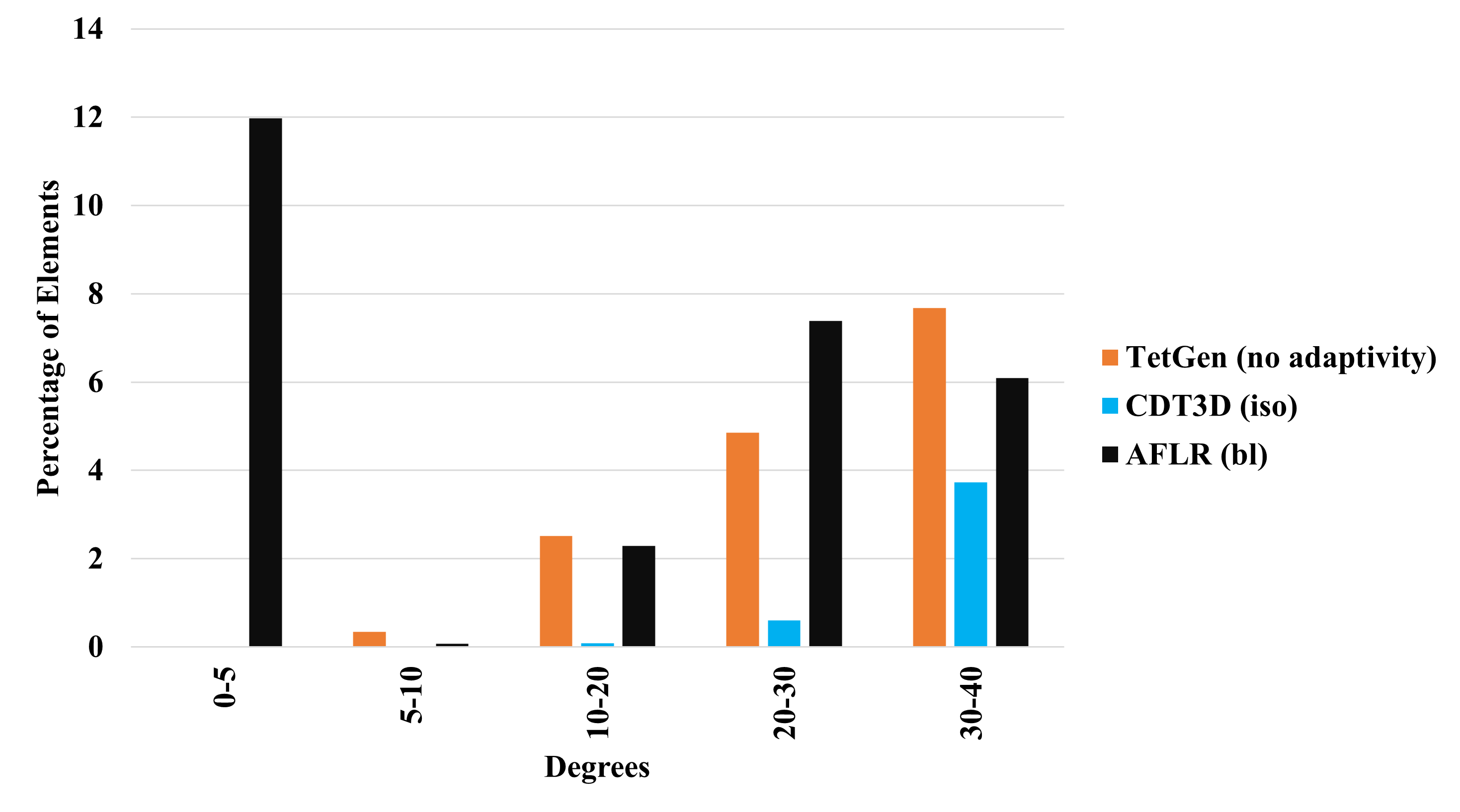}}
\subfigure[]{\label{fig:artery2_dihedral_large}\includegraphics[width=0.49\textwidth]{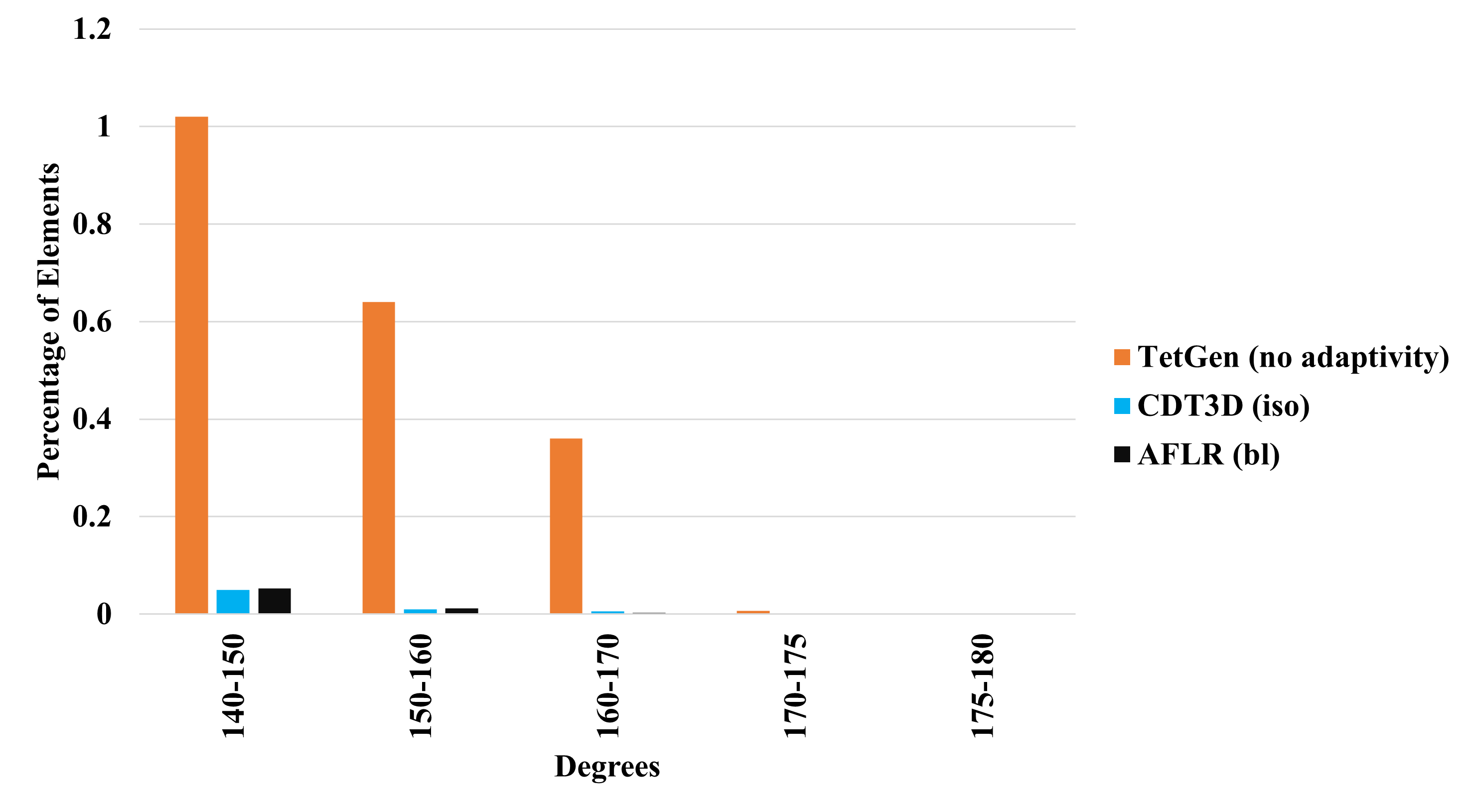}}
\caption{Quality statistics are shown comparing the dihedral angles of the isotropic meshes generated by TetGen, CDT3D, and AFLR (boundary layer) for the second (middle cerebral artery bifurcation) aneurysm case. (a) shows the full distribution of element dihedral angles. (b) shows the distribution of elements with dihedral angles between 0 and 40 degrees. (c) shows the distribution of elements with dihedral angles between 140 and 180 degrees.}
\label{fig:artery2_dihedrals}
\end{figure}

\begin{figure}[!htb]
\centering
\subfigure[]{\label{fig:arteries_MR}\includegraphics[width=0.49\textwidth]{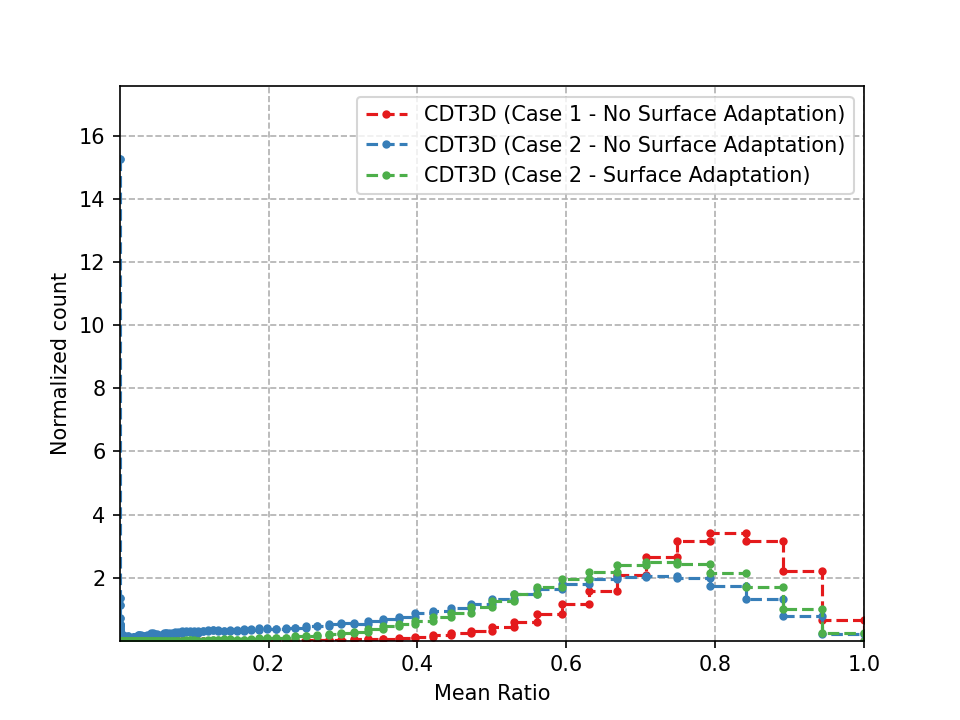}}
\subfigure[]{\label{fig:arteries_MR_log}\includegraphics[width=0.49\textwidth]{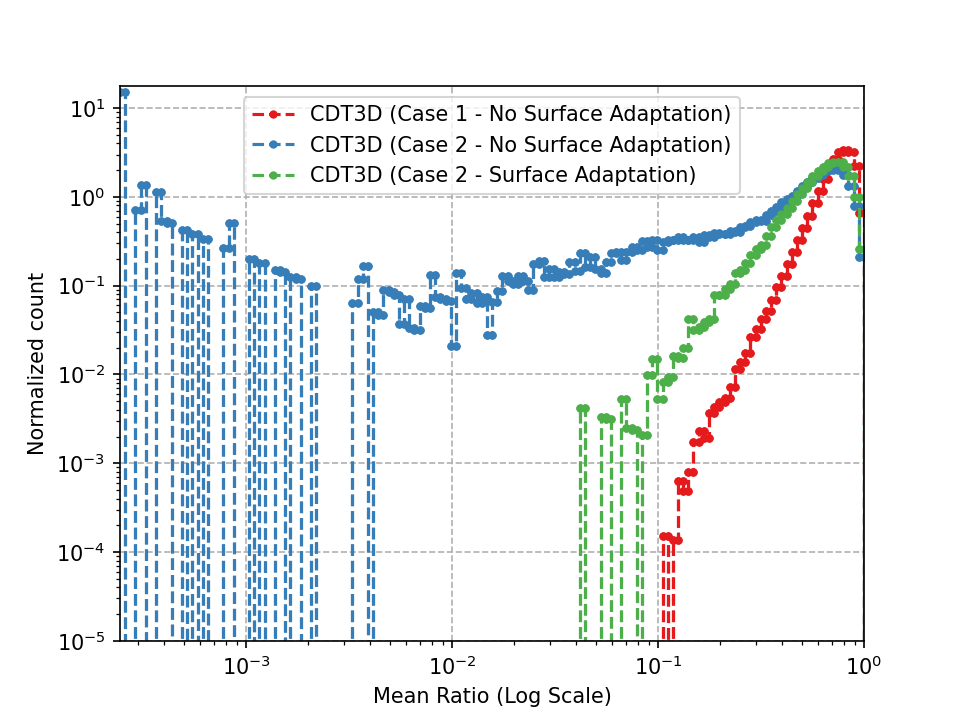}}
\subfigure[]{\label{fig:arteries_EL}\includegraphics[width=0.49\textwidth]{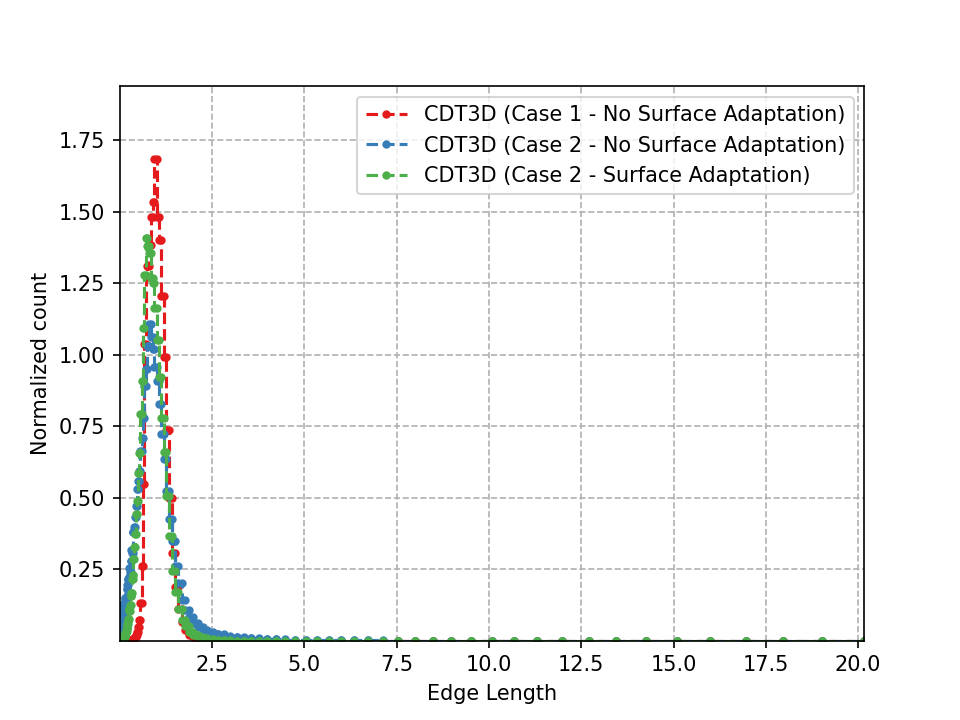}}
\subfigure[]{\label{fig:arteries_EL_log}\includegraphics[width=0.49\textwidth]{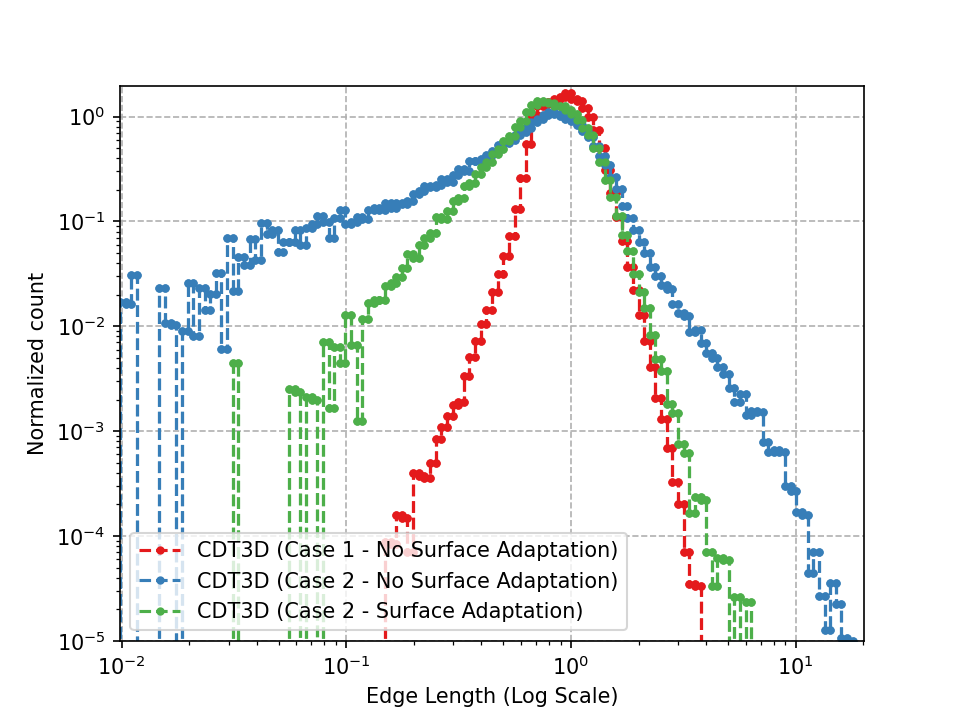}}
\caption{Shown are distributions of quality statistics for the mean ratio, (a) and (b), and edge lengths, (c) and (d), of elements within the anisotropic meshes generated by CDT3D for both aneurysm cases (in linear and logarithmic scales).}
\label{fig:arteries_CDT3D_MR_EL}
\end{figure}

Figure \ref{fig:artery1_dihedrals} compares the quality of the isotropic meshes generated by each method for case 1 and Figure \ref{fig:artery2_dihedrals} compares those generated for case 2. Table \ref{table_dihedrals} shows the smallest and largest dihedral angles found in the isotropic meshes generated by each method. The CDT3D meshes contain better quality compared to the TetGen and PODM meshes. This is particularly seen in (b) and (c) of Figure \ref{fig:artery1_dihedrals} and Figure \ref{fig:artery2_dihedrals}, which show that the TetGen and PODM meshes contain more elements with small and large angles. Although the AFLR boundary layer grids contain many elements with small angles, the grids have a good overall distribution of angle quality. Recall that qualitative results are examined with respect to their metric conformity for the adapted anisotropic meshes. Figure \ref{fig:arteries_CDT3D_MR_EL} shows that CDT3D maintains good quality when generating an anisotropic mesh for the first case. Although there are few elements with a mean ratio measure less than 0.1, the majority of elements have a high mean ratio and edge lengths close to 1. With regards to the second case, the anisotropic mesh generated by CDT3D exhibits poor quality when the surface of the input mesh is not adapted (containing more elements with a mean ratio less than 0.1). However, when CDT3D is permitted to adapt the surface, the final mesh quality is much better. When adapting the surface, the same parameters as those in Table \ref{table_input_parameters} are utilized with surface adaptation turned on. It should be noted that CDT3D converges and completes anisotropic adaptation in less time when surface adaptation is permitted (generating 90K tetrahedra in about 22 seconds when utilizing 20 cores) as opposed to its performance when restricting the method to preserve the input surface. The quality of the CDT3D mesh generated for the first case shows that CBC3D indeed generates a mesh of good quality when converting the first case image, given that surface adaptation is turned off for CDT3D and it still generates a final volume of good quality. In order to provide a volume that conforms to the boundary layer grid generated by AFLR, surface adaptation is turned off and CDT3D is therefore limited by the quality of the input surface mesh for the second aneurysm case. It should be noted that surface adaptation was attempted first, where the CDT3D-adapted surface was provided as input to AFLR. However, AFLR failed to process the CDT3D-adapted surface and generate a boundary layer grid. We also attempted surface refinement and volume generation using AFLR, to then subsequently use as a background mesh for CDT3D's anisotropic adaptation (which could potentially be merged with a boundary layer grid generated over the AFLR-refined surface). AFLR however fails to refine the input PLC surface of the second case. Nevertheless, the majority of elements in both anisotropic meshes generated by CDT3D for the second aneurysm case (with and without surface adaptation) have a high mean ratio and edge lengths close to 1.

\begin{table}[htb]\footnotesize
\caption{Smallest and largest dihedral angles in isotropic meshes generated by each method for both aneurysm cases.}
\centering
\begin{tabular}{lc|cc}
\hline
Methods & Angle & Aneurysm 1 & Aneurysm 2 \\
\hline
\multirow{2}{*}{TetGen (no adaptivity)} & Smallest & 1.67 & 2.71 \\
 & Largest & 176.73 & 171.3 \\
\multirow{2}{*}{TetGen (w/ adaptivity)} & Smallest & 3.99 & - \\
 & Largest & 171.66 & - \\
 \multirow{2}{*}{AFLR (bl)} & Smallest & 1.48 & 2.8 \\
 & Largest & 176.69 & 171.83 \\
\multirow{2}{*}{PODM (no vol. adaptivity)} & Smallest & 4.71 & - \\
 & Largest & 170.15 & - \\
\multirow{2}{*}{PODM (w/ vol. adaptivity)} & Smallest & 4.65 & - \\
 & Largest & 170.14 & - \\
\multirow{2}{*}{CDT3D (iso)} & Smallest & 3.99 & 3.54 \\
 & Largest & 165.25 & 172.93 \\
\hline
\end{tabular}
\label{table_dihedrals}
\end{table}

\subsection{Hierarchical Load Balancing} \label{HLB_results}
\begin{figure}[!htb]
\begin{center}
\includegraphics[width=1\textwidth]{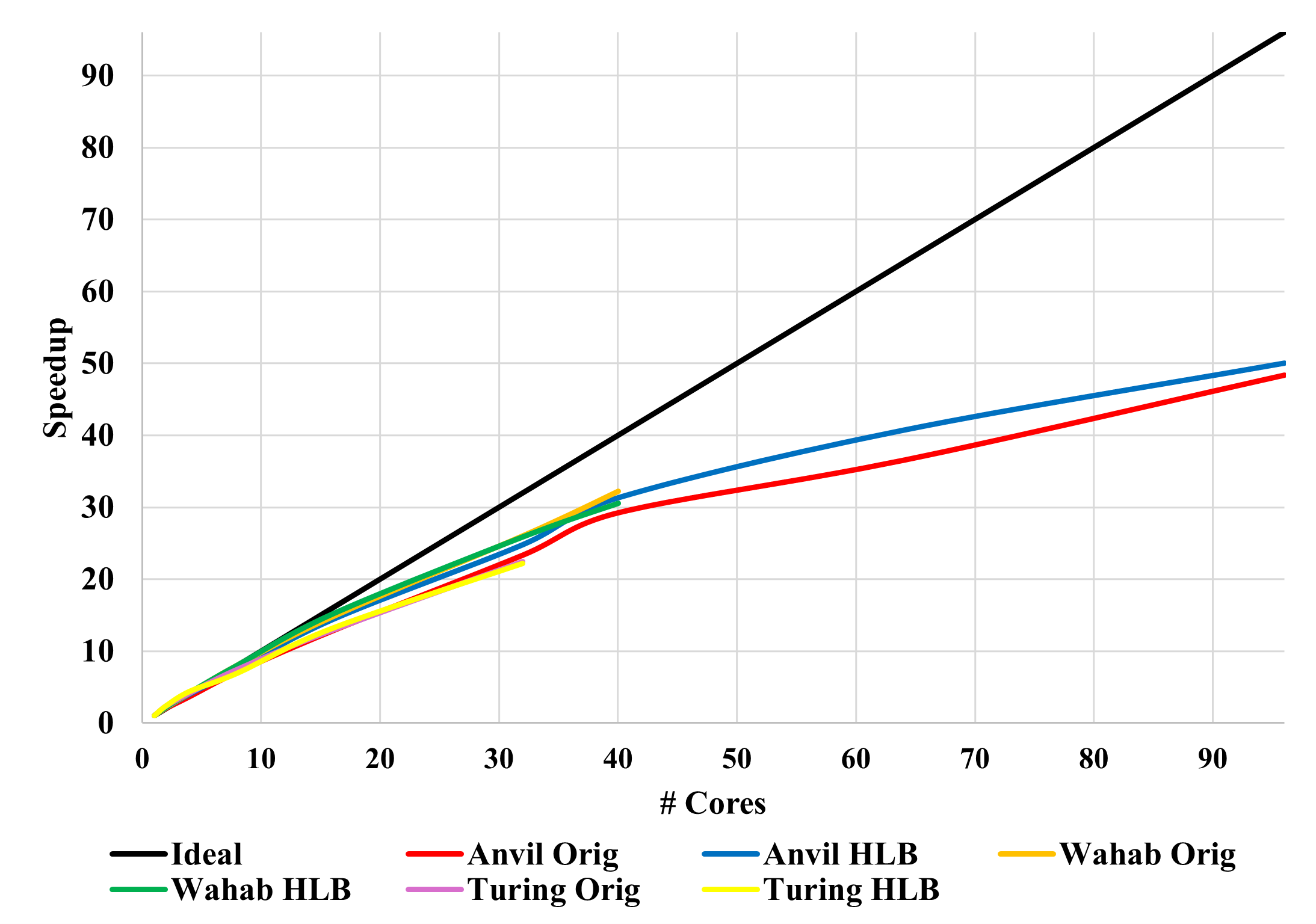}
\end{center}
\caption{A strong scaling speedup performance comparison is given of the original (Orig) and hierarchical load balancing (HLB) versions of CDT3D when generating approximately 100 million elements.}
\label{fig:HLB_Speedup}
\end{figure}

We evaluate the effectiveness of the hierarchical load balancing technique using much larger test cases (i.e., more work per thread) than those seen in the previous section. Figure \ref{fig:HLB_Speedup} compares the speedup achieved with and without the updated load balancing model on three supercomputers - Purude's Anvil, ODU's Wahab, and ODU's Turing supercomputers. The purpose is to gauge any potential improvements when utilizing smaller machines like the Turing supercomputer (with older specifications) versus utilizing larger machines with newer specifications (as described in section \ref{I2M_experimental_setup}). This strong scaling experiment is tested utilizing the CDT3D-generated isotropic volume mesh of the second aneurysm case from the previous section as input. The velocity-based metric tensor field is scaled from a complexity of about 20 thousand to a complexity of 50 million (generating approximately 100 million elements). We first observe that the hierarchical load balancing model gives a noticeable improvement on the Anvil supercomputer (up to about 5\% additional speedup when utilizing 64 cores). On the other hand, the new model does not improve performance much on Turing and when utilizing 40 cores on Wahab. CDT3D's original load balancing model performs slightly better than the new model on these smaller machines. A detailed analysis of the hierarchical load balancing model and its performance results is provided in section \ref{sec:discussion}.

\subsection{Optimized Local Reconnection} \label{local_reconnection_results}
\begin{figure}[!htb]
\centering
\subfigure[]{\label{fig:LR_opt_artery1_large_MR}\includegraphics[width=0.49\textwidth]{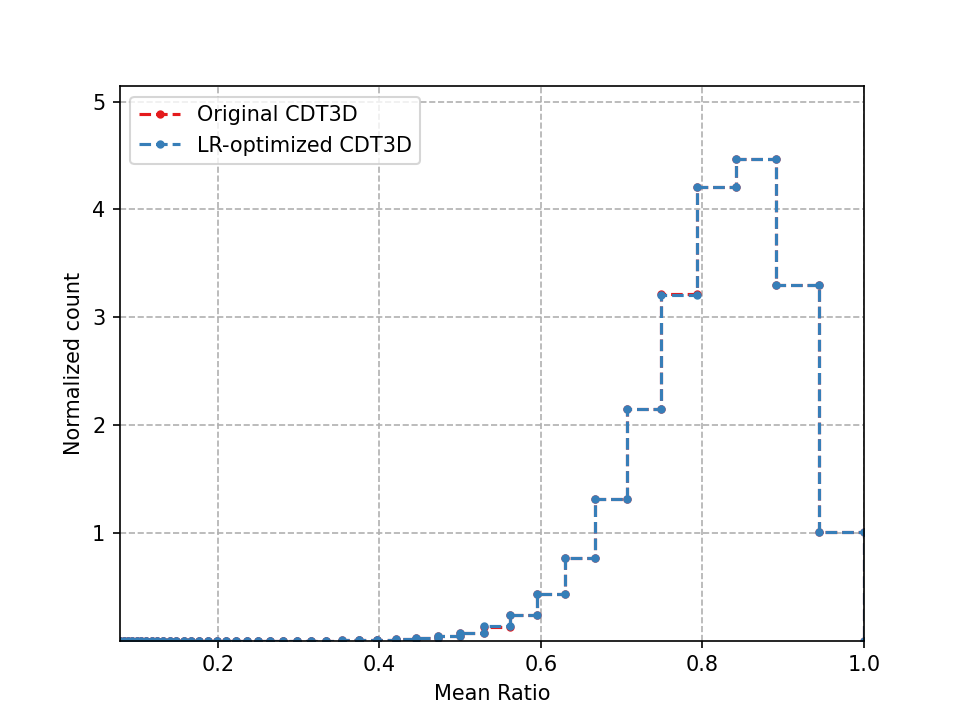}}
\subfigure[]{\label{fig:LR_opt_artery1_large_MR_log}\includegraphics[width=0.49\textwidth]{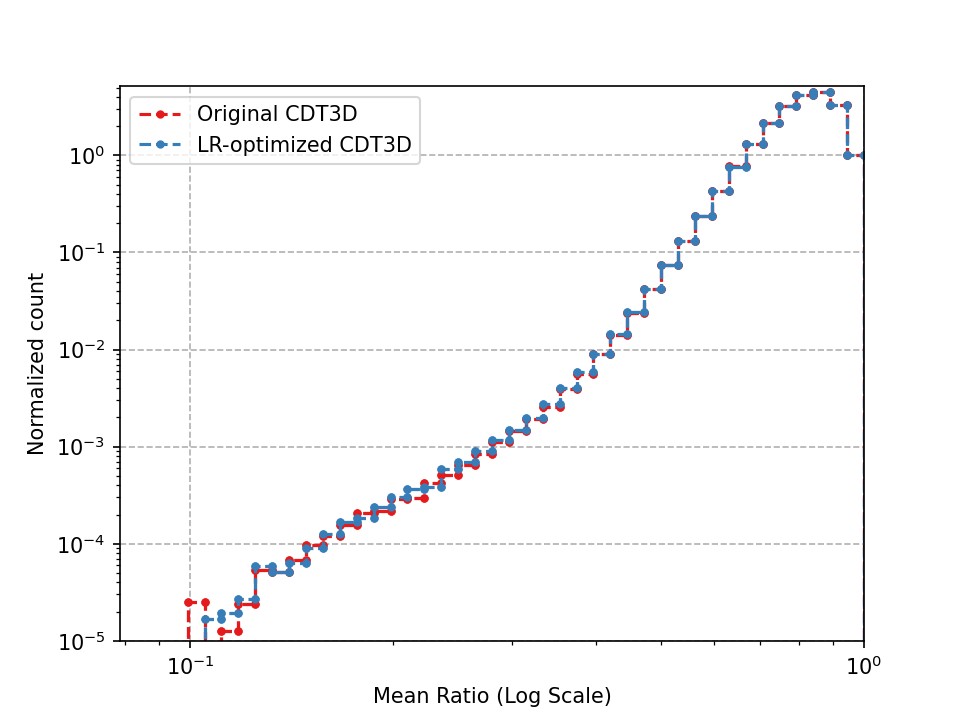}}
\subfigure[]{\label{fig:LR_opt_artery1_large_EL}\includegraphics[width=0.49\textwidth]{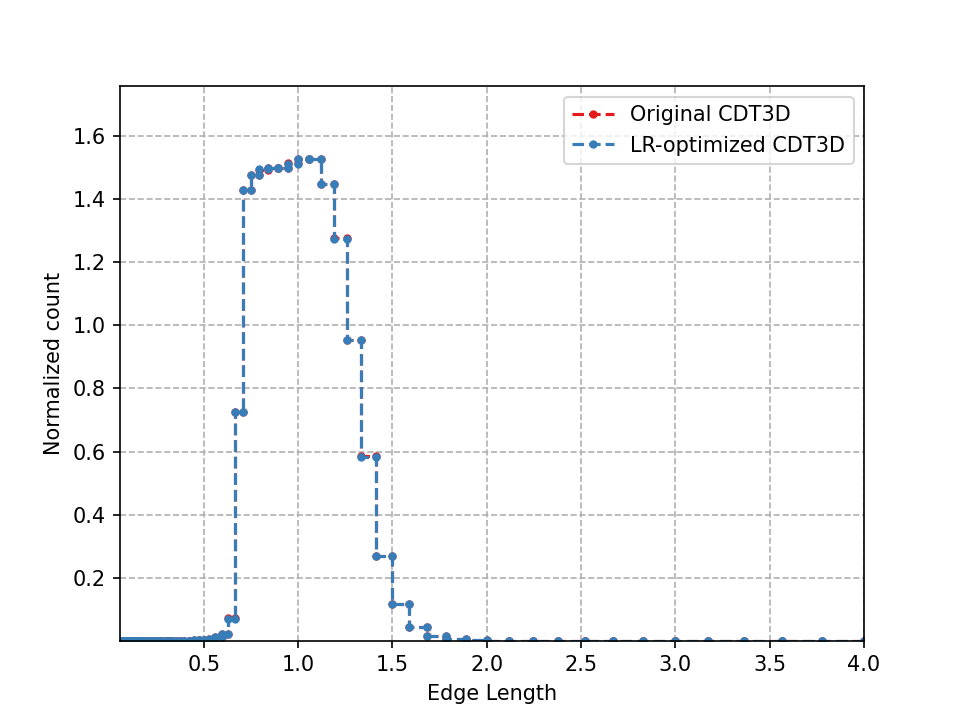}}
\subfigure[]{\label{fig:LR_opt_artery1_large_EL_log}\includegraphics[width=0.49\textwidth]{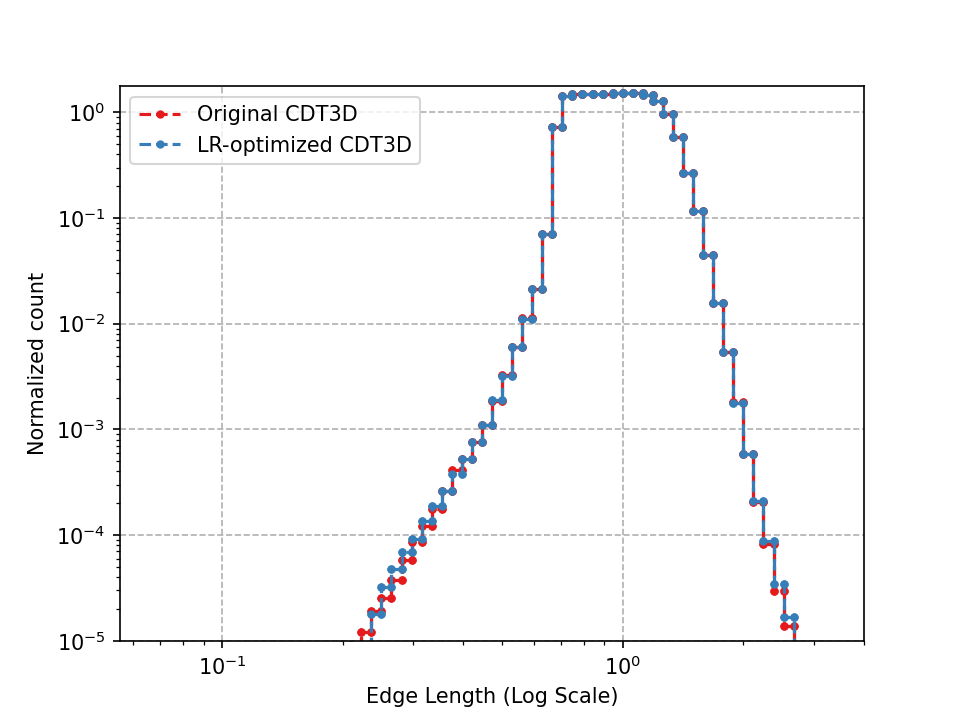}}
\caption{Shown are distributions of quality statistics for the mean ratio, (a) and (b), and edge lengths, (c) and (d), of elements within the meshes generated by the local reconnection-optimized CDT3D for the first (carotid cavernous) aneurysm case at 10 million complexity (in linear and logarithmic scales).}
\label{fig:artery1_large_CDT3D_LR_opt_MR_EL}
\end{figure}

\begin{figure}[!htb]
\centering
\subfigure[]{\label{fig:LR_opt_artery3_large_MR}\includegraphics[width=0.49\textwidth]{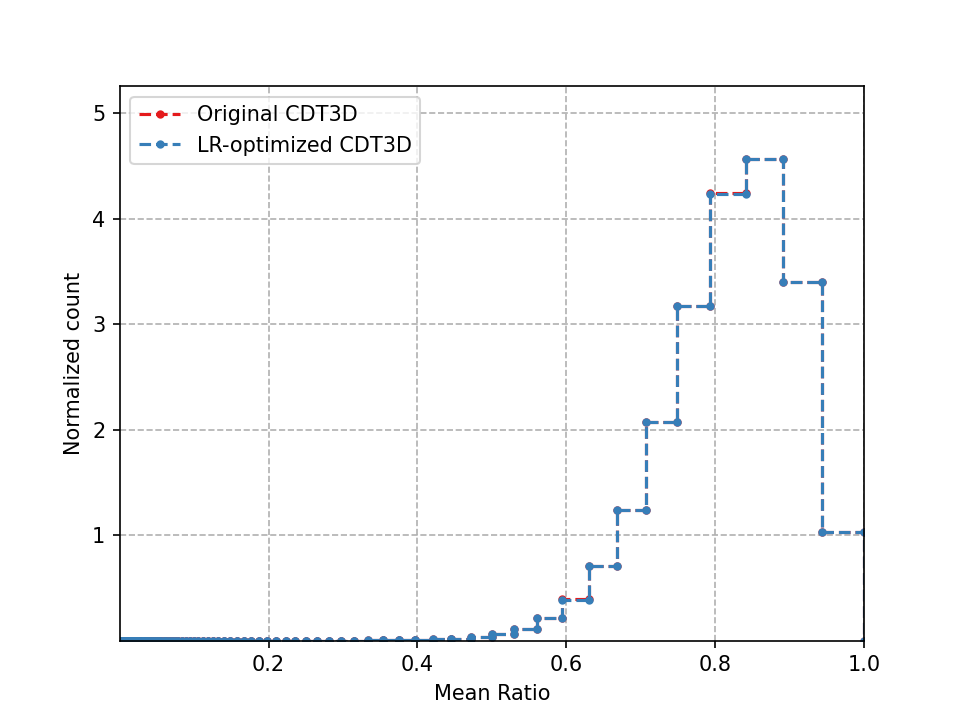}}
\subfigure[]{\label{fig:LR_opt_artery3_large_MR_log}\includegraphics[width=0.49\textwidth]{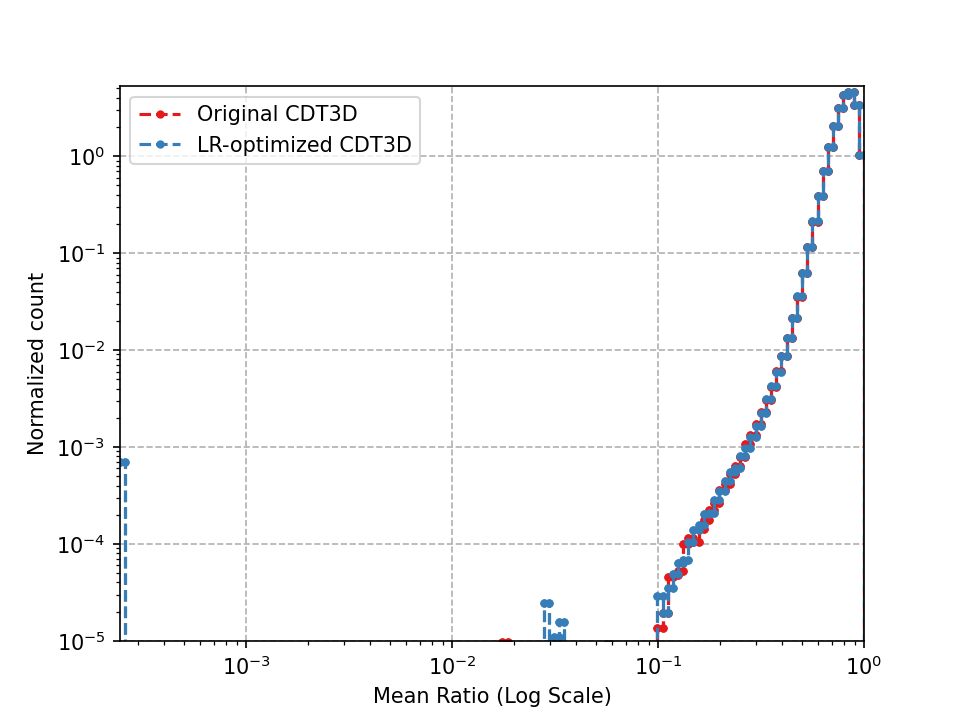}}
\subfigure[]{\label{fig:LR_opt_artery3_large_EL}\includegraphics[width=0.49\textwidth]{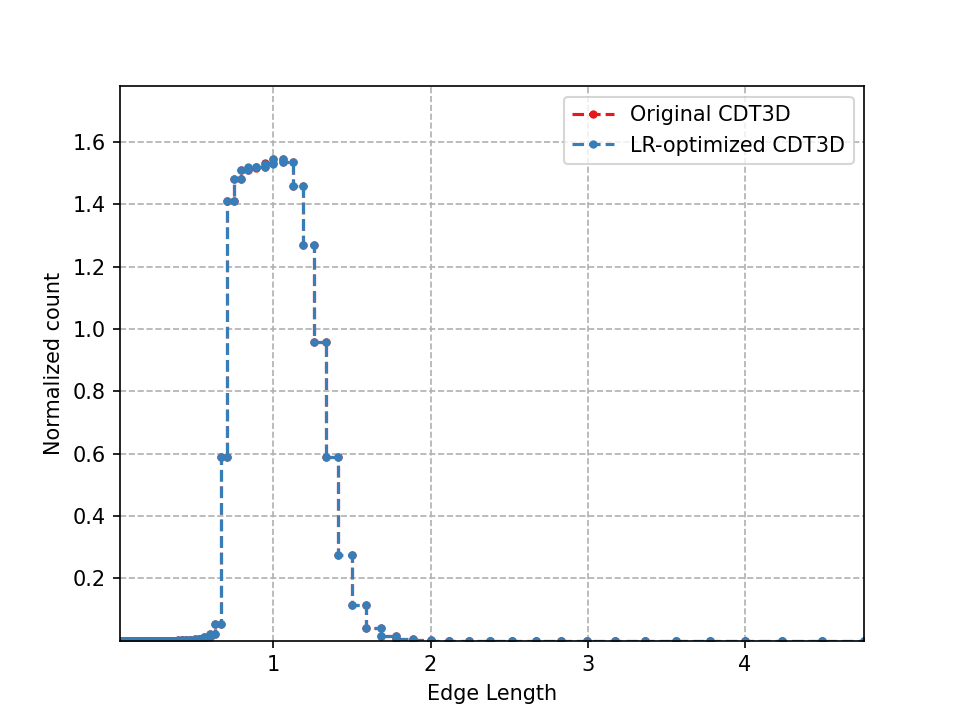}}
\subfigure[]{\label{fig:LR_opt_artery3_large_EL_log}\includegraphics[width=0.49\textwidth]{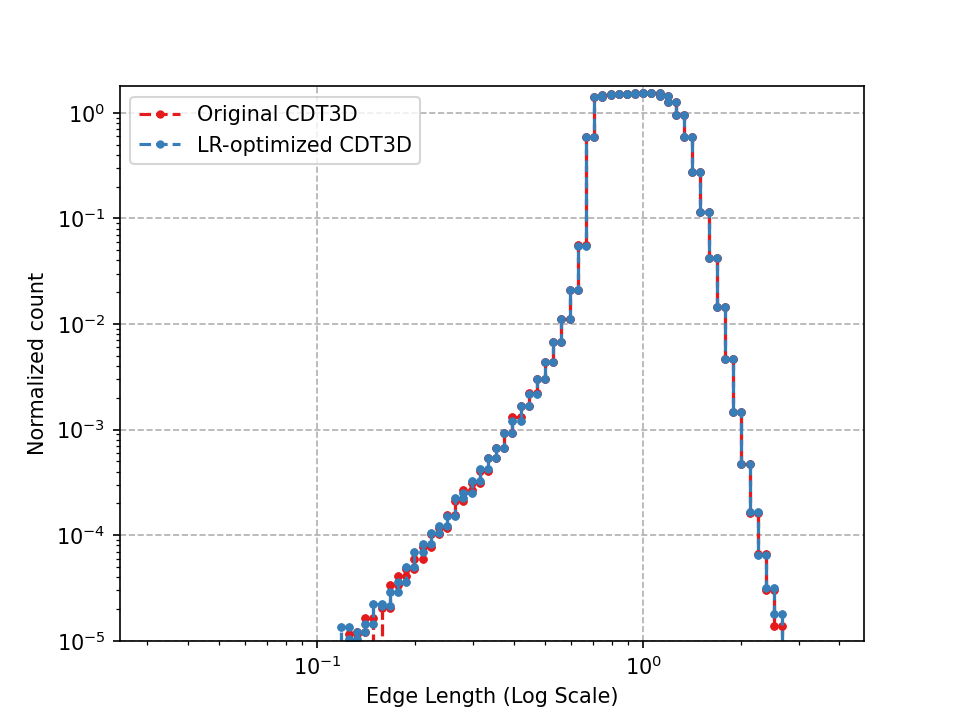}}
\caption{Shown are distributions of quality statistics for the mean ratio, (a) and (b), and edge lengths, (c) and (d), of elements within the meshes generated by the local reconnection-optimized CDT3D for the second (middle cerebral artery bifurcation) aneurysm case at 50 million complexity (in linear and logarithmic scales).}
\label{fig:artery3_large_CDT3D_LR_opt_MR_EL}
\end{figure}

We gauge the performance improvement given by the optimized local reconnection algorithm in CDT3D, discussed in section \ref{local_reconnection}, by testing the updated method with both aneurysm cases on the Wahab supercomputer. We scale the complexity of the metric tensor field for the carotid cavernous aneurysm from about 180 thousand to 10 million (generating approximately 50 million elements using the CBC3D-generated volume mesh as input). We again scale the complexity of the middle cerebral artery bifurcation aneurysm case to 50 million (generating approximately 100 million elements). Threads are pinned to CPU cores for each of these executions (using CDT3D's respective input parameter, shown in table \ref{table_input_parameters}). Table \ref{table_LR_artery1} shows the runtimes for the first (carotid cavernous) aneurysm case while table \ref{table_LR_artery2} shows the results for the second case. Recall that there are two phases of CDT3D's operation pipeline, within which the local reconnection operation is executed (during mesh adaptation, MA, and during final quality improvement, QI). For both aneurysm cases, the MA local reconnection time remains about the same. This is because the mesh adaptation phase (while it does improve quality) focuses on satisfying point spacing criteria in the metric field. This means that this operation is executed over elements until little to no new points are created/inserted into the grid. On the other hand, we see a significant performance improvement for the Quality Improvement phase's local reconnection operation (by about 3 times). When executed sequentially on Wahab, this saves 1 (for the first case) to 3 hours (for the second case) in end-to-end time compared to the original CDT3D. When utilizing 40 cores, there is still a 3x reduction in QI local reconnection time but there isn't as big of an improvement in end-to-end time. This is because the QI local reconnection time occupies only about 15\% of end-to-end runtime in the original CDT3D. This is reduced to about 5\% with its optimization. The results seen specifically for the QI local reconnection time shows the performance enhancement that such an algorithm can provide for local reconnection-based methods. Figures \ref{fig:artery1_large_CDT3D_LR_opt_MR_EL} and \ref{fig:artery3_large_CDT3D_LR_opt_MR_EL} show that the quality of the meshes generated by the local reconnection-optimized CDT3D is preserved, as they maintain comparable quality to that generated by the original CDT3D.

\begin{table}[htb]\small
\caption{A runtime comparison (approximately in \textbf{minutes}) of the original CDT3D and its optimized Local Reconnection (LR) version is shown when generating a mesh for the first (carotid cavernous) aneurysm at 10 million complexity on the Wahab supercomputer. `MA' means the Mesh Adaptation phase of CDT3D while `QI' means the Quality Improvement phase of CDT3D.}
\centering
\begin{tabular}{cc|ccccc}
\hline
\multirow{2}{*}{Method} & & \multicolumn{5}{c}{CPU Cores} \\
& & 1 & 5 & 10 & 20 & 40 \\
\hline
\multirow{3}{*}{Original CDT3D} & MA Local Reconnection & 232 & 41 & 21 & 11 & 7 \\
& QI Local Reconnection & 111 & 20 & 10 & 5 & 2.5 \\
& End-to-end & 501 & 94 & 49 & 25 & 15 \\
\hline
\multirow{3}{*}{Optimized CDT3D} & MA Local Reconnection & 228 & 40 & 21 & 11 & 7 \\
& QI Local Reconnection & 38 & 6 & 3 & 1.6 & 0.8 \\
& End-to-end & 421 & 79 & 42 & 22 & 13 \\
\hline
\end{tabular}
\label{table_LR_artery1}
\end{table}

\begin{table}[htb]\small
\caption{A runtime comparison (approximately in \textbf{minutes}) of the original CDT3D and its optimized Local Reconnection (LR) version is shown when generating a mesh for the second (middle cerebral artery bifurcation) aneurysm at 50 million complexity on the Wahab supercomputer. `MA' means the Mesh Adaptation phase of CDT3D while `QI' means the Quality Improvement phase of CDT3D.}
\centering
\begin{tabular}{cc|ccccc}
\hline
\multirow{2}{*}{Method} & & \multicolumn{5}{c}{CPU Cores} \\
& & 1 & 5 & 10 & 20 & 40 \\
\hline
\multirow{3}{*}{Original CDT3D} & MA Local Reconnection & 524 & 96 & 52 & 36 & 18 \\
& QI Local Reconnection & 249 & 42 & 21 & 10 & 5 \\
& End-to-end & 1097 & 210 & 110 & 67 & 33 \\
\hline
\multirow{3}{*}{Optimized CDT3D} & MA Local Reconnection & 507 & 95 & 51 & 33 & 18 \\
& QI Local Reconnection & 91 & 13 & 6 & 3 & 1.6 \\
& End-to-end & 916 & 181 & 95 & 57 & 30 \\
\hline
\end{tabular}
\label{table_LR_artery2}
\end{table}

\begin{figure}[!htb]
\centering
\subfigure[]{\label{fig:podm_large_dihedral_overall}\includegraphics[width=0.9\textwidth]{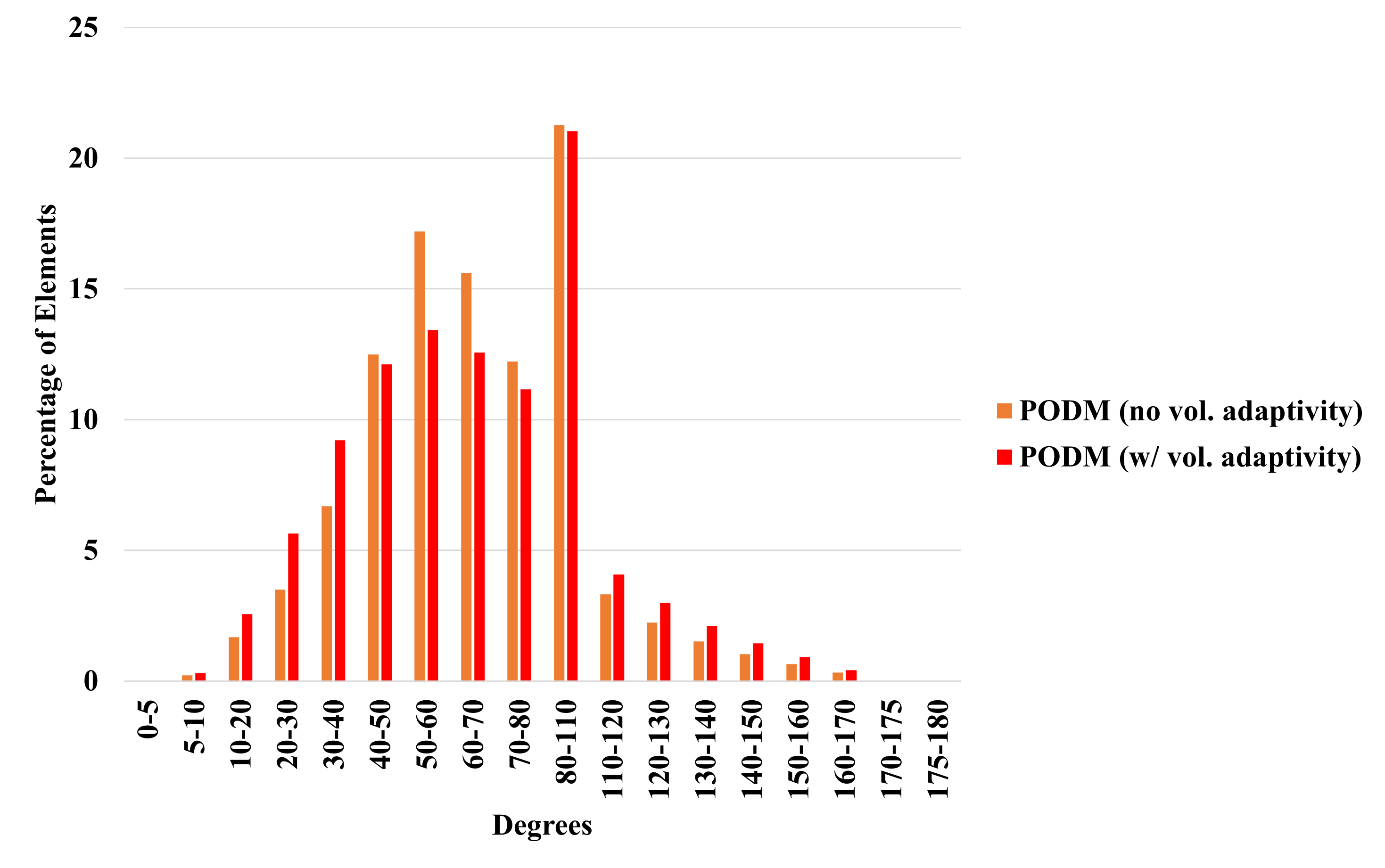}}
\subfigure[]{\label{fig:podm_large_dihedral_small}\includegraphics[width=0.49\textwidth]{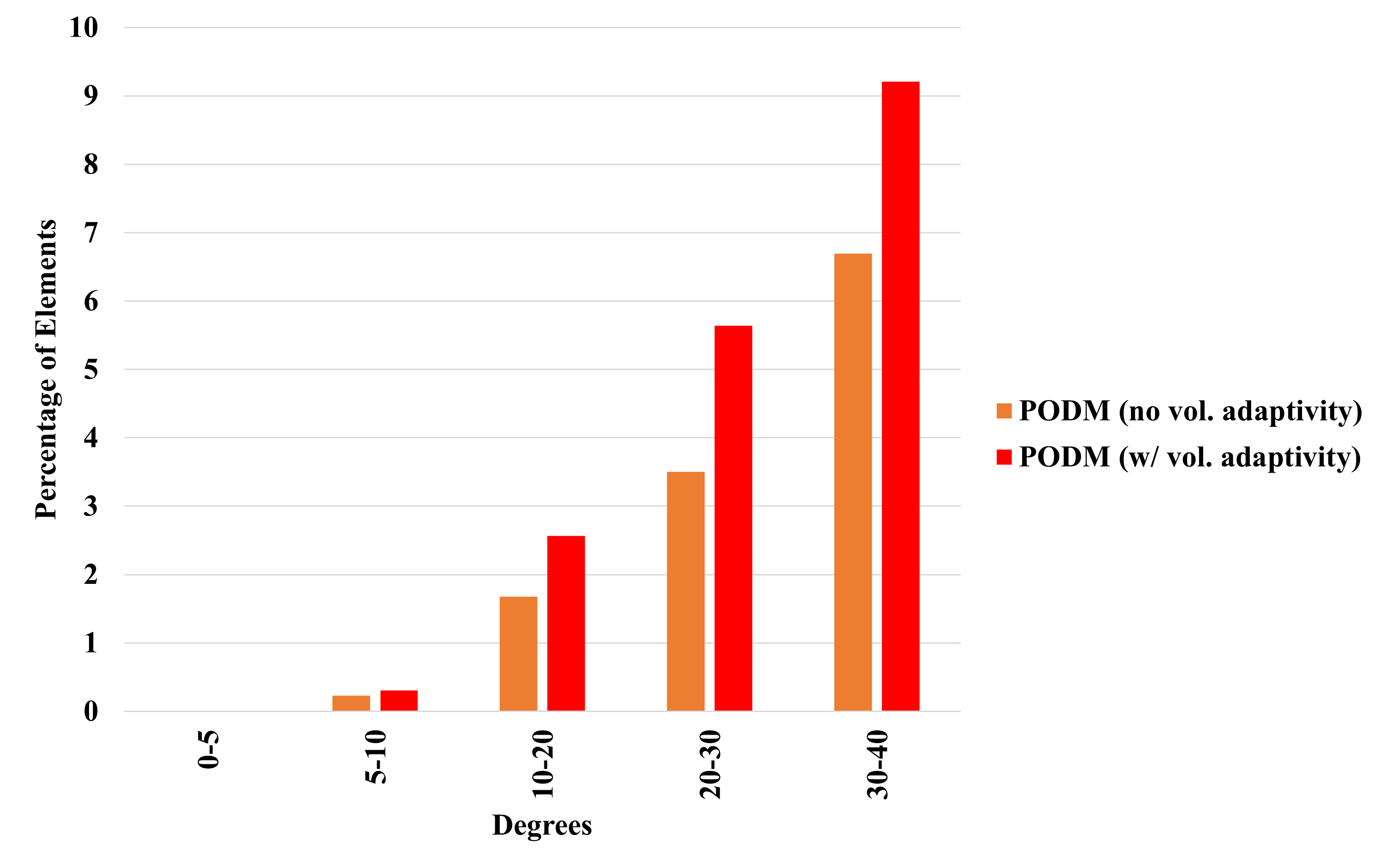}}
\subfigure[]{\label{fig:podm_large_dihedral_large}\includegraphics[width=0.49\textwidth]{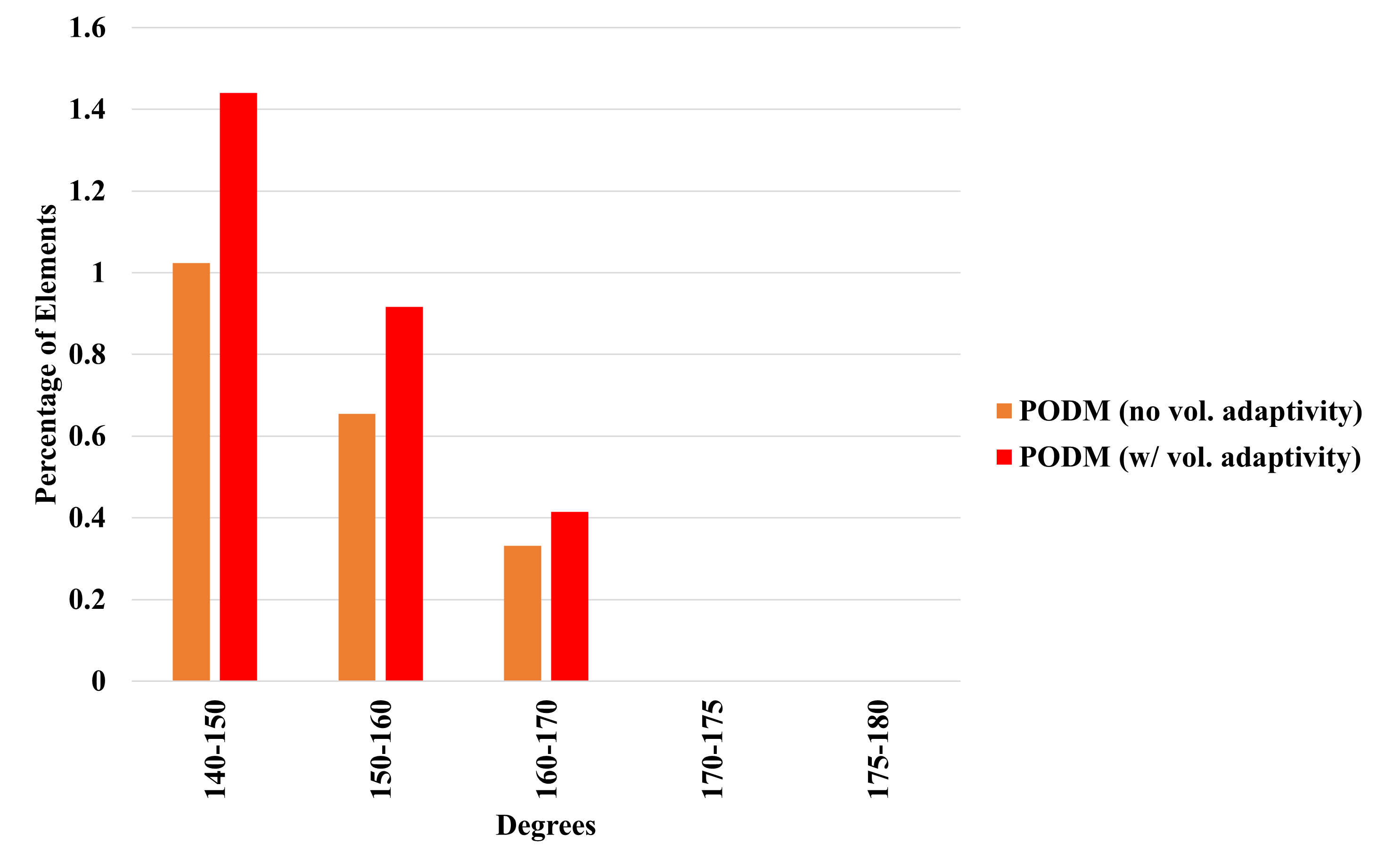}}
\caption{Quality statistics are shown comparing the dihedral angles of the large (approx. 50 million element) isotropic meshes generated by PODM for the first (carotid cavernous) aneurysm case. (a) shows the full distribution of element dihedral angles. (b) shows the distribution of elements with dihedral angles between 0 and 40 degrees. (c) shows the distribution of elements with dihedral angles between 140 and 180 degrees.}
\label{fig:podm_large_dihedrals}
\end{figure}

Finally, we show the runtimes when generating large meshes on the Anvil supercomputer by the optimized CDT3D (including both its hierarchical load balancing model and local reconnection optimization), PODM's Function Approximation (where its $\delta$ parameter is adjusted to 0.127 for volume adaptivity and 0.25 when executed without volume adaptivity to generate more elements for the first aneurysm case), and \textit{refine} \cite{ParkRefine}. Although it is originally designed and used heavily for aerospace cases \cite{TsolakisEvaluation}, \textit{refine} is ultimately a state-of-the-art open-source, distributed memory adaptive anisotropic mesh generation method. We gauge its capability to be utilized for a medical case. \textit{refine} utilizes a combination of edge split, edge collapse, and point smoothing operations to generate a unit grid that adheres to a given metric field. In our evaluation, the default parameters for \textit{refine} are utilized with the exception of the domain decomposition method. We use ParMETIS, as this resulted in the best runtimes for \textit{refine} among its domain decomposition options. Both CDT3D and \textit{refine} utilize the scaled metric tensor field(s) to generate large meshes. Threads are pinned to CPU cores for both CDT3D's and PODM's executions (using their respective input parameters). It should be noted that PODM's hierarchical load balancing model is not utilized because its source code was designed and written specifically for the Pittsburgh Supercomputer Center's Blacklight architecture. We instead utilize PODM's random work-stealing algorithm for its load balancing \cite{foteinos2014high}. Table \ref{table_artery1_large_performance} shows the runtimes for PODM and CDT3D when generating approximately 50 million elements for the first (carotid cavernous) aneurysm case. Table \ref{table_artery1_large_number_of_elements} shows the approximate mesh sizes generated by each method for the first case. PODM generates its isotropic meshes in less than a minute when using 96 cores while CDT3D generates its anisotropic meshes in about 5 minutes. PODM exhibits excellent end-user productivity by generating an adaptive mesh at a rate of about 1.17M elements/sec. The optimized CDT3D exhibits an adaptation rate of about 143K elements/sec for the first aneurysm case. Figure \ref{fig:podm_large_dihedrals} shows the dihedral angle distribution of the large meshes generated by PODM. The method generates elements with the same relative distribution of element quality as with the smaller cases. Table \ref{table_podm_large_dihedrals} shows that the minimum and maximum angles in both meshes remain about the same. PODM maintains good quality when performing volume adaptivity to generate large meshes as well as the small ones.

\begin{table}[htbp]\footnotesize
\caption{The time spent (approximately in \textbf{minutes}) when generating large-size meshes (about 50 million elements) for the first (carotid cavernous) aneurysm case on the Anvil supercomputer is shown.}
\centering
\begin{tabular}{l|cccccccc}
\hline
& \multicolumn{8}{c}{CPU Cores} \\
Methods & 1 & 2 & 4 & 8 & 16 & 32 & 64 & 96 \\\hline
PODM (no adaptivity) & 9.6 & 4.8 & 2.4 & 1.3 & 0.7 & 0.4 & 0.3 & 0.2 \\
PODM (w/ adaptivity) & 50 & 25 & 13 & 6.6 & 3.4 & 1.8 & 1.1 & 0.7 \\
CDT3D (Original) & 294 & 142 & 75 & 39 & 21 & 12 & 8 & 5.9 \\
CDT3D (Optimized) & 259 & 122 & 65 & 35 & 19 & 10 & 7.8 & 5.5 \\
\hline
\end{tabular}
\label{table_artery1_large_performance}
\end{table}

\begin{table}[htb]\footnotesize
\caption{Number of tetrahedra and points generated by each method when generating large meshes for the first (carotid cavernous) aneurysm case at 10 million complexity. `Tets' is short for tetrahedra. `M' means million.}
\centering
\begin{tabular}{l|cc}
\hline
Methods & Tets & Points \\
\hline
PODM (no adaptivity) & 48.2M & 8.6M \\
 \hline
PODM (w/ adaptivity) & 49.3M & 12.2M \\
\hline
CDT3D (Original) & 47.4M & 8.1M \\
\hline
CDT3D (Optimized) & 47.3M & 8.1M \\
\hline
\end{tabular}
\label{table_artery1_large_number_of_elements}
\end{table}

\begin{table}[htb]\footnotesize
\caption{Smallest and largest dihedral angles in the large isotropic meshes generated by PODM for the first (carotid cavernous) aneurysm case}
\centering
\begin{tabular}{l|cc}
\hline
Methods & Smallest & Largest\\
\hline
PODM (no vol. adaptivity) & 4.35 & 170.29 \\
PODM (w/ vol. adaptivity) & 4.25 & 170.26 \\
\hline
\end{tabular}
\label{table_podm_large_dihedrals}
\end{table}

\begin{figure}[!htb]
\centering
\subfigure[]{\label{fig:LR_opt_refine_artery3_large_MR}\includegraphics[width=0.49\textwidth]{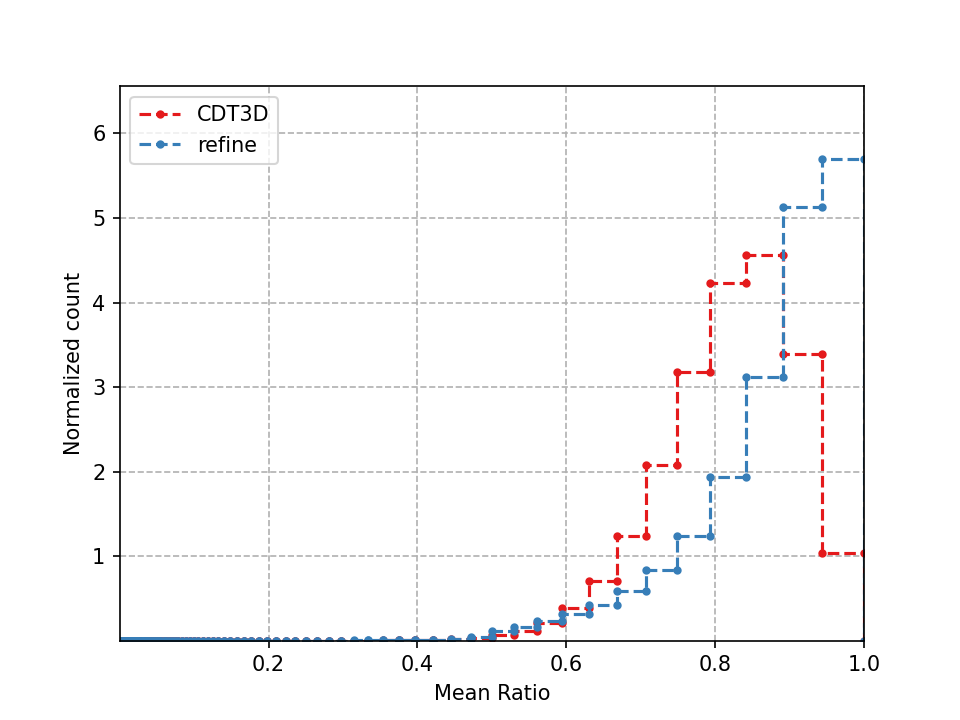}}
\subfigure[]{\label{fig:LR_opt_refine_artery3_large_MR_log}\includegraphics[width=0.49\textwidth]{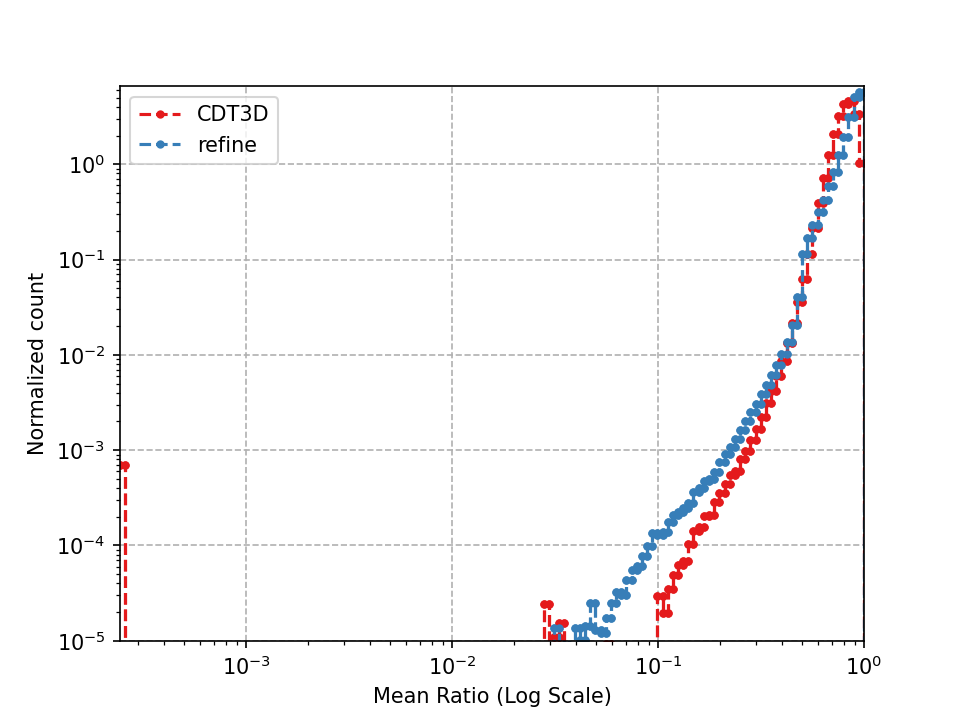}}
\subfigure[]{\label{fig:LR_opt_refine_artery3_large_EL}\includegraphics[width=0.49\textwidth]{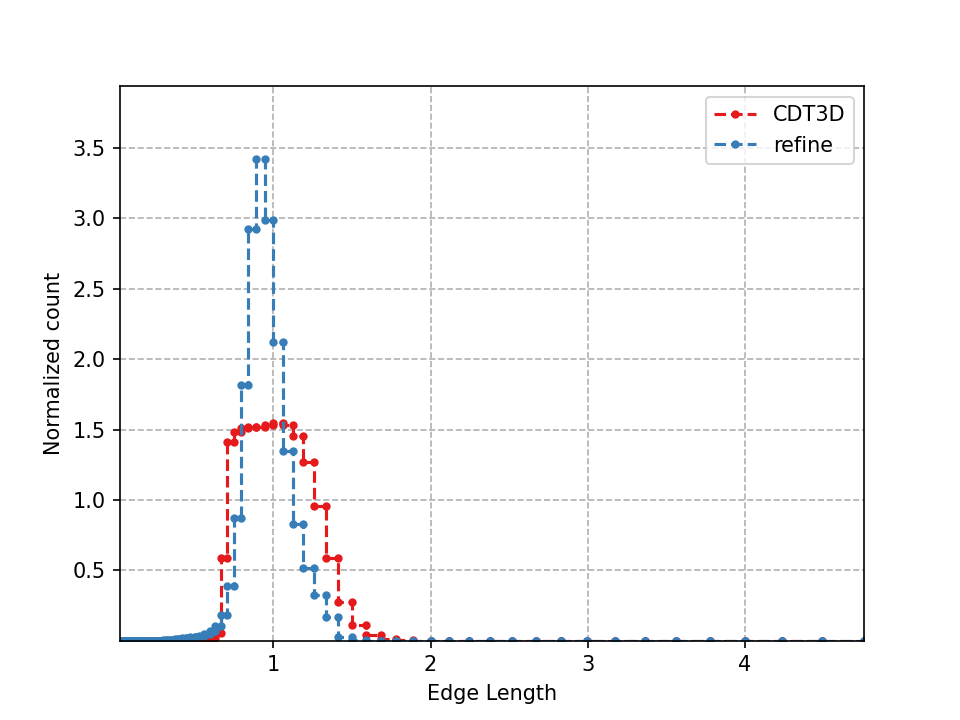}}
\subfigure[]{\label{fig:LR_opt_refine_artery3_large_EL_log}\includegraphics[width=0.49\textwidth]{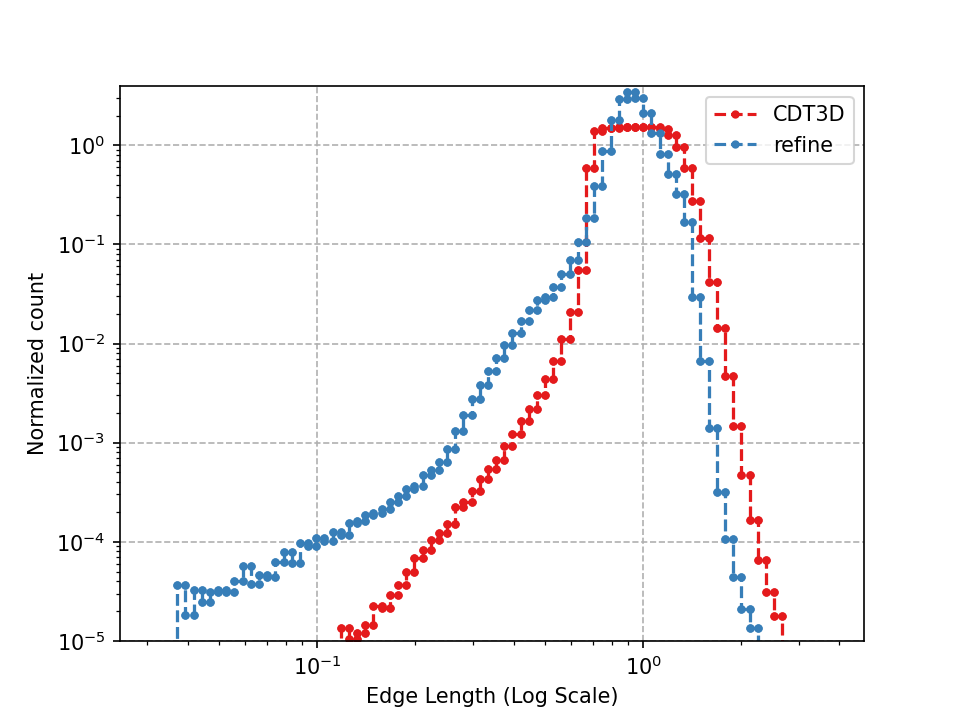}}
\caption{Shown are distributions of quality statistics for the mean ratio, (a) and (b), and edge lengths, (c) and (d), of elements within the meshes generated by the optimized CDT3D and \textit{refine} for the second (middle cerebral artery bifurcation) aneurysm case at 50 million complexity (in linear and logarithmic scales).}
\label{fig:artery3_large_CDT3D_LR_opt_refine_MR_EL}
\end{figure}

Table \ref{table_artery2_large_performance} shows the runtimes for CDT3D and \textit{refine} when generating approximately 100 million elements for the second (middle cerebral artery bifurcation) aneurysm case. CDT3D exhibits good scalability for both aneurysm cases. Some of the runtimes are not reported for \textit{refine} because the distributed method crashes due to running out of memory when executed with 16 cores or less. 
Table \ref{table_artery2_large_number_of_elements} shows the approximate mesh sizes generated by each method for the second case. Figure \ref{fig:artery3_large_CDT3D_LR_opt_refine_MR_EL} shows the quality of the meshes generated by both CDT3D (optimized) and \textit{refine} for the second case. While the CDT3D mesh exhibits good overall quality, the mesh generated by \textit{refine} exhibits better quality. However, this is at the cost of generating about 60 million additional elements compared to CDT3D to satisfy the same level of metric conformity. \textit{refine} also requires much more time, as it is about 9 times slower than CDT3D when using 96 cores. CDT3D also exhibits a faster adaptation rate of about 120K elements/sec when using 96 cores while \textit{refine} has an adaptation rate of about 23K elements/sec. Consequently, if the user is limited to utilizing a smaller machine (i.e., less CPU cores), \textit{refine} may be suitable for predictive simulations if time is not a primary concern and high mesh quality is desired for better solution accuracy with a solver. There is a tradeoff between quality and mesh adaptation speed when compared to CDT3D.

\begin{table}[htbp]\footnotesize
\caption{The time spent (approximately in \textbf{minutes}) when generating large-size meshes for the second (middle cerebral artery bifurcation) aneurysm case at 50 million complexity on the Anvil supercomputer is shown.}
\centering
\begin{tabular}{l|cccccccc}
\hline
& \multicolumn{8}{c}{CPU Cores} \\
Methods & 1 & 2 & 4 & 8 & 16 & 32 & 64 & 96 \\\hline
CDT3D (Original) & 658 & 325 & 178 & 93 & 51 & 28 & 18 & 13 \\
CDT3D (Optimized) & 554 & 265 & 146 & 81 & 43 & 24 & 16 & 12 \\
\textit{refine} & - & - & - & - & - & 399 & 131 & 113 \\
\hline
\end{tabular}
\label{table_artery2_large_performance}
\end{table}

\begin{table}[htb]\footnotesize
\caption{Number of tetrahedra and points generated by each method when generating large meshes for the second (middle cerebral artery bifurcation) aneurysm case at 50 million complexity. `Tets' is short for tetrahedra. `M' means million.}
\centering
\begin{tabular}{l|cc}
\hline
Methods & Tets & Points \\
\hline
CDT3D (Original) & 98.4M & 16.6M \\
\hline
CDT3D (Optimized) & 97.9M & 16.5M \\
\hline
\textit{refine} & 158.4M & 27.1M \\
\hline
\end{tabular}
\label{table_artery2_large_number_of_elements}
\end{table}

\section{Discussion} \label{sec:discussion}
Although they only generate isotropic meshes, TetGen and PODM exhibit good quality while generating fewer elements compared to CDT3D's anisotropic mesh generation (for the small test cases of our evaluation). Additionally, TetGen and PODM meet the real-time requirement for small-size problems (and large-size problems with PODM). CDT3D generates isotropic meshes with better quality but at the cost of generating more elements. This further showcases that local reconnection-based techniques may still offer better output mesh quality than Delaunay-based techniques. Depending on the accuracy needed for numerical simulations, this tradeoff between element count and quality may be acceptable, particularly when combined with CDT3D's anisotropy (seen in Figures \ref{fig:Mesh_Cuts_2_artery1} and \ref{fig:Mesh_Cuts_artery2}). Both the Build\_Sizing/Metric and CDT3D methods show good scalability on a single multicore node, particularly when generating an anisotropic mesh. Due to the fact that CDT3D is designed to handle much larger problems (such as that in Figure \ref{fig:artery2_images} or the scaled complexity benchmarks), its performance suffers from the overhead of parallelizing adaptation for such small meshes. Nevertheless, it also provides near real-time performance when utilizing a full multicore node (for anisotropic mesh adaptation) and when utilizing a single CPU core for isotropic mesh generation. CDT3D exhibits much faster performance when generating isotropic meshes from the PLCs of the first and second cases.

Several factors can be attributed to the observations made in section \ref{HLB_results} about the hierarchical load balancing model and its performance. The latency introduced by remote memory accesses between numa nodes can still become a performance bottleneck on the Anvil supercomputer. This is because the machine is much larger (8 numa nodes per shared memory node of 128 cores) than Turing (2 numa nodes per shared memory node of 32 cores) and Wahab (4 numa nodes per shared memory node of 40 cores). When executing the command \texttt{numactl -H} at the command line on a cc-NUMA high-performance computing machine, one can see the mappings of CPU cores to numa nodes and more specifically, the distance between numa nodes. Indeed, the distance between numa nodes on Anvil is greater for some nodes than the farthest distance between numa nodes on Wahab and Turing (given that they have fewer nodes and cores than Anvil). Consequently, the hierarchical load balancing model makes an impact on a larger machine like Anvil. 

To verify that the model indeed serves its purpose of reducing remote memory accesses between numa nodes, we also tested our method using Intel's vtune profiler\footnote{\url{https://www.intel.com/content/www/us/en/developer/tools/oneapi/vtune-profiler.html}}. Given that it requires special access to system files, we only tested our method with vtune on the Wahab supercomputer (using 40 cores) when adapting the same artery case at 50 million complexity. Table \ref{table_memory_accesses} shows several statistics given by vtune's memory access analysis of both versions of CDT3D. The local and remote access counts refer to the total local and remote accesses among the two sockets (recall that there are two sockets in a single shared memory node in these machines). The hierarchical load balancing model indeed reduces the number of remote memory accesses by almost 50\%. 

\begin{table}[htb]
\caption{The approximate memory accesses of the original CDT3D's load balancing model vs. the hierarchical load balancing (HLB) model is shown - 'M' means million.}
\centering
\begin{tabular}{c|cc}
\hline
& Original & HLB \\
\hline
Local Memory Accesses & 11.4M & 12.6M \\
Remote Memory Accesses & 50.7M & 26.4M \\
Remote Cache Accesses & 46.4M & 31.5M \\
\hline
\end{tabular}
\label{table_memory_accesses}
\end{table}

There is however an overhead from determining an idle thread with the shortest numa node distance every time a busy thread attempts to give work to another. The worst case time complexity for each thread (based on algorithm \ref{alg:nws}) is $O(t\log(b))$ where t is the number of idle threads and b is the number of buckets assigned to this particular thread. When utilizing 40 cores on Wahab, this accounts for approximately 6\% overhead. Coupled with the fact that there are few numa nodes on this machine (and consequently, a less expensive distance between its 4 numa nodes compared to Anvil's 8 numa nodes), the hierarchical model barely offers improvement in terms of speedup on Wahab. While offering an improvement on Anvil, it should be noted that the hierarchical load balancing model does not offer as much an improvement as seen in its implementation from \cite{foteinos2014high}. Simply put, this is attributed to the difference in the newer specifications of Anvil as opposed to the Pittsburgh Supercomputing Center's Blacklight machine utilized in \cite{foteinos2014high}. Some of these differences include (1) DDR4 memory supported by Anvil (6 memory channels per CPU, offering higher memory bandwidth and lower latency) as opposed to the Blacklight's DDR3 memory (4 memory channels per CPU) and (2) the Ultra Path Interconnect (UPI) links used by Anvil (up to 3 per CPU, 10.4 GT/s) compared to Blacklight's QuickPath Interconnect (QPI) links (up to 4 per CPU but at 6.4 GT/s). The newer interconnect links reduce inter-socket latency. Overall, the hierarchical load balancing model offers improvement in terms of memory access and runtime on larger machines, but this improvement is minimized on a newer, single shared memory node given its advancements in architecture compared to the older architecture utilized in \cite{foteinos2014high}. 

\section{Future Work} \label{i2m_future_work}
\begin{figure}[!htb]
\begin{center}
\includegraphics[width=1\textwidth]{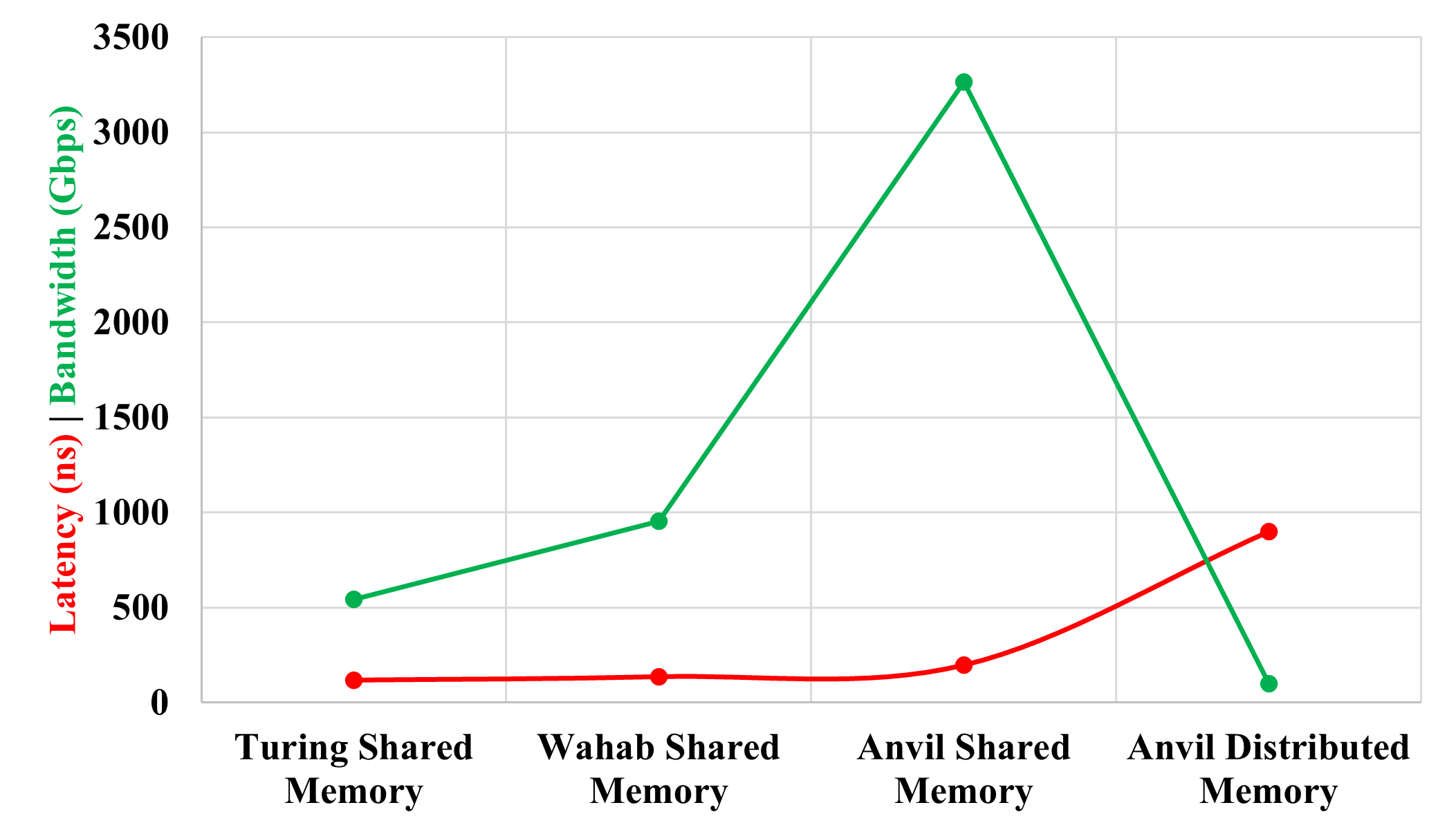}
\end{center}
\caption{Shown is the remote memory access latency (collected 
using Intel's lmbench program) and the bandwidth of each supercomputer \cite{TuringSpecs, WahabSpecs, WahabSpecs2, AnvilProcessorSpecs, AnvilSpecifications}.}
\label{fig:RMA_Latency}
\end{figure}

Figure \ref{fig:RMA_Latency} shows the latency of remote memory accesses and memory bandwidth on each of the supercomputers of our evaluation. We observe that latency increases as the size of a shared memory node increases (i.e., the number of cores). When transitioning from an Anvil shared memory node to Anvil's distributed memory environment, average message latency between numa nodes increases from about 190 nanoseconds to 900 nanoseconds between distributed memory nodes while bandwidth significantly decreases (memory bandwidth within a single node vs. the network interconnect bandwidth between distributed nodes). These remote memory access latencies were obtained by running Intel's version of the lmbench benchmark program\footnote{\url{https://github.com/intel/lmbench}} on a node within each supercomputer. Bandwidth specifications are also available for the CPUs utilized within Turing \cite{TuringSpecs}, Wahab \cite{WahabSpecs}, and Anvil \cite{AnvilProcessorSpecs, AnvilSpecifications}. 
Given that our hierarchical load balancing model grants additional speedup even when executed on a shared memory node (i.e., Anvil) that has low latency and high bandwidth, the potential benefits of such a model are expected to be exacerbated where limiting the distance of data migration becomes paramount due to the much higher latency and lower network interconnect bandwidth in distributed memory. Although focused on generating structured hexahedral meshes for explosion shock wave simulations, such a model is implemented in \cite{HierarchicalDistributed2025} using MPI-3 Remote Memory Access to create a virtual window for shared memory data transfer between distributed memory nodes. The method is able to utilize a static load balancing technique with low communication overhead because the neighborhood relationships of structured meshes remain fixed. Although overhead increases as the number of nodes utilized increases (due to implicit global synchronization caused by shared variables between nodes), the model indeed reduces average communication latency (by prioritizing local memory accesses) and improves the parallel efficiency of the method compared to when it utilizes traditional memory access techniques. This virtual window for shared memory access between distributed nodes could potentially be applied to anisotropic meshing; however, this is outside the scope of this paper and will be addressed in future work.

Given its performance when adapting the aneurysm geometries at scaled complexities, we predict that CDT3D would be well suited for processing a geometry generated from an image of a large vascular structure (such as in Figure \ref{fig:artery2_images}). This would likely require many more elements compared to the smaller test cases of this study (e.g., millions to potentially hundreds of millions of elements which in the future may be needed for aneurysm cases requiring even finer mesh resolution such as microvascular aneurysms). This will be explored in future work. It should be noted that a routine within CDT3D, which attempts to remove boundary slivers \cite{DrakopoulosCDT3D}, was turned off for the scaled second aneurysm case. This likely contributed to the low quality elements seen in Figure \ref{fig:artery3_large_CDT3D_LR_opt_MR_EL}. Both the original and optimized versions of the method sometimes crash when attempting to remove boundary slivers for this particular geometry at 50 million complexity. It should also be noted that although an Anvil shared memory node contains 128 cores, we tested CDT3D (and subsequently the other methods) with only 96 cores. CDT3D hangs during runtime when the method is executed with 128 cores. As these problems require further investigation, they will be remedied in future work. Nevertheless, the meshes generated by CDT3D for both large cases (and when utilizing up to 96 CPU cores) exhibit good overall quality. Additionally, slivers are indeed removed successfully when generating the smaller meshes (and when generating approximately 50 million elements for the carotid cavernous aneurysm case). 

The mesh deformation scheme employed by CBC3D in our I2M pipeline produces meshes with a smoothness of $C^0$. This smoothness is sufficient for aneurysm geometries such as those in our evaluation \cite{EwCCBC3D}; however, future work is required to improve the smoothness of geometries reconstructed from micro-CT images (i.e., some stent geometries). Our approach must be updated to provide a smoothness of $C^1$. Given CDT3D's ability to process CAD models, this approach remains to be tested within a medical numerical simulation that includes CAD models of stents (testing potential avenues of treatment for other aneurysm cases). 

Eventually, the tools utilized in the pipeline of the proposed approach should be integrated into a single software. It should be noted that conversions are required between the input and output files of the methods utilized. For example, several of the methods use the VTK library \cite{VTKWebsite}. Because they were each developed at different time periods, one method produces output utilizing a newer version of the library while another (which utilizes an older version) needs to read this as input. Although the conversion times for the small test cases in our evaluation are negligible, this discrepancy between the methods will likely negatively impact performance when the pipeline is tested on large cases. Integrating these methods into a single software will not only eliminate this problem, but will remove the complexity of compiling all codes separately and running them one after the other in the pipeline. Rather than separating image-to-mesh surface adaptation from anisotropic volume generation (and freezing the surface during this step), it would be more effective to perform both simultaneously when converting the segmented image (similar to PODM's approach \cite{foteinos2014high}). If integrated with a boundary layer generation method, this can help prevent the inhibition of mesh quality improvement (such as that shown for the second aneurysm case in Figure \ref{fig:arteries_CDT3D_MR_EL}). Finally, the sequential methods must be parallelized (or replaced with other parallel methods) in order to maximize potential performance, especially for large cases. As shown in \cite{GarnerThesis}, parallelizing AFLR is a non-trivial task (and similar challenges would be expected if one wished to generate boundary layer grids in parallel using the method). Parallelizing the sequential components of the pipeline is outside the scope of this paper however and should be addressed in future work.

Most importantly, the next step is to integrate both proposed methods into a numerical simulation. Such a study will focus on comparing the accuracy and error of critical simulation metrics (such as Wall Shear Stress and velocity fields) to determine if the PODM-generated isotropic meshes are sufficient for clinical-grade predictions, or if the CDT3D-generated anisotropic meshes are necessary to accurately predict an aneurysm's potential rupture. We will also measure the performance of each method within the context of the simulation's end-to-end execution time to validate their near real-time performance capabilities. This future work is essential to confirm the end-user productivity of these approaches.

\section{Conclusion} \label{sec:conclusion}
Presented are two techniques that are designed to help streamline the discretization of complex vascular geometries within the numerical modeling process. The first approach combines multiple software tools into a single pipeline to generate an adaptive anisotropic mesh from a segmented image of a carotid cavernous aneurysm, providing near real-time performance, good quality, fidelity, smoothness and robustness. The adaptive anisotropic mesh generation method is also tested using a surface mesh of a middle cerebral artery bifurcation aneurysm. The anisotropic method leverages a hierarchical load balancing model while also using an optimized local reconnection algorithm that is three times faster than its previous implementation from past studies \cite{EwCCDT3D}. The second presented technique focuses on generating adaptive isotropic meshes while utilizing a user-defined sizing function, also providing real-time performance, good quality, and fidelity. When utilizing 96 CPU cores on a single, multicore node on Purdue University's Anvil supercomputer \cite{Anvil2022}, the adaptive isotropic method (PODM) generates about 50 million elements in less than a minute while the adaptive anisotropic method (CDT3D) generates approximately the same amount in about 5 minutes. Such performance encourages an investment into a multi-core, shared memory machine that can potentially be utilized within a clinical setting (e.g., a single AMD processor with up to 96 CPU cores can cost about \$10K). Alternatively, end users may benefit from leveraging the concurrency offered by a virtual machine (such as Google's Compute Engine \cite{GoogleComputeEngine}) when integrating the presented methods into their numerical simulation workflows. 

Both techniques (converting medical images to either isotropic or anisotropic meshes) are presented as feasible options for end users, depending on the simulation and accuracy (i.e., tradeoff between speed and the mesh type) needed. For example, isotropic meshes of cerebral vasculature have been utilized within hemodynamic simulations to accurately predict flow conditions of stents post-deployment \cite{LargeMeshesStents2019}. This study notes the expensive computational cost of generating meshes with up to 200 million elements, which could potentially be generated quickly using a method like PODM. On the other hand, the study in \cite{AnisotropicCerebralAneurysms2013} shows that anisotropic meshes may offer more accurate results when including directional information (important for metrics like blood flow velocity). Just as the parallel anisotropic adaptation procedure (CDT3D) was tested within aerospace CFD simulations using CAD models \cite{EwCCDT3D}, the method is expected to provide accurate results for medical numerical simulations involving CAD-based stent models in near real-time when executed on large multicore cc-NUMA (shared memory) architectures. Having shown the feasibility of these approaches with regards to meeting medical image-to-mesh conversion requirements, the next step is to test both methods' integration into a medical numerical simulation (while utilizing an error-based sizing function for PODM) to validate the end-user productivity of these approaches. 

\section{Acknowledgements}
{
This research was sponsored by the Richard T. Cheng Endowment at Old Dominion University, SURA grant No. C2024-FEMT-011-01, and the National Institute of General Medical Sciences of the National Institutes of Health under Award Number 1T32GM140911. This study was also partially supported by a pilot award from the Long Island Network for Clinical and Translational Science. The content is solely the authors’ responsibility and does not necessarily represent the official views of the National Institutes of Health. Additionally, this work was supported by the Anvil research computing cluster of Purdue University and the Wahab and Turing clusters at Old Dominion University. The authors would like to thank Dr. Fotis Drakopoulos and Dr. Christos Tsolakis for helping with the tools utilized in the adaptive pipeline of our approach. Additionally, we'd like to thank Dr. David Marcum for sharing the AFLR software utilized in this study, and Evangelia Frastali for helping to determine some optimal parameters for the pipeline methods tested in our evaluation. We'd also like to thank Spiros Tsalikis for discussions regarding the sizing function.
}



\bibliographystyle{elsarticle-num} 
\bibliography{ltexpprt_references}






\end{document}